\documentclass[iop]{emulateapj}

\usepackage{natbib}
\usepackage[bookmarks,breaklinks]{hyperref}
   \hypersetup{
     colorlinks,
     citecolor=blue,
     filecolor=magenta,
     linkcolor=blue,
     urlcolor=cyan% color of external links
}
\usepackage{multirow}

\shorttitle{Star-formation in radio-AGN}
\shortauthors{Karouzos et al.}

\begin{document}


\title{A tale of two feedbacks: star-formation in the host galaxies of radio-AGN}


\author{M. Karouzos and M. Im}
\affil{CEOU - Astronomy Program, Department of Physics \& Astronomy, Seoul National University, Gwanak-gu, Seoul, Korea}
\email{mkarouzos@astro.snu.ac.kr}

\author{M. Trichas}
\affil{EADS Astrium Ltd., Gunnels Wood Road, Stevenage SG1 2AS, UK}

\author{A. Ruiz}
\affil{Istituto Nazionale di Astrofisica (INAF), Osservatorio Astronomico di Brera, via Brera 21, 20121 Milano, Italy}
\affil{Inter-University Centre for Astronomy and Astrophysics (IUCAA), Post Bag 4, Ganeshkhind, 411 007 Pune, India}

\author{T. Goto}
\affil{Dark Cosmology Centre, Niels Bohr Institute, University of Copenhagen, Juliane Maries Vej 30, 2100 Copenhagen, Denmark}

\author{M. Malkan}
\affil{Division of Astronomy and Astrophysics, 3-714 UCLA, LA CA 90095-1547, USA}

\author{Y. Jeon, J.H. Kim, H.M. Lee,  and S.J. Kim}
\affil{CEOU - Astronomy Program, Department of Physics \& Astronomy, Seoul National University, Gwanak-gu, Seoul, Korea}

\author{N. Oi, H. Matsuhara, T. Takagi, K. Murata, T. Wada, and K. Wada}
\affil{Institute of Space and Astronautical Science, JAXA, Yoshino-dai 3-1-1, Sagamihara, Kanagawa 229-8510, Japan}

\author{H. Shim}
\affil{Department of Earth Science Education, Kyungpook National University, Daegu 702-701, Republic of Korea}

\author{H. Hanami}
\affil{Physics Section, Faculty of Humanities, Iwate University, Ueda 3 chome, 18-34 Morioka, Morioka, Iwate 020-8550, Japan}

\author{S. Serjeant}
\affil{Department of Physics \& Astronomy, The Open University, Walton Hall, Milton Keynes, UK}

\author{G.J. White and C. Pearson}
\affil{Department of Physics \& Astronomy, The Open University, Walton Hall, Milton Keynes, UK}
\affil{RALSpace, Science and Technology Facilities Council, Rutherford Appleton Laboratory, Chilton, Didcot, OX11 0NL, UK}

\author{Y. Ohyama}
\affil{Academia Sinica, Institute of Astronomy and Astrophysics (ASIAA), No. 1, Sec. 4, Roosevelt Rd., Taipei 10617, Taiwan, R.O.C.}


%\altaffiltext{1}{Visiting Astronomer, Cerro Tololo Inter-American Observatory.
%CTIO is operated by AURA, Inc.\ under contract to the National Science


\begin{abstract}
Several lines of argument support the existence of a link between activity at the nuclei of galaxies, in the form of an accreting supermassive black hole, and star-formation activity in these galaxies. The exact nature of this link is still under debate. Radio jets have long been argued to be an ideal mechanism that allows AGN to interact with their host galaxy, either by depositing energy in the inter-stellar medium (ISM) and effectively suppressing or even quenching star-formation, or by driving shocks through the ISM, compressing molecular gas, and setting the stage for triggering star-formation. In this context, we are using a sample of radio sources in the North Ecliptic Pole (NEP) field to study the nature of the putative link between AGN activity and star-formation. This is done by means of spectral energy distribution (SED) fitting. We use the excellent spectral coverage of the AKARI infrared space telescope together with the rich ancillary data available in the NEP to build SEDs extending from UV to far-IR wavelengths. Through SED fitting we constrain both the AGN and host galaxy components. We find a significant AGN component in our sample of relatively faint radio-sources ($<$mJy), that increases in power with increasing radio-luminosity. At the highest radio-luminosities, the presence of powerful jets dominates the radio emission of these sources. A positive correlation is found between the luminosity of the AGN component and that of star-formation in the host galaxy, independent of the radio luminosity. By contrast, for a given redshift and AGN luminosity, we find that increasing radio-luminosity leads to a decrease in the specific star-formation rate. The most radio-loud AGN are found to lie on the main sequence of star-formation for their respective redshifts. For the first time, such a two-sided feedback process is seen in the same sample. We conclude that radio jets do suppress star-formation in their host galaxies but appear not to totally quench it. Our results therefore support the maintenance nature of ``radio-mode'' feedback from radio-AGN jets.
\end{abstract}



\section{Introduction}
\label{sec:intro}
The discovery of a number of scaling relations ($M_{BH}$-$\sigma$, \citealt{Ferrarese2000}, \citealt{Gebhardt2000}, \citealt{Merritt2001}, \citealt{Tremaine2002}; $M_{BH}$-$M_{bulge}$, \citealt{Magorrian1998},\citealt{McLure2002}; and $M_{BH}$-$L_{bulge}$, \citealt{Kormendy1995}, \citealt{Marconi2003}, also see \citealt{Kormendy2013}) connecting the properties of the nuclear regions of galaxies to their global, host galaxy, characteristics has led to a still ongoing debate about the possible physical processes that could give rise to this connection. This link has been argued to be either coincidental, or a result of stochastic processes like a series of random galactic mergers (e.g., \citealt{Jahnke2011}, \citealt{Graham2012}), or causal.

In the later family of scenarios, it is believed that there is a form of ``cross-talk'' between the central super-massive black hole (SMBH) and its surrounding host galaxy, also known as the co-evolution scenario. A robust, physical interpretation of that ``cross-talk'' is still widely debated. One candidate physical process that could be responsible for the co-evolution of galactic nuclei and their host galaxies is generally labelled as feedback and entails a large number of potentially different mechanisms that affect and effectively regulate the growth of either the host galaxy, or the central SMBH, or both. Such regulation has been shown to be important in large cosmological simulations (e.g., \citealt{Springel2005}, \citealt{Croton2006}).

One of the main ingredients of feedback models is the presence of outflows that inadvertently affect the host galaxy. The second point usually common across the spectrum of feedback realizations is the presence of activity within the nucleus of the galaxy (AGN), i.e., the ongoing accretion of matter onto the central SMBH of the galaxy (e.g., \citealt{Fabian2012}). Alternatively, the aforementioned outflows, or winds, can be launched by supernovae explosions (e.g., \citealt{Dekel1986}, \citealt{Efstathiou2000}, \citealt{Scannapieco2006}). Here we are interested in the role of AGN as regulators of star-formation and we are thus going to focus on AGN-related feedback mechanisms.

The study of star-formation in the host galaxies of AGN has always been of great interest, due to a number of factors among which (a) the coincidence of the peaks of cosmic star-formation and nuclear activity in terms of redshift (e.g., \citealt{Hopkins2006c}, \citealt{Kistler2013}, and \citealt{Richards2006}, \citealt{Aird2010}, respectively), (b) the apparently necessary role of a feedback mechanism to halt the growth of halos above a given mass limit (e.g., \citealt{Croton2006}, \citealt{Sijacki2009}), and (c) the scaling relations, which imply that supermassive black holes and their host galaxies grow in tandem, establishing yet another link between cosmic black hole accretion and star-formation. In the past, AGN have traditionally been identified and their host galaxies studied using either optical selection (through their blue colors, e.g., \citealt{Richards2002}, \citealt{Smith2005}, or their spectral emission lines, e.g., \citealt{Kauffmann2003}, \citealt{Richards2006}) or X-ray selection (e.g., \citealt{Comastri2003}, \citealt{Brandt2005}, \citealt{Treister2011}). To a lesser extent, active galaxies have been identified through their mid-IR excess emission (e.g., \citealt{Lacy2004}, \citealt{Stern2005}, \citealt{Messias2012}, \citealt{Hanami2012}) or their radio excess emission (e.g., \citealt{Drake2003}, \citealt{Draper2011}, \citealt{DelMoro2013}). Nevertheless, significant problems in studying the host galaxies of powerful AGN arise due to the fact that emission is severely contaminated by the AGN at most wavelengths (far-IR being potentially the least affected, e.g., \citealt{Hatziminaoglou2010}). 

Radio-loud active galaxies provide ideal star-formation quencher candidates as they are characterized by powerful, well-collimated outflows that are known to be able to deposit large quantities of mechanical energy in their surroundings (e.g., \citealt{McNamara2005}, \citealt{Wagner2011}). The role of radio-loud AGN and their jets in the evolution of galaxies, in particular with respect to star-formation, has been studied intensively. Results can be divided in two broad families, one advocating for radio-jets effectively inhibiting global star-formation (e.g., \citealp{Best2005,Best2007,Best2012}, \citealt{Croton2006}), with the other supporting a positive star-formation feedback (e.g., \citealt{Bicknell2000}, \citealt{Silk2010}, \citealt{Kalfountzou2012}, \citealt{Gaibler2012}, \citealt{Best2012}). Results coming out of the SDSS have shown that radio-selected, optically bright AGN (the SDSS spectroscopy optical magnitude limit is ~19.5 in the r-band) reside in massive galaxies and exhibit generally old stellar populations and weak signs of ongoing star-formation activity (e.g.,\citealt{Kauffmann2003}, \citealt{Best2005}). In a similar manner, a detailed analysis of some of the most luminous radio-AGN on the sky (from the 2 Jy and 3CRR samples) has found no strong star-formation signatures in most of them, with only a handful being detected as actively star-forming galaxies by Spitzer IRS spectra (\citealt{Dicken2012}). \citet{Herbert2010} studied the stellar populations of  a sample of intermediate redshift ($z\sim0.5$), high luminosity ($>10^{25}$ W Hz$^{-1}$ sr$^{-1}$) low-frequency radio-AGN, finding high-excitation, high-luminosity sources showing evidence for younger stellar populations compared to their low-excitation, low-luminosity counterparts (similar results where presented using Herschel data from \citealt{Hardcastle2013}). On a single source basis, individual radio-AGN have been found to exhibit suppressed star-formation, or to show none at all, with evidence for their molecular gas being ionized or even blown out by radio-jets (e.g., \citealt{Nesvadba2010}, \citealt{Morganti2013}). On the other side of the fence, there has also been a number of studies supporting jet-induced star-formation, both using statistical methods and also looking at individual sources (e.g., \citealt{Bicknell2000}, \citealt{Kalfountzou2012}, \citealt{Zinn2013}). The latter results can be explained by shocks driven by the radio-jets in the interstellar medium (ISM) that compress the ISM and eventually lead to enhanced star-formation efficiency. On the other hand, negative feedback could be explained either through the warming-up and ionization of the ISM and hence leading to less efficient star-formation (e.g., \citealt{Pawlik2009}, \citealt{Nesvadba2010}), or through direct expulsion of the molecular gas from the galaxy, effectively removing the ingredient for stars to form (e.g., \citealp{Nesvadba2006,Nesvadba2011}, \citealt{Morganti2010}).

It is therefore apparent, that although some form of feedback is needed to explain the observational results supporting co-evolution of central spheroids and their galaxies, much still remains unclear. Is there a direct, causal, connection between the growth of SMBH and their host galaxies? If so, when and how does this growth ``regulation'' happen? To that effect, we are interested in studying the star-formation properties of a radio-AGN sample and look for the putative link between the two. In the process, we also want to investigate whether this link is a positive or a negative one.

To do this, we investigate the broadband spectral energy distributions (SEDs) of a sample of radio sources and try to decouple, by means of SED template fitting, the AGN and star-formation components. In contrast to previous studies we use data from a very deep radio survey, therefore including faint radio-AGN which were missed by studies using surveys like FIRST of NVSS. As a result we are probing a broader range of radio-jet powers. In addition, we employ data from the AKARI Infrared Satellite that offer an excellent spectral coverage and allows a detailed treatment of the different emission components in the infrared (namely, cold dust heated by star-formation and warm/hot dust heated by the nuclear activity). In effect, and uniquely to our study, we can constrain both the star-formation and AGN components of these radio-sources simultaneously through our SED-fitting procedure. Combining the above, we are able to expand the study of AGN feedback in radio-AGN in terms of radio-power, AGN luminosity, and star-formation luminosity.

This paper is organized as follows: in Section \ref{sec:NEP} we describe the North Ecliptic Pole (NEP) field and the observations carried out in this field by the AKARI Infrared Satellite, in Section \ref{sec:sample} we introduce the sample of sources we use and describe the data available, Section \ref{sec:sed} contains the method used for the data analysis, while in Sections \ref{sec:agn}, \ref{sec:SFR}, and \ref{sec:sSFR} we present our results pertaining to the AGN content of the radio-sources, the absolute star-formation in the host galaxies of radio-AGN, and finally their specific star-formation and evidence of any feedback mechanism, respectively. In Section \ref{sec:discuss} we summarize our results, comparing them with other similar studies, discuss their importance, and go through possible caveats and shortcomings of our analysis. Finally in Section \ref{sec:conc} we list our conclusions. Throughout the paper we assume the cosmological parameters $H_{0}=71$ $km s^{-1} Mpc ^{-1}$, $\Omega_{M}=0.27$, and $\Omega_{\Lambda}=0.73$ (from the latest WMAP release; \citealt{Komatsu2011}).

\section{The AKARI North Ecliptic Pole (NEP) field}
\label{sec:NEP}
The AKARI space telescope (\citealt{Murakami2007}) was a satellite telescope launched by ISAS/JAXA in 2006 carrying two main instruments, the InfraRed Camera (IRC, \citealt{Onaka2007}) and the Far-Infrared Surveyor (FIS, \citealt{Kawada2007}). One of the AKARI legacy fields is the North Ecliptic Pole (NEP) field with its center at [RA,DEC]=[18:00:00,+66:33:38]. Using the IRC instrument aboard the AKARI, a two-tier survey was conducted in the NEP field, with the first, wide, tier covering a total area of $\sim5.4$ deg$^{2}$ (NEP-Wide, NEPW; \citealt{Kim2012}), while the second tier focused on a smaller area of $\sim0.67$ deg$^{2}$ (NEP-Deep, NEPD; \citealt{Takagi2012}). Of particular note is the spectral coverage of the IRC. With a total of 9 spectral bands, ranging from 2.4$\mu$m (N2 band) to 24$\mu$m (L24 band), the instrument continuously covers the whole wavelength range, including the prominent wavelength gap (9-20$\mu$m) that characterized observations with the IRAC and MIPS instruments aboard Spitzer.

\begin{figure}[htbp]
\begin{center}
\includegraphics[width=0.5\textwidth,angle=0]{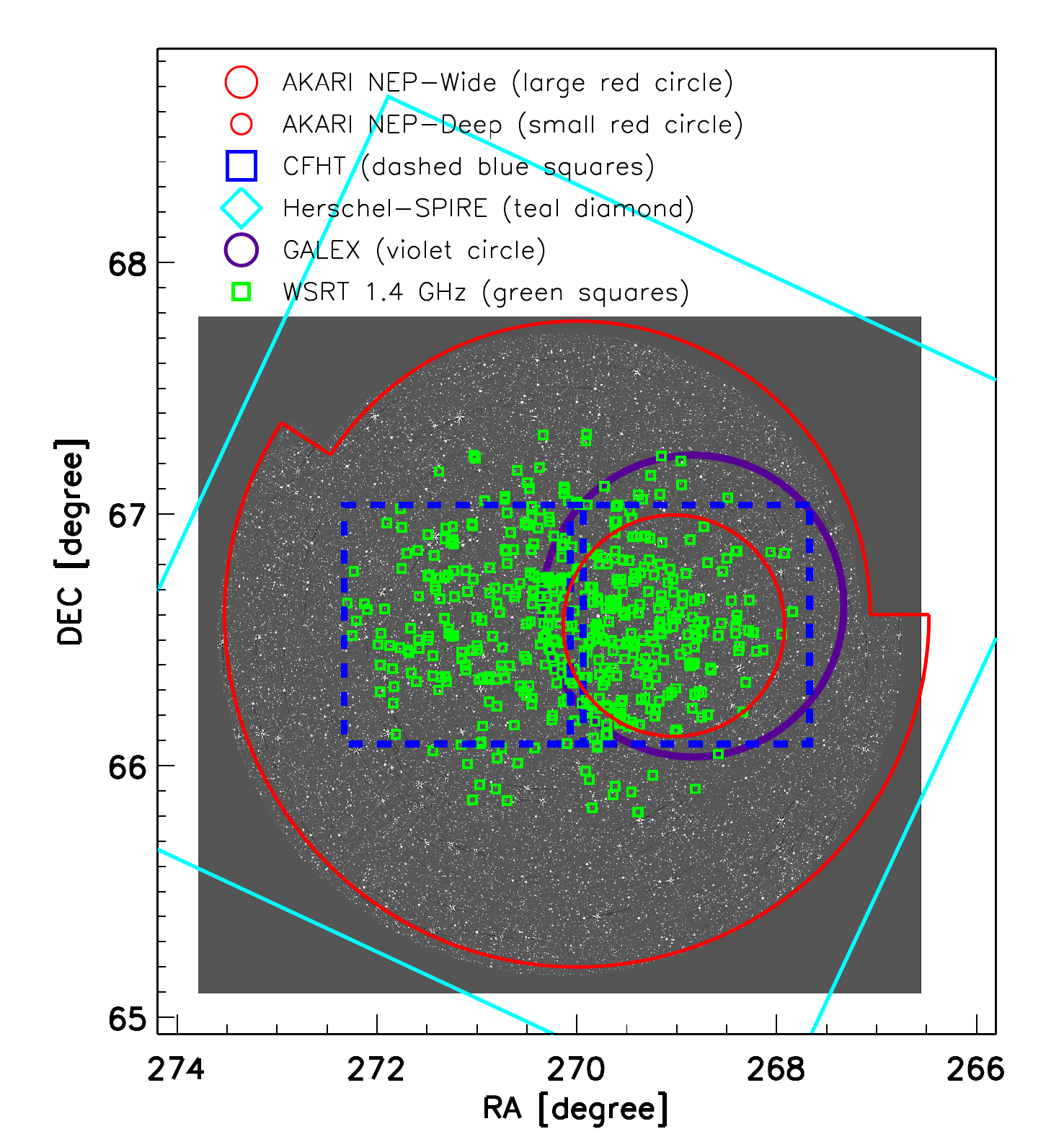}
\caption{Area coverage of the main AKARI surveys (NEP-Wide: large red circle, NEP-Deep: small red circle) together with some of the ancillary datasets (GALEX: violet circle, CFHT: dashed blue squares, Herschel-SPIRE: teal diamond). The WSRT 1.4 GHz catalogue sources are shown with green squares. The background image is from the AKARI IRC N2 band (2.4 $\mu$m).}
\label{fig:coverage}
\end{center}
\end{figure}

Several parallel efforts at different wavelengths have provided a rich ancillary dataset to complement the observations by AKARI. Of particular note and relevance to this work for the NEPW field are deep GALEX observations of a part of the NEPW field (Malkan, private communication), deep optical observations using the CFHT (\citealt{Hwang2007}) and SNUCAM (\citealt{Im2010}) on the 1.5m telescope of Maidanak observatory (\citealt{Jeon2010}), near-IR observations with the FLAMINGOS instrument on the 2.1m telescope at KPNO (Jeon et al., in prep.), and radio observations at 1.4 GHz with the WSRT (\citealt{White2010}). In 2012, the NEP field was also observed with the SPIRE (\citealt{Griffin2010}) instrument aboard the Herschel Space Observatory, at wavelengths between 250 and 500 $\mu$m. Finally, several spectroscopic campaigns have also taken place, with the NEPW mainly being covered by WIYN and MMT observations (\citealt{Shim2013} ), as well as deep observations with the DEIMOS multi-fiber spectrograph on the Keck telescope, centered on the NEPD field (Takagi et al., in prep.). In total more than 2000 spectroscopic redshifts are available in the NEPW field. In addition, for the NEPD field, deep optical observations by Subaru telescope and deep optical and near-IR observations by the CFHT telescope are also available (Oi et al. 2013, in prep.). In Fig. \ref{fig:coverage} the area coverage of the main AKARI surveys is shown together with some of the ancillary datasets (CFHT, GALEX, Herschel-SPIRE).

\section{Sample definition and data}
\label{sec:sample}
As we described in Section \ref{sec:intro}, we are interested in investigating the star-formation properties of radio-AGN and in particular looking for possible feedback signatures in the radio-AGN host galaxies. To do this we need to define first a sample of radio sources, identify AGN-dominated systems, decompose the contribution of star-formation in their broadband energy distributions, and finally investigate possible links to their nuclear properties. To do this, we employ the multi-wavelength data available in the NEP field (see Sect. \ref{sec:NEP}). For a summary of all the data used in this paper, see Tables \ref{tab:multi_NEPW} and \ref{tab:multi_NEPD}.

\begin{table}
\caption{Information about the photometric data available for the NEP-Wide field and in particular for the IR-radio cross-matched sample. Column (1) gives the name of the instrument, Cols. (2), (3), and (4) the waveband name, central wavelength, and sensitivity, respectively, and Col. (5) the number of sources in our cross-matched sample detected in that band. \\$\dagger$ For the SPIRE bands sensitivity is defined at the 3$\sigma$ level and it is given in units of mJy. For the WSRT the frequency in GHz and the sensitivity in $\mu$Jy beam$^{-1}$ are given instead of wavelength in $\mu$m and sensitivity in AB mag.}
\begin{center}
\begin{tabular}{|c|c|c|c|c|}
\hline
\multicolumn{5}{|c|}{N$_{NEPW}$=214} \\
\hline
Instrument						&	Band	&	Wavelength	&	Sensitivity		&	N	\\
								&			&	($\mu$m)		&	(AB mag)		&		\\
\hline
\multirow{2}{*}{GALEX}				&	FUV		&	0.15			&		26.1		&	22	\\
			 					&	NUV		&	0.23			&		26.7		&	40	\\
\hline
\multirow{5}{*}{MegaCam (CFHT)}		&	u*		&	0.37			&		26.0		&	136	\\
								&	g		&	0.49			&		26.1		&	146	\\
								&	r		&	0.63			&		25.6		&	149	\\
								&	i 		&	0.78			&		24.7		&	149	\\
								&	z 		&	1.17			&		23.7		&	137	\\
\hline
\multirow{3}{*}{Maidanak}				&	B		&	0.43			&		23.4		&	14	\\
								&	R		&	0.66			&		23.1		&	23	\\
								&	I		&	0.80			&		22.3		&	22	\\
\hline
\multirow{2}{*}{\scriptsize{FLAMINGOS} (KPNO)}	&	J		&	1.23			&		21.4		&	146	\\
								&	H		&	1.67			&		21.4		&	146	\\
\hline
\multirow{9}{*}{IRC (AKARI)}			&	N2		&	2.4			&		20.9		&	182	\\
								&	N3		&	3.2			&		21.1		&	208	\\
								&	N4		&	4.1			&		21.1		&	200	\\
								&	S7		&	7.0			&		19.5		&	71	\\
								&	S9W		&	9.0			&		19.3		&	77	\\
								&	S11		&	11.0			&		19.0		&	67	\\
								&	L15		&	15.0			&		18.6		&	53	\\
								&	L18W	&	18.0			&		18.7		&	56	\\
								&	L24		&	24			&		18.0		&	40	\\
\hline
\multirow{4}{*}{WISE}				&	W1		&	3.4			&		19.9		&	182	\\
								&	W2		&	4.6			&		19.1		&	177	\\
								&	W3		&	12			&		16.7		&	123	\\
								&	W4		&	22			&		14.6		&	79	\\
\hline
\multirow{3}{*}{SPIRE (Herschel)}		&	SP1		&	250			&		27.0		&	39	\\
								&	SP2		&	350			&		22.5		&	38	\\
								&	SP3		&	500			&		32.4		&	17	\\
\hline
WSRT$\dagger$					&	L		&	1.4 			&		21.0		&	214	\\
\hline
%\cline{2-5}
\end{tabular}
\end{center}
\label{tab:multi_NEPW}
\end{table}%

\begin{table}
\caption{Same as Table \ref{tab:multi_NEPW} but for the NEPD.}
\begin{center}
\begin{tabular}{|c|c|c|c|c|}
\hline
\multicolumn{5}{|c|}{N$_{NEPD}$=95}\\
\hline
Instrument						&	Band	&	Wavelength	&	Sensitivity		&	N	\\
								&			&	($\mu$m)		&	(AB mag)		&		\\
\hline
\multirow{2}{*}{GALEX}				&	FUV		&	0.15			&		26.1		&	33	\\
			 					&	NUV		&	0.23			&		26.7		&	55	\\
\hline
\multirow{5}{*}{MegaCam (CFHT)}		&	u*		&	0.37			&		26.0		&	70	\\
								&	g		&	0.49			&		27.1		&	84	\\
								&	r		&	0.63			&		26.3		&	85	\\
								&	i 		&	0.78			&		25.6		&	91	\\
								&	z 		&	1.17			&		24.5		&	95	\\
\hline
\multirow{4}{*}{\scriptsize{SuprimeCam (Subaru)}}			&	B		&	0.43			&		28.4		&	36	\\
								&	V		&	0.54			&		28.0		&	36	\\
								&	R		&	0.65			&		27.4		&	36	\\
								&	I		&	0.80			&		27.0		&	36	\\
								&	z		&				&		26.2		&	36	\\
\hline
\multirow{3}{*}{WIRCam (CFHT)}		&	Y		&	1.02			&		23.9		&	73	\\
								&	J		&	1.25			&		23.5		&	74	\\
								&	K		&	2.15			&		23.0		&	79	\\
\hline
\multirow{9}{*}{IRC (AKARI)}			&	N2		&	2.4			&		21.4		&	88	\\
								&	N3		&	3.2			&		21.7		&	92	\\
								&	N4		&	4.1			&		22.1		&	89	\\
								&	S7		&	7.0			&		19.7		&	88	\\
								&	S9W		&	9.0			&		19.5		&	83	\\
								&	S11		&	11.0			&		19.3		&	87	\\
								&	L15		&	15.0			&		18.7		&	71	\\
								&	L18W	&	18.0			&		18.7		&	72	\\
								&	L24		&	24			&		17.8		&	46	\\
\hline
\multirow{4}{*}{WISE}				&	W1		&	3.4			&		19.9		&	93	\\
								&	W2		&	4.6			&		19.1		&	91	\\
								&	W3		&	12			&		16.7		&	75	\\
								&	W4		&	22			&		14.6		&	59	\\
\hline
\multirow{3}{*}{SPIRE (Herschel)}		&	SP1		&	250			&		27.0		&	29	\\
								&	SP2		&	350			&		22.5		&	27	\\
								&	SP3		&	500			&		32.4		&	13	\\
\hline
WSRT$\dagger$					&	L		&	1.4 			&		21		&	95	\\
\hline
%\cline{2-5}
\end{tabular}
\end{center}
\label{tab:multi_NEPD}
\end{table}%

\subsection{Radio-IR Source Cross-matching}
Our core sample is taken from the radio survey of the NEP at 1.4 GHz. To define our radio-source sample we utilize the IR AKARI catalogs of \citet{Kim2012}, for the NEPW, and of  Oi et al. 2013 (in prep.), for the NEPD, which we cross-match with the radio catalog at 1.4 GHz of \citet{White2010}. The original IR band-merged AKARI NEPW catalog contains 114794 sources at an N2 (2.4$\mu$m) band AB magnitude limit of $\sim21$ and a resolution of $\sim4$ arcseconds. By assuming a 5$\sigma$ in the near-IR bands (N2 and N3) and by using a (J-N2)-(g-i) color-color diagram to exclude stars (following \citet{Baldry2010}), we end up with a base near-IR catalog for the NEPW that includes 61165 sources. The NEPD on the other hand originally contains 10313 sources at a deeper flux limit of 9.6 $\mu$Jy at N2 band. We exclude sources with N2 or N3 detections below the 5$\sigma$ flux limit (see Table \ref{tab:multi_NEPD}) and sources with stellar colors, in the same manner as for the NEP-Wide. After this step we end up with 5523 sources that form the base near-IR NEPD catalog. For the purpose of our cross-matching we therefore use in total a sample of 66688 near-IR detected sources.  

The radio catalog covers an area of 1.7 deg$^{2}$ (compared to the total of 5.4 deg$^{2}$ of the total NEP-Wide AKARI survey, see Fig. \ref{fig:coverage}) with a beam size of 17 arcseconds and a sensitivity of 21 $\mu$Jy beam$^{-1}$. The final catalog contains 462 radio sources at a 5$\sigma$ detection limit. Two main ways to match catalogs in different wavelengths have been widely used in the literature: the Poisson probability based method of \citet{Downes1986} (e.g., \citealt{Ivison2007}, \citealt{Hodge2013}) and the Bayesian based likelihood ratio method, as described in \citet{Sutherland1992} (e.g., \citealt{Rodighiero2010a}, \citealt{Jarvis2010}). The latter requires the assumption (based usually on the data at hand) of the global magnitude distribution of a given ``type'' of source, according to which the probability for a candidate source to be a background source and a true counterpart is calculated. The ratio of the two is used to identify true matches. The former follows the opposite direction, in that the proximity and magnitude of each individual candidate source are used to calculate the probability that the candidate source is not a background source. Hence, although both methods use the proximity and apparent magnitude of a candidate source to identify true counterparts, the \citet{Downes1986} method does so in a more straightforward way. For very large samples, where global distributions can be more robustly constrained from the data, the likelihood ratio method potentially yields better results. For the case of the matching of the WSRT source with the AKARI catalogues we use the Poisson-probability based method of \citet{Downes1986}. In short, around each radio source we calculate the Poisson probability for each near-IR source to be within a circle of radius $r_{c}$. This Poisson probability is defined as:
$$P^{*}=1-e^{-\pi r^{2}N_{m}},$$
where r is the distance of the candidate counterpart from the multi-wavelength source, and $N_{m}$ is the surface number density within a radius r and limiting near-IR magnitude m. $r_{c}$ is defined through the positional uncertainties of the AKARI IRC instrument (assumed here $\sigma_{IRC}$= 0.2 arcsec, e.g., \citealt{Onaka2007}) and that of the WSRT array ($\sigma_{WSRT}$=5 arcsec, e.g., \citealt{White2010}). The expected number of events (i.e., near-IR sources) with $P\leqslant P^{*}$ can then be approximated (for a finite search radius $r_{c}$) as 
$$E=P_{c}=\pi r_{c}^{2}N_{T},$$ 
for $P^{*}\geqslant P_{c}$, and 
$$E=P^{*}(1+\ln{P_{c}/P^{*}}),$$ 
for $P^{*}<P_{c}$. $P_{c}$ is a critical Poisson probability, defined by the surface number density, $N_{T}$, at the limiting magnitude of the NIR sample.

\begin{figure}[htbp]
\begin{center}
\includegraphics[width=0.5\textwidth,angle=0]{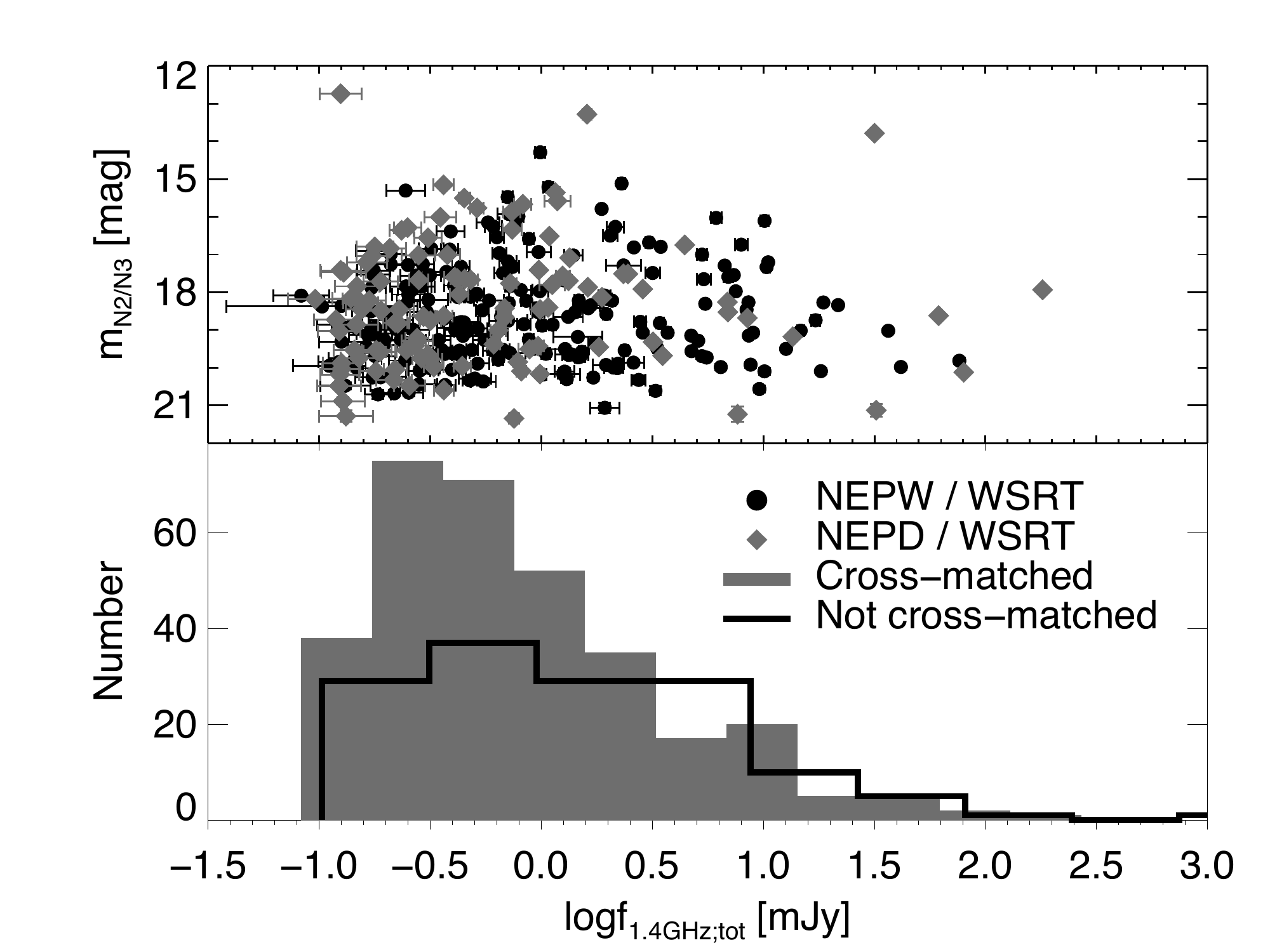}
\caption{The apparent NIR AB magnitude (N2, or N3 for sources undetected in N2) and 1.4 GHz radio flux of the NIR-IR cross-matched sample are shown. In total 321 sources are cross-matched between the NEPW and NEPD near-IR catalogs and the WSRT 1.4 GHz radio catalog. Upper: NEP-Wide (black circles) and NEP-Deep (gray diamonds) radio cross-matched sources are shown separately. Lower: The filled histogram (gray) shows the 1.4 GHz flux distribution of our cross-matched sample, while the empty histogram (black) shows the 1.4 GHz flux distribution of the radio sources that were not cross-matched to any AKARI source.}
\label{fig:radio_optical}
\end{center}
\end{figure}

Finally, the probability of a chance cross-identification of the source can be calculated as $1-e^{-E}$. The near-IR candidate with the lowest such probability is chosen to be the true counterpart. The above is done iteratively, first matching sources between the NEPW and the WSRT catalogues and then a further cross-matching is done between NEPD and the WSRT catalogues. Given the overlap of the catalogues, there is a number of radio sources matched with both an NEPD and an NEPW source. In these cases we keep the NEPD match, as the depth and quality of ancillary data for the NEPD is higher than for the NEPW.\\
Following this procedure, the final IR-radio catalog contains 321 cross-matched sources. In Fig. \ref{fig:radio_optical} the radio 1.4 GHz fluxes and near-IR AKARI N2 band AB magnitudes of the cross-matched sample are shown. In addition we show the radio fluxes of the radio-sources that were not cross-matched with an AKARI source. \\

\subsection{Photometric Redshifts}
For the purpose of our study, the estimation of the redshift of our sources is necessary. As was described in Section \ref{sec:NEP}, a number of spectroscopic surveys were conducted within the NEP field. Given the generally faint optical nature of radio sources, most of the radio sources do not have spectroscopic redshifts. A small subsample of 32 cross-matched radio-IR sources have spectra. Photometric redshifts have been calculated previously for the NEPD field (e.g., \citealt{Negrello2009}, \citealt{Takagi2010}, \citealt{Hanami2012}), using different SED fitting methods. We extend these studies calculate photometric redshfits using the publicly available LePhare code (\citealt{Arnouts1999}, \citealt{Ilbert2006}) and the latest catalogues available for the NEPW. For the NEPD, we use the photometric redshifts from Oi et al. (2013, in prep.). While the full photometric redshift catalog for the whole NEP-Wide field will be presented elsewhere, we give here a short summary of the methodology.

For the photometric redshift estimation of NEPW sources we use the near-UV GALEX band as well as the full optical bands and near-IR bands, extending out to the W2 WISE band (4.6 $\mu$m). In particular we use the following bands: NUV, u*, g, r, i, z, B, R, I, J, H, N2, W1, and W2 (refer to Table \ref{tab:multi_NEPW} for the respective central wavelengths and sensitivities). Given the overlap between the radio and CFHT observations, 149 of the cross-matched NEPW sources have CFHT data, while the remaining 65 are covered by our Maidanak observations. The limited wavelength coverage of the Maidanak optical data (3 bands in the optical) leads to very high photometric redshift uncertainties and a large fraction of catastrophic outliers. We therefore focus on the sources with CFHT data, thus narrowing down the number of available sources in the NEPW. 

\begin{figure}[htbp]
\begin{center}
\includegraphics[width=0.5\textwidth,angle=0]{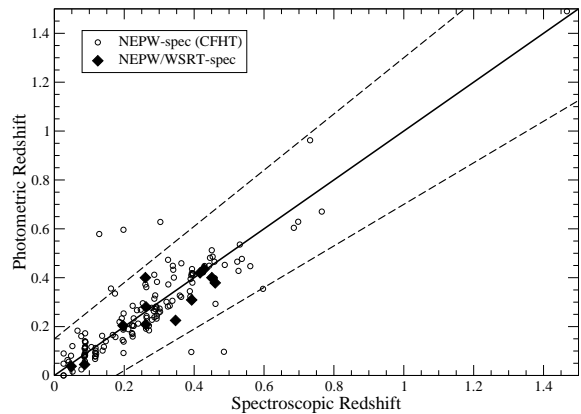}
\caption{Photometric redshift as a function of spectroscopic redshift for NEPW sources classified as galaxies through their optical spectra (open circles) and for radio-IR cross-matched sources in the NEPW with good quality optical spectra (filled diamonds). The solid line denotes the equality line, while sources outside the dashed lines are defined as outliers. Most radio-IR sources show very good quality photometric redshifts.}
\label{fig:spec_photoz}
\end{center}
\end{figure}

For the calculation of photometric redshifts we use the set of CFHT galaxy SED templates from \citet{Ilbert2006} and the \citet{Polletta2007} AGN templates. Several stellar template libraries are used to exclude stellar contamination to our sample. We use the dataset of existing spectroscopic redshifts in the NEP-Wide field to calibrate possible photometric offsets using the adaptive method described in, e.g., \citet{Ilbert2006}. By concentrating on the overlap between the CFHT and NEPW coverage, we focus on 143 NEPW sources with good quality optical spectra and classified, through their optical spectra, as normal galaxies (i.e., optical AGN are excluded), to train our photometric redshift code. The largest offset is found for the NUV band at 0.26 AB mag, with an average offset for all bands of $\sim$0.1 AB mag. In Fig. \ref{fig:spec_photoz} we show a comparison between the NEPW spectroscopic and photometric redshifts, for sources with good quality optical spectra and classified as galaxies (open circles) and for our sub-sample of NEPW radio-sources with optical spectra (filled diamonds). The resulting photometric uncertainty for the NEPW sample is $\sigma_{NMAD}=0.043$ and a catastrophic outlier fraction of $\sim5\%$. For the NEPD the uncertainty is slightly lower at $\sigma_{NMAD}=0.040$ but with a higher outlier fraction at $\sim8.7\%$.

\begin{figure}[htbp]
\begin{center}
\includegraphics[width=0.5\textwidth,angle=0]{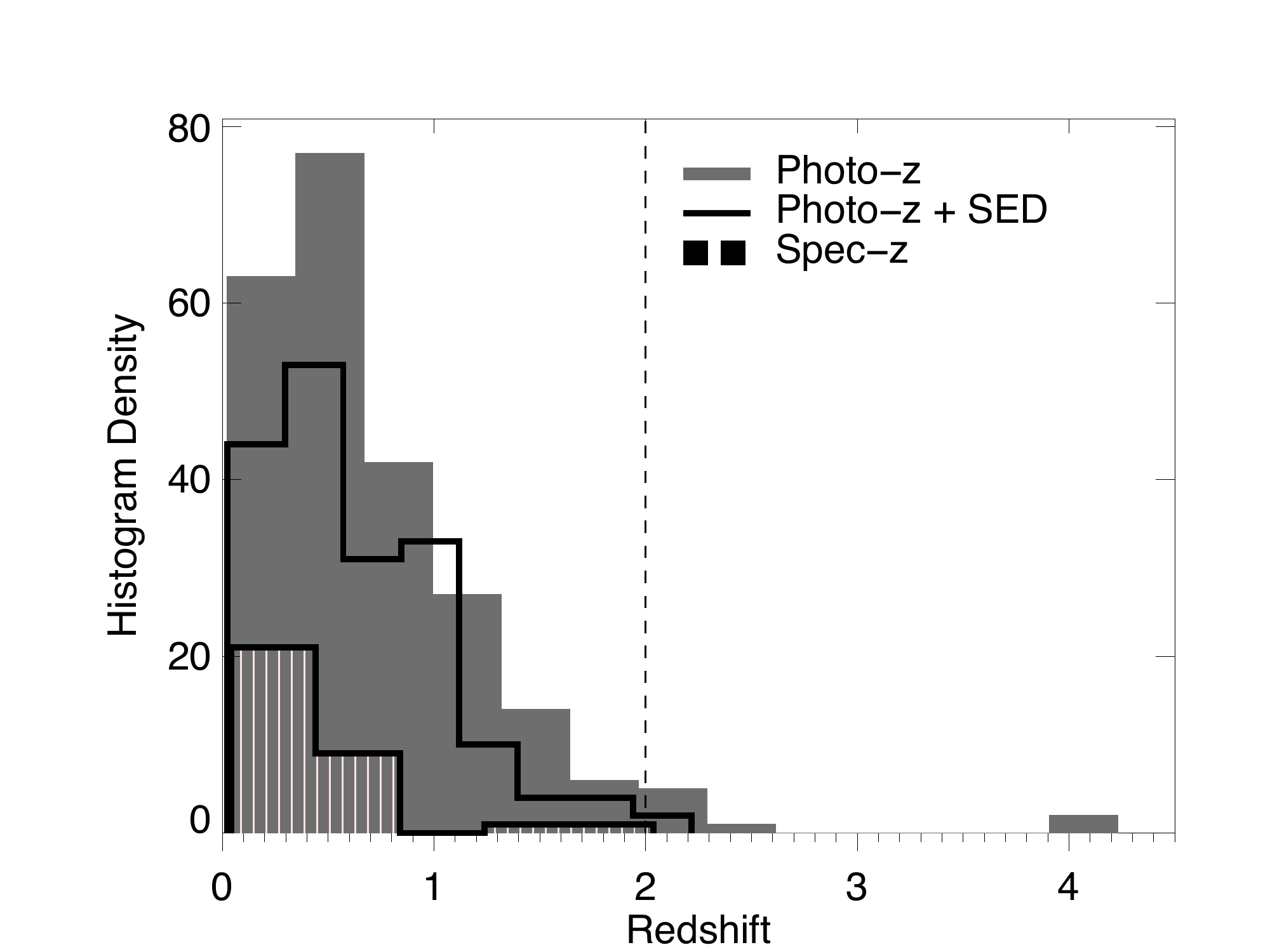}
\caption{Redshift distribution for our main sample (shaded gray), for the sub-sample of good and acceptable SED fits (see Section \ref{sec:sed}, open black), and for spectroscopically observed sources (stripped black). The vertical dashed line at z=2 gives a qualitative upper limit up to which photometric redshifts are reliable and denotes the limit above which sources are excluded from this study.}
\label{fig:z}
\end{center}
\end{figure}

In total, 237 sources out of the 321 originally matched ($\sim74\%$) with photometric and/or spectroscopic redshifts form the basic sample which we will use in the following. In Fig. \ref{fig:z} we show the redshift distribution for our sample, as well as a number of sub-samples.\\

\subsection{Radio Loudness}
\begin{figure}[htbp]
\begin{center}
\includegraphics[width=0.5\textwidth,angle=0]{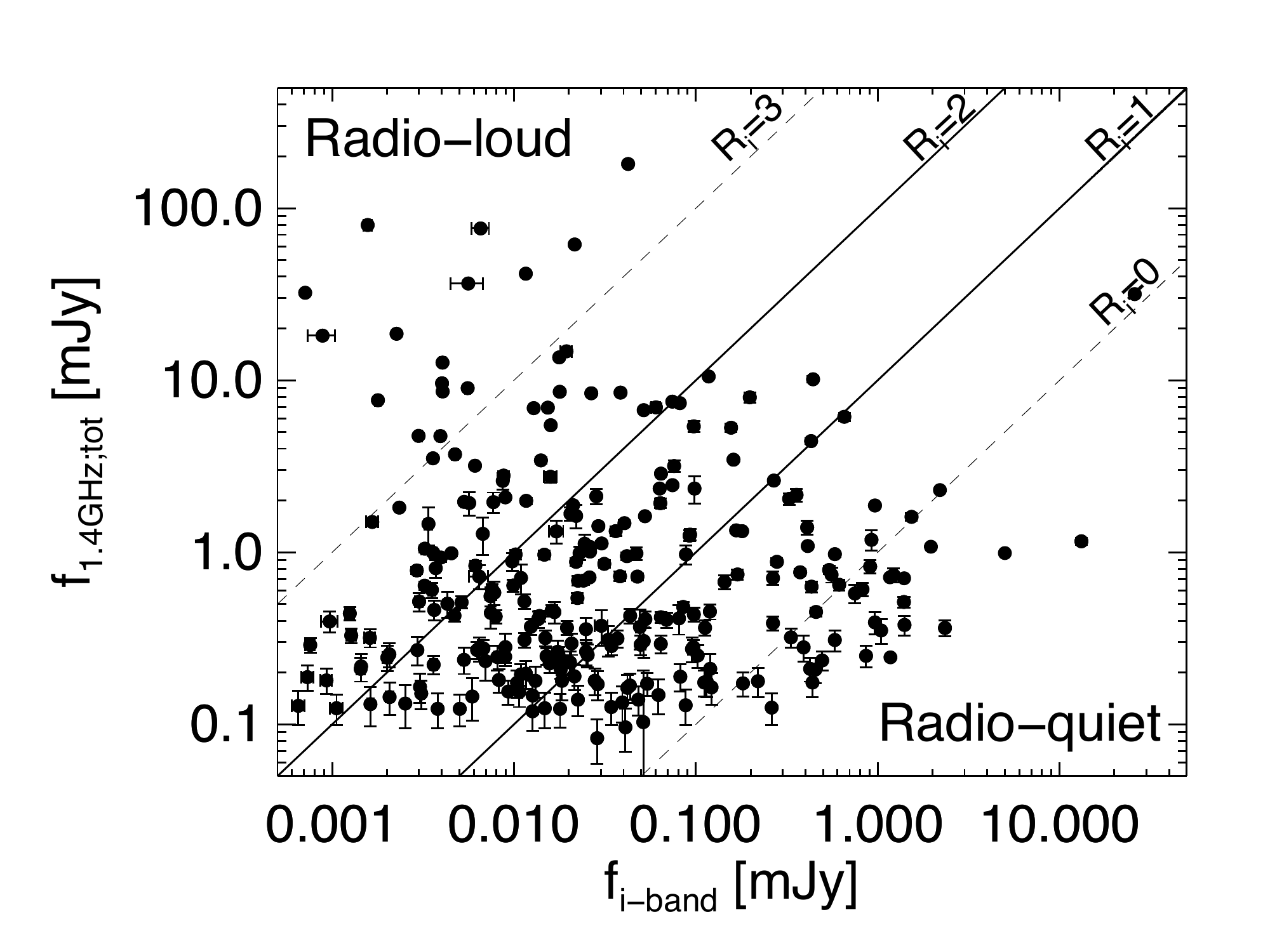}
\caption{Distribution of extinction-corrected optical i-band flux and radio 1.4 GHz flux for the sample of 237 IR-radio sources for which an acceptable photometric redshift could be calculated or a spectroscopic redshift is available. The diagonal lines denote constant radio-loudness values, with the solid lines marking the limits for radio-loud and radio-quiet sources. A most radio-loud object would be in the upper-left corner of the diagram, while a totally radio-quiet object would lie in the lower-right corner.}
\label{fig:opt_radio}
\end{center}
\end{figure}

Radio loudness, as first defined by \citet{Kellermann1989}, is a measure of the power of the radio-jet in an AGN and its dominance over the overall energy output of the nucleus. Traditional radio-AGN, with prominent radio-jets, are classified as radio-loud, while radio-quiet sources are thought to miss or have a very weak jet component. The original definition of radio-loudness, $R_{i}$, was the ratio between the luminosity at 5 GHz and the luminosity in the optical (4000$\AA$). Under that definition, radio-loud sources are those with $R_{i}>10$. Given the available data for the NEP field, here we use an alternative definition of radio-loudness from \citet{Ivezic2002},
$$R_{i}=log\left(\frac{f(1.4GHz)}{f(7480\AA)}\right),$$ 
where $\lambda=7480\AA$ is the central wavelength of the i-band in the optical. Under that definition, we classify sources with $R_{i}>2$ as radio-loud, while for $R_{i}<1$ a source is classified as radio-quiet. For the calculation of the radio-loudness we are using i-band fluxes corrected for Galactic extinction (using E(B-V) values from \citealt{Schlegel1998}). In Fig. \ref{fig:opt_radio} we show the distribution of optical i-band and radio 1.4 GHz fluxes, together with constant radio-loudness lines. A comparison with the sample of \citet{Ivezic2002} from FIRST and SDSS shows qualitative agreement in the $f_{i-band}-f_{1.4GHz}$ parameter space (e.g., Fig. 19 of \citealt{Ivezic2002}). In a similar fashion, \citet{White2012} look at the relation between the flux at 1.4 GHz and far-IR wavelengths (at 65 and 90 $\mu$m).

According to our definitions above and the defined redshift limit of $z_{lim}=2$, we have in total 60 radio-loud (RL) and 88 radio-quiet (RQ) sources. The RL sources have an average redshift of $z_{avg;RL}=1.12\pm0.07$, while for RQ sources this is $z_{avg;RQ}=0.323\pm0.026$.\\ 

We note that we do not use any K-correction for either the radio or i-band fluxes. This leads to an uncertainty of our radio-loudness measure of:
$$dR_{i}=(\alpha_{opt}-\alpha_{radio})\cdot\log{(1+z)}.$$
For a nominal radio spectral index of 0.8 (typical of the radio emission of star-forming galaxies) and a range of redshift from 0 to 2, we get an uncertainty of $0.04\cdot (\alpha_{i}-0.8)$. The radio-IR cross-matched sources show a wide range of optical spectral indices (derived from a linear fit to their rest-frame optical bands) leading to potential under-estimation of up to 0.47 for a source at z=2 and an optical spectral index of $\alpha_{i}=1.8$. In practice, we find that by applying a fixed K-correction for an assumed $\alpha_{radio}$ the classification of RL and RQ sources does not change for any of our sources. We thus for the following use the observed radio-loudness, foregoing a generic K-correction, given the absence of measured radio spectral indices and for many sources of optical spectral indices as well.

\section{Broadband energy distributions}
\label{sec:sed}
\begin{figure}[htbp]
\begin{center}
\includegraphics[width=0.5\textwidth,angle=0]{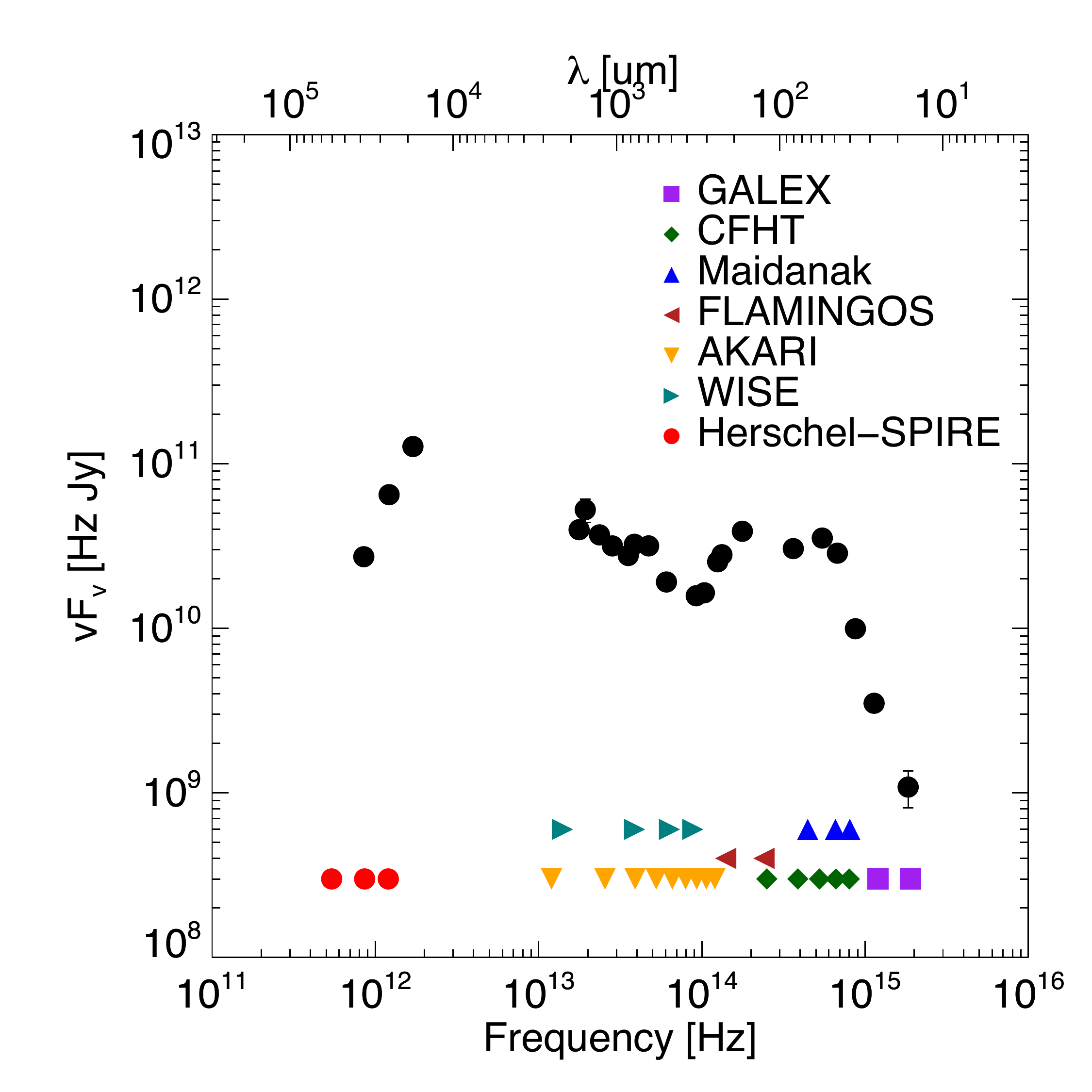}
\caption{Example of a broadband SED from our sample. With different symbols we denote the different bands. Plotted frequencies are in the rest-frame of the source. Radio 1.4 GHz flux is not plotted as it is not used for the SED fitting. This is an NEPW source with ID 38402 and its best fit SED is shown in Fig. \ref{fig:SED_fit_example}.}
\label{fig:sed}
\end{center}
\end{figure}

In order to first identify the AGN in our samples and then decouple the nuclear emission from a possible star-formation component, we want to study the broadband SEDs of our IR-radio sources. Given the richness of the ancillary data for the NEP field and the dense and homogeneous coverage of the IR wavebands from the AKARI IRC instrument, we can build detailed SEDs for our sources. To that end, we utilize all the bands mentioned in Tables \ref{tab:multi_NEPW} and \ref{tab:multi_NEPD}, ranging from the far-UV to the far-IR. The radio flux is not used in the SED fitting process as templates that cover the whole range between radio and far-UV wavelengths with consistent quality are scarce. In addition, given the limited wavelength coverage of the radio regime, we can extract little physical information by including the radio flux in our SED fitting. In Fig. \ref{fig:sed} we present an example of such an SED.\\

\begin{figure*}[htbp]
\begin{center}
\includegraphics[width=0.33\textwidth,angle=0]{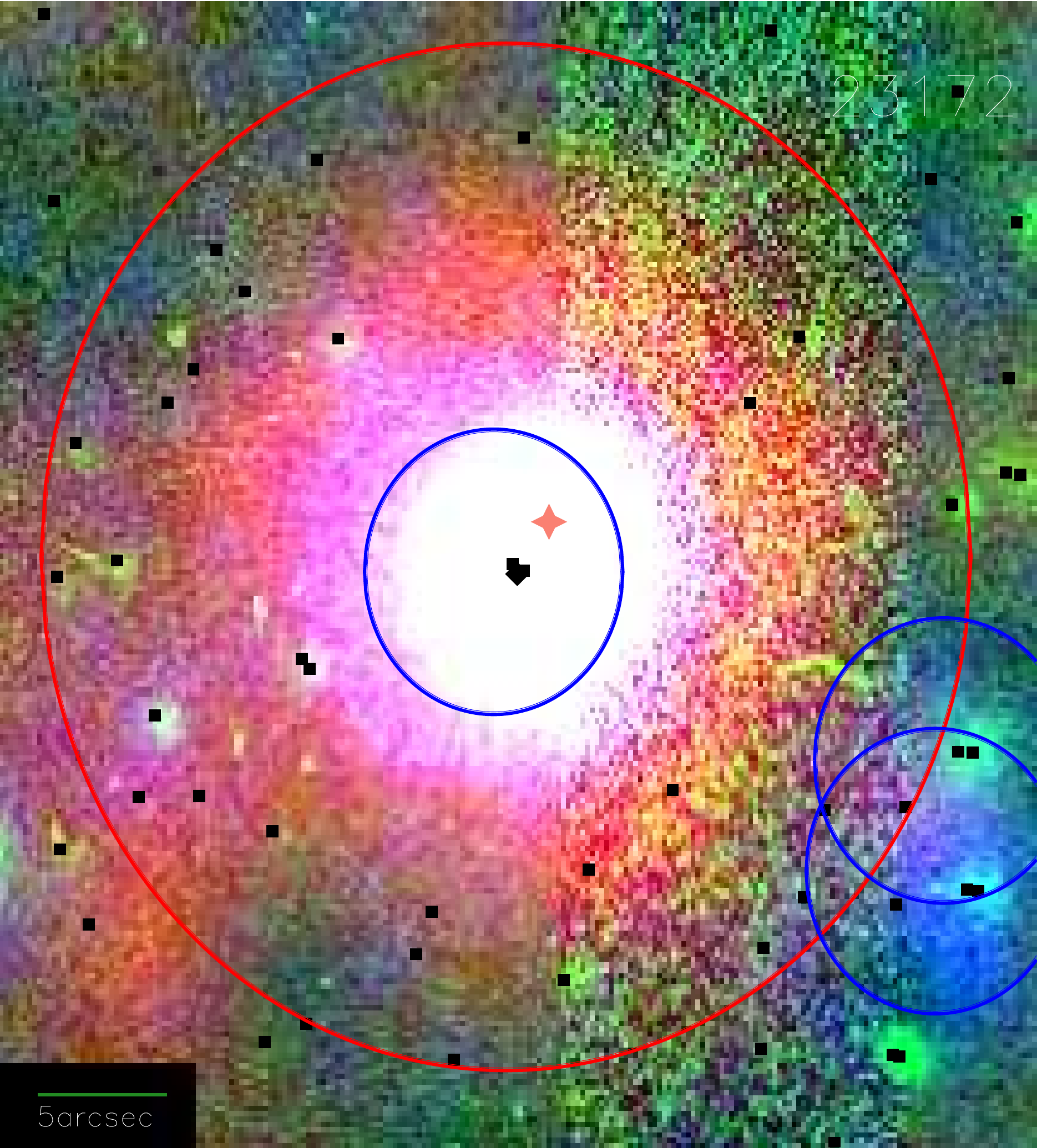}
\includegraphics[width=0.33\textwidth,angle=0]{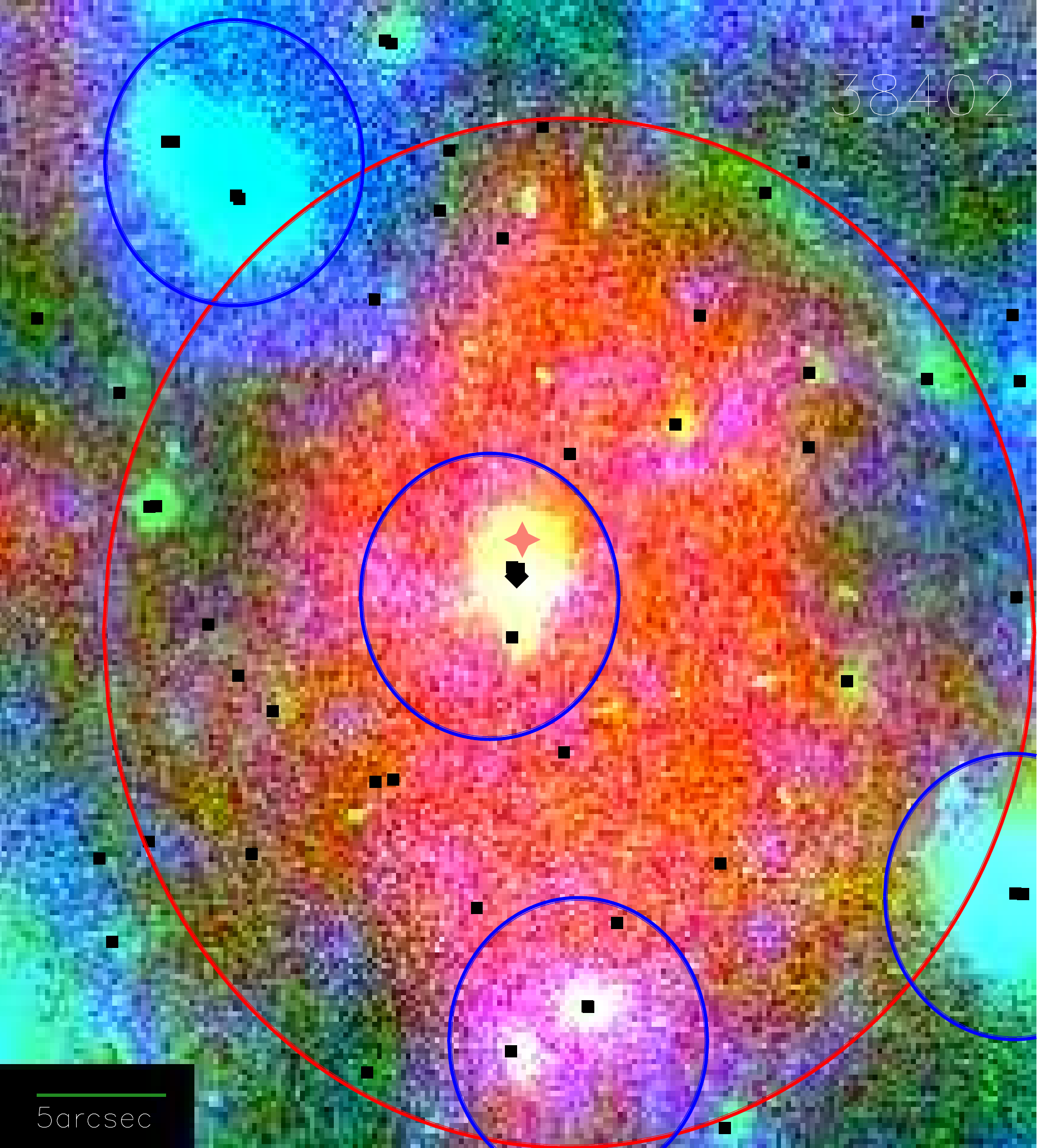}
\includegraphics[width=0.33\textwidth,angle=0]{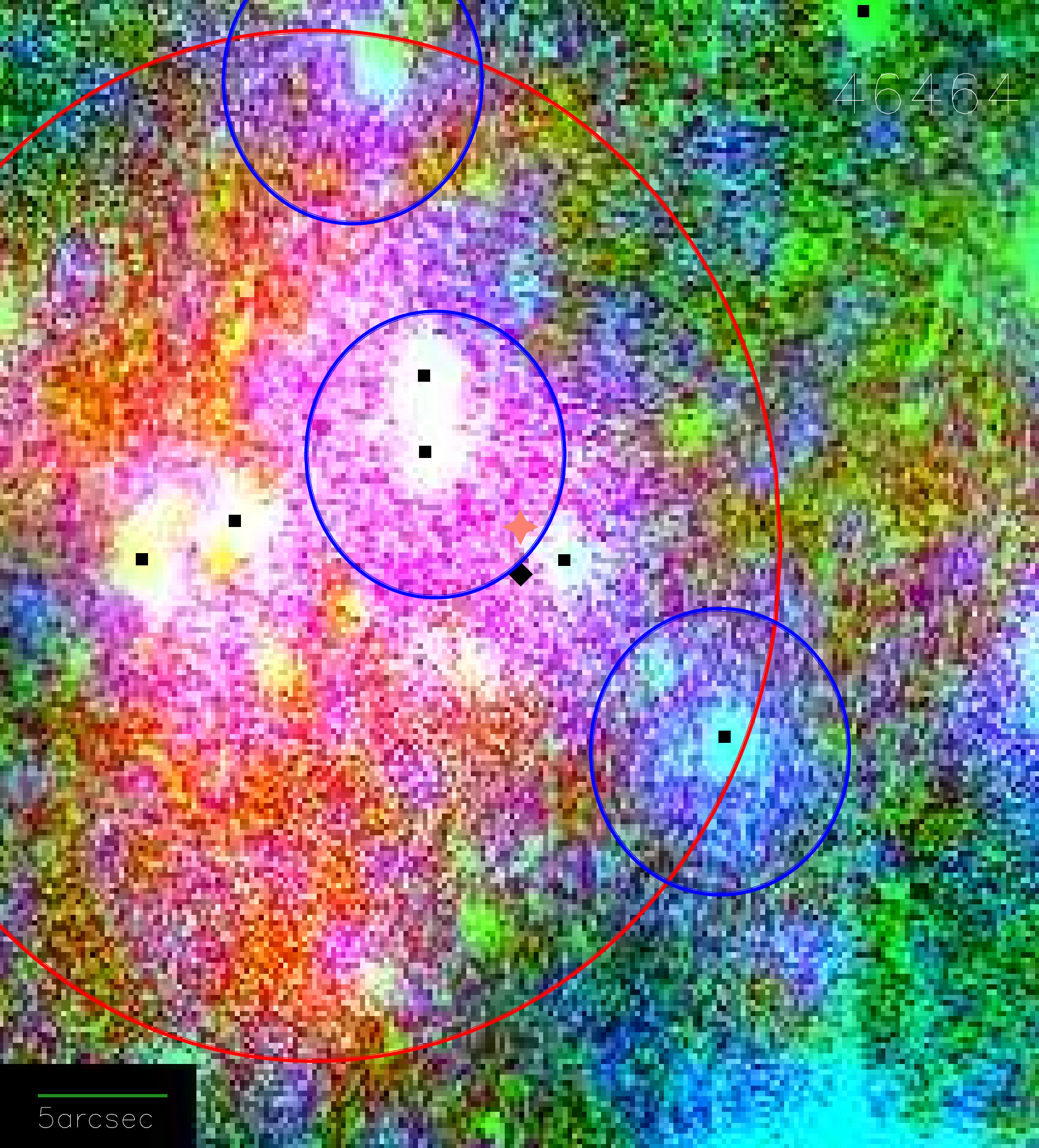}
\caption{Three examples of cross-matching between an IR-radio cross-matched source, a GALEX source, and a Herschel source.  Images are RGB composites of the Herschel-SPIRE 250$\mu$m band (R), the CFHT r-band (G), and the GALEX NUV band (B). The orange star denotes the position of the radio source, the black diamond shows the position of the cross-matched AKARI source, and black squares note the position of other detected optical sources. Blue circles mark the position and beam size of GALEX, while red circles denote Herschel-SPIRE sources. Left: All three sources are well on top of each other leading to a near-perfect match. All flux points are used as detections. Middle: UV and IR-radio sources on top of each other, while slightly offset to the Herschel source. Several optical sources within the Herschel-SPIRE beam and at least one more UV source. As the IR-radio source is detected in the long mid-IR bands of AKARI, the cross-match is considered true and all flux points are used as detections. Right: IR-radio source is offset to both the GALEX and Herschel sources. Besides several optical sources, at least three UV sources are within the Herschel-SPIRE beam. The IR-radio source is not detected in the long mid-IR AKARI bands. As a result this is considered a false match. The UV source is considered to not be associated with the IR-radio source, while the Herschel fluxes are used as upper limits.}
\label{fig:composite_example}
\end{center}
\end{figure*}

For the cases of the GALEX and Herschel data a two-step process was followed to cross-match the individual catalogs with the IR-radio cross-matched catalog. First a proximity search was carried out using the positions of the AKARI near-IR sources (a 5 arcseconds radius was used in both cases). Next, we constructed composite thumbnail images of all of our sources using the optical (CFHT or Subaru), UV (GALEX), and far-IR (Herschel-SPIRE) data. We visually inspected each thumbnail to decide whether a match in proximity was likely a true match. Through visual inspection we assigned a ``Matched'' and a ``Confusion'' flag to each source, noting both whether this appears to be a true match and whether there is a high probability for this source to be confused (i.e., several optical and/or UV sources included within the larger beam of the SPIRE instrument). For sources with obvious mismatches we assign upper limits according to the depth of the SPIRE survey of the NEP (27, 22.5, and 32.4 mJy for the 250, 350, and 500 $\mu$m bands, respectively). Sources with heavy confusion\footnote{As heavily confused sources we assumed sources for which more than 4 optical or 2 UV sources are enclosed within the beam of the SPIRE instrument. For the check we used the size of the 250 $\mu$m beam-size of 18 arc seconds.} are assigned the flux value of the initially cross-matched sources but as upper limit rather than detection. For ambiguous cases for the Herschel cross-matching (light confusion and/or off-center emission), the mid-IR properties of the IR-radio cross-matched source were checked. If the source is detected in at least two of the three long mid-IR bands of AKARI, then the cross-match is assumed true. In the opposite case, the flux of the initially cross-matched source is set as an upper limit. In Fig. \ref{fig:composite_example} we show examples of the above mentioned cases.

In addition to detections and the Herschel-SPIRE upper limits for the confused sources, we also used AKARI and Herschel-SPIRE upper limits for all non-detected sources to further constrain the SED fitting (by definition the whole NEP field is covered by the AKARI IRC, while the Herschel-SPIRE NEP campaign has as well covered the field in its entirety). In effect, the SPIRE data provide the bottleneck of the SED fitting, defining the sub-sample of sources that have full SED coverage. In total 68 of the radio-IR sources are detected by Herschel-SPIRE. Of these, 64 have photometric and/or spectroscopic redshifts. Through the visual inspection described above, 13 sources are classified as false matches or heavily confused sources and are therefore only assigned only Herschel-SPIRE upper limits. Out of the remaining 51 Herschel-SPIRE detected sources, 44 have full coverage from UV to far-IR, while the remaining 7 miss UV data. In conclusion of the 237 sources with redshifts, $\sim18\%$ have full UV-to-far-IR coverage, an additional $\sim24\%$ have full UV-to-mid-IR coverage with additional upper limits for the far-IR, $\sim12\%$ have full optical-to-mid-IR coverage and far-IR upper limits, while the remaining $\sim46\%$ is detected in a combination of optical and IR bands with upper limits for the undetected IR bands.

Having built our broadband SEDs following the method outlined above, we have applied a template-fitting method for the SED modeling similar to those used in the SWIRE Photometric Redshift Catalogue (\citealt{Rowan2008}) or the Imperial IRAS Faint Source Catalogue (\citealt{Wang2009}) but uses the model by \citet{Ruiz2010} and \citet{Trichas2012}. On one hand we fitted the optical/NIR SED (wavelengths shortward of 3.2 um) using a set of nine templates. On the other hand we fitted the IR SED using a set of four templates (to take into account the stellar contribution at short wavelengths, the best-fit optical template is included in the IR modeling). In both cases we used a $\chi^2$ minimization technique. Finally, the optical and IR best-fit models were added to obtain a complete model of the SED covering the entire wavelength range.

Several physical components contribute to the emission that comprise the broadband SED of these objects, including stellar emission and star-formation heated dust from the host galaxy, emission from the AGN torus, and nuclear emission from the accretion disk. In particular, the IR emission is a result of both the AGN (mainly in the near and mid-IR) and the host galaxy (far-IR being dominated by dust heated by massive young stars). As a result, the SED fitting of such objects presents a complicated task and often requires multi-component fits (e.g., \citealt{Lacy2007}, \citealt{Seymour2008}, \citealt{Barthel2012}). The optical to near-IR emission is mostly dominated by either a strong old stellar component, or an optical AGN, while the mid-IR to far-IR emission is mostly dominated by either a Type-2 AGN or a star-formation component. Although emission from either component extends within the other half of the SED, these processes are physically distinct and can be studied quasi-separately. It is this physical motivation together with the lack of adequately good templates spanning the full wavelength range of our data that dictates the two-step SED fitting process employed here. As a result we are able to effectively separate the AGN and the galaxy component contributions for each object.

Our set of optical templates includes one elliptical galaxy, four spirals, one SB and three QSO (see \citealt{Rowan2008} for a complete description of these templates). We included an additional extinction component to this model (using the \citealt{Calzetti2000} extinction curve), with the extinction $A_V$ as a free parameter.

\begin{figure*}[htbp]
\begin{center}
\includegraphics[width=0.245\textwidth,angle=0]{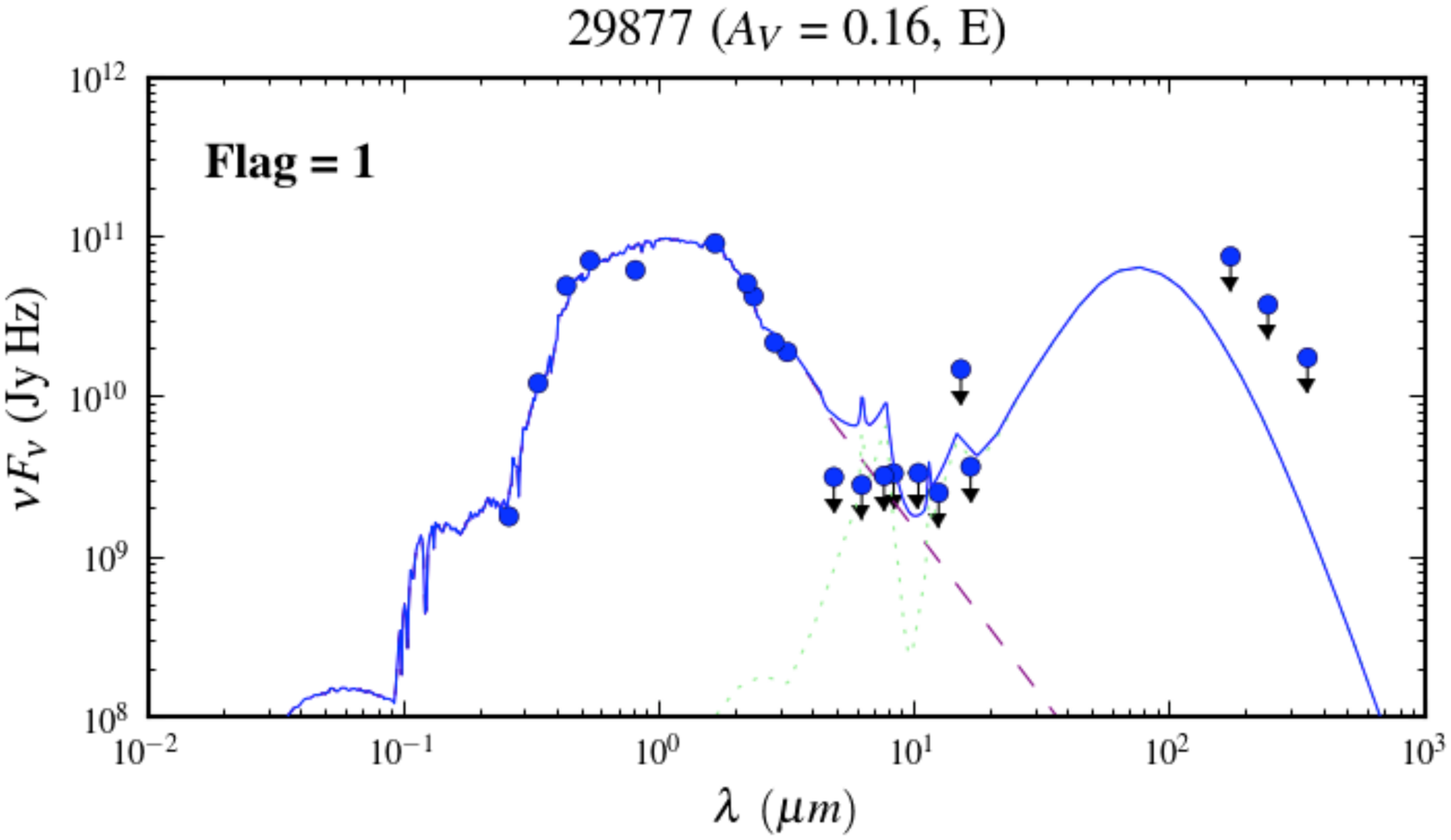}
\includegraphics[width=0.245\textwidth,angle=0]{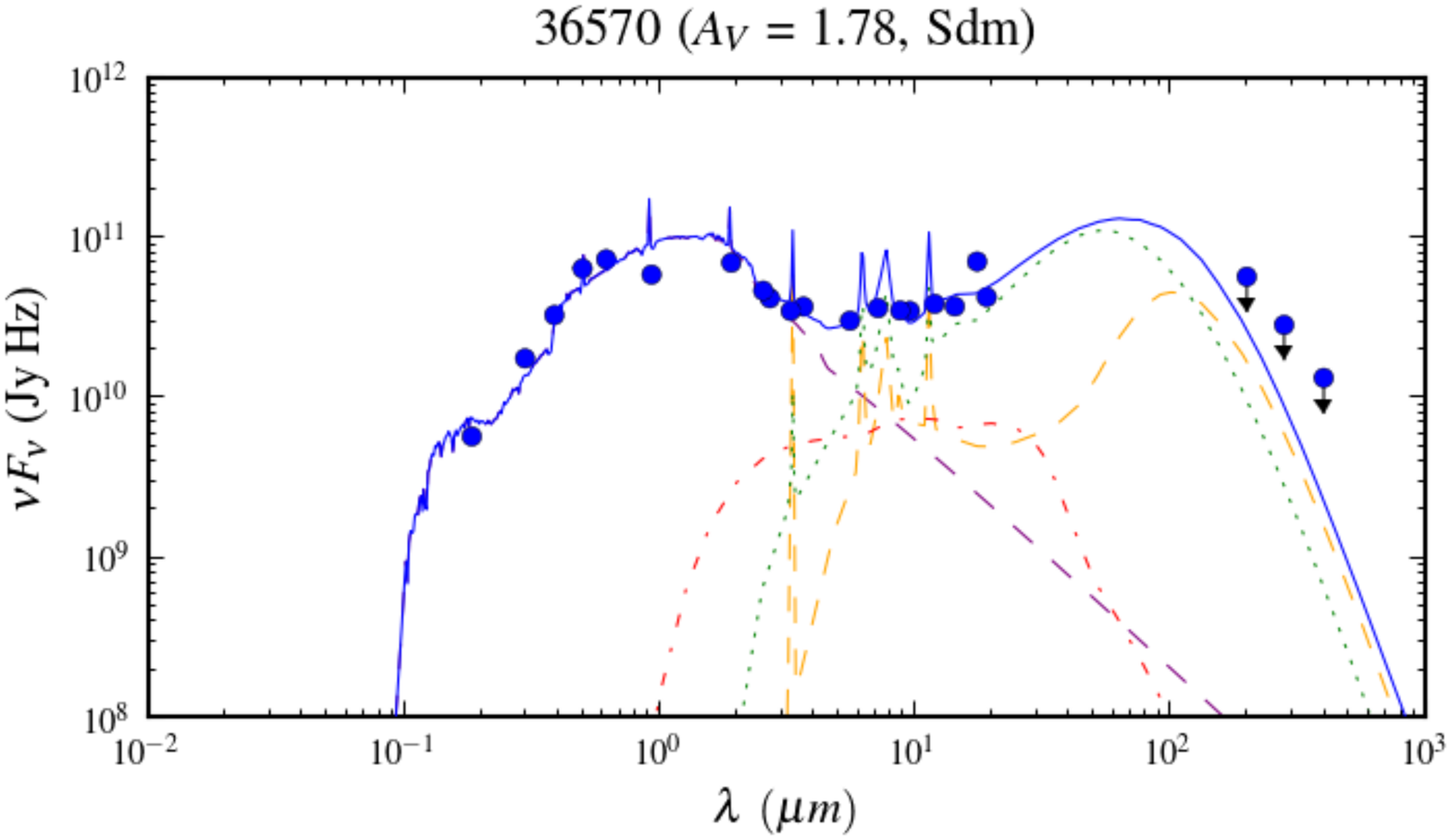}
\includegraphics[width=0.245\textwidth,angle=0]{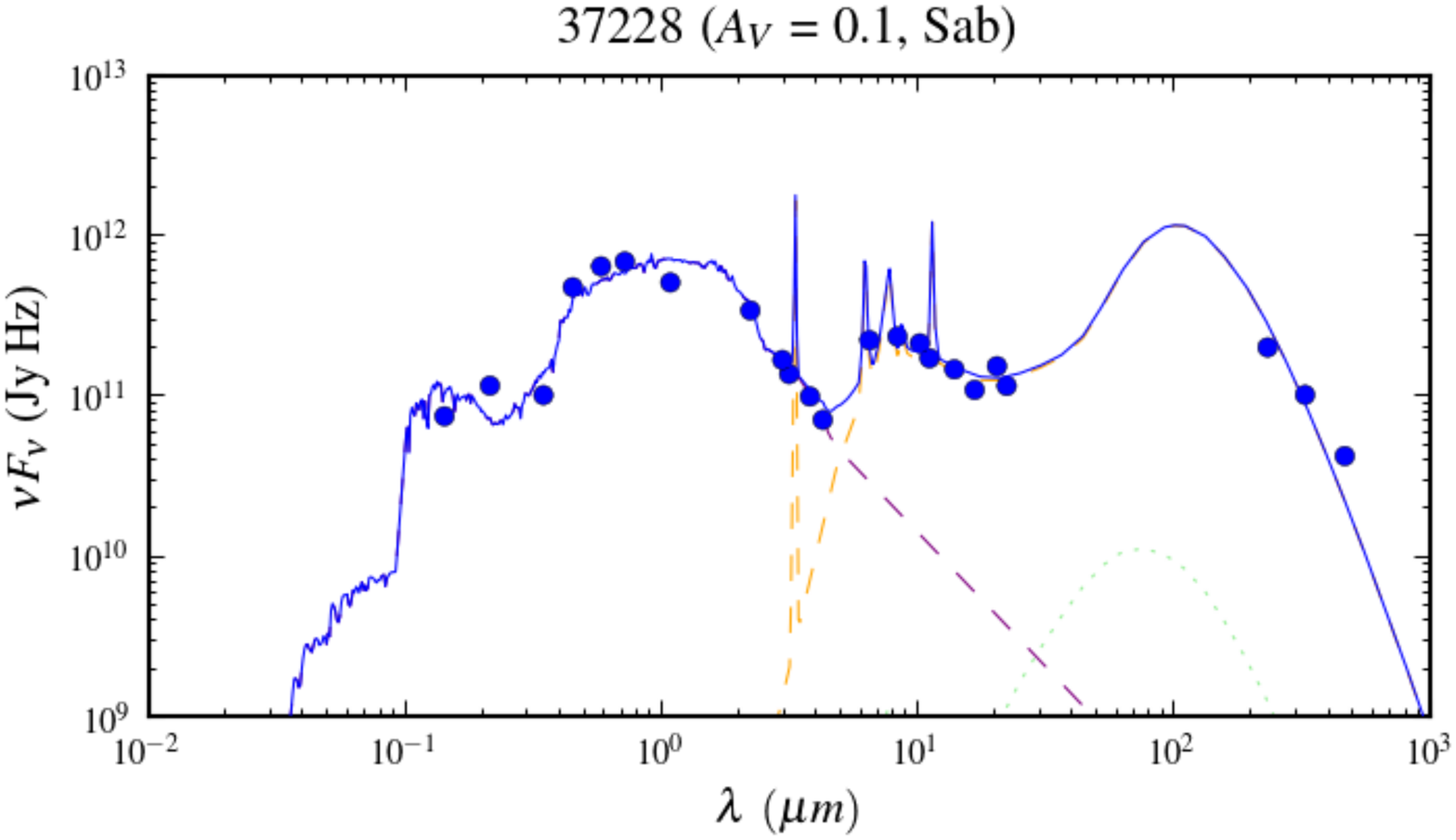}
\includegraphics[width=0.245\textwidth,angle=0]{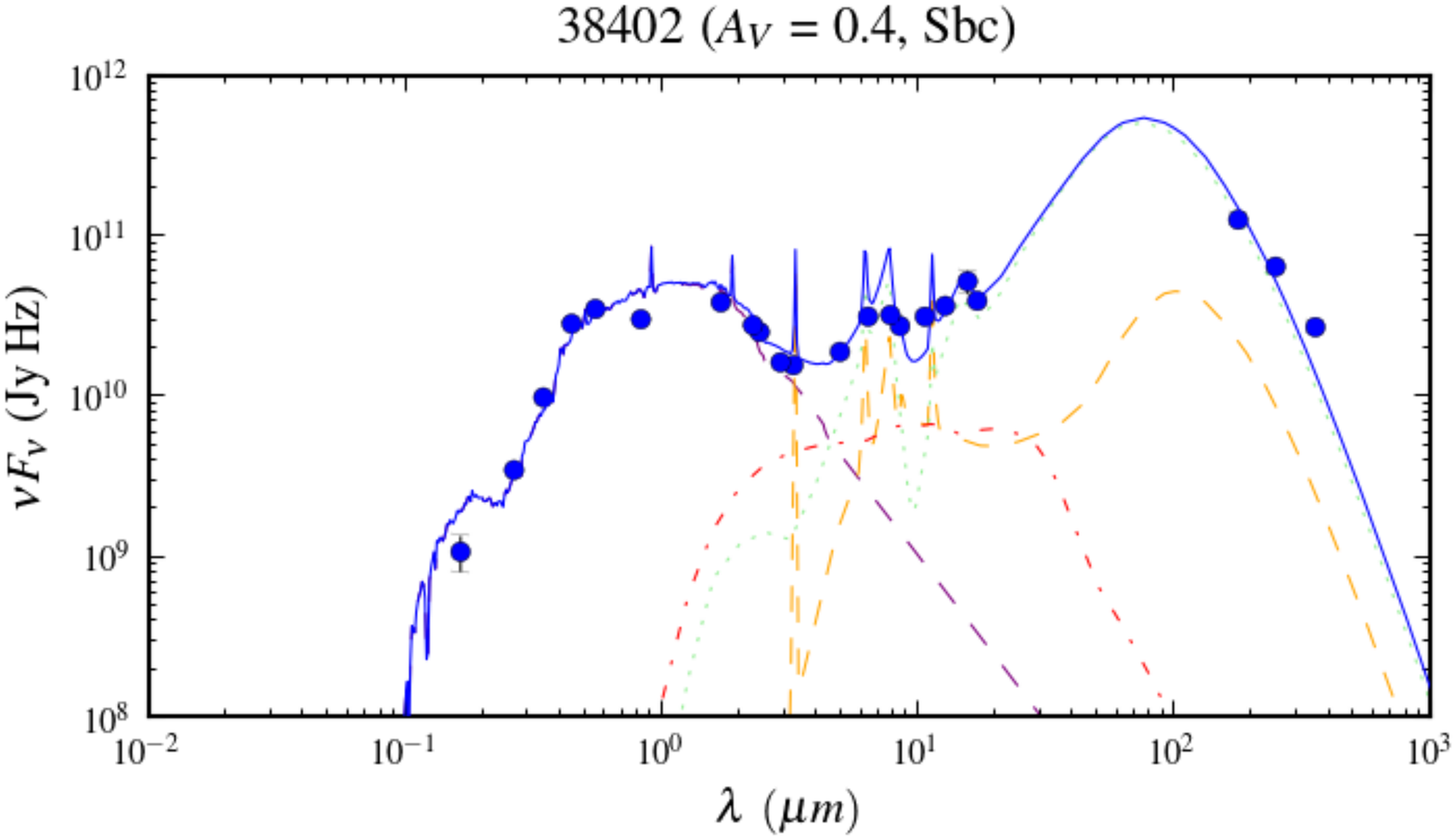}\\
\includegraphics[width=0.245\textwidth,angle=0]{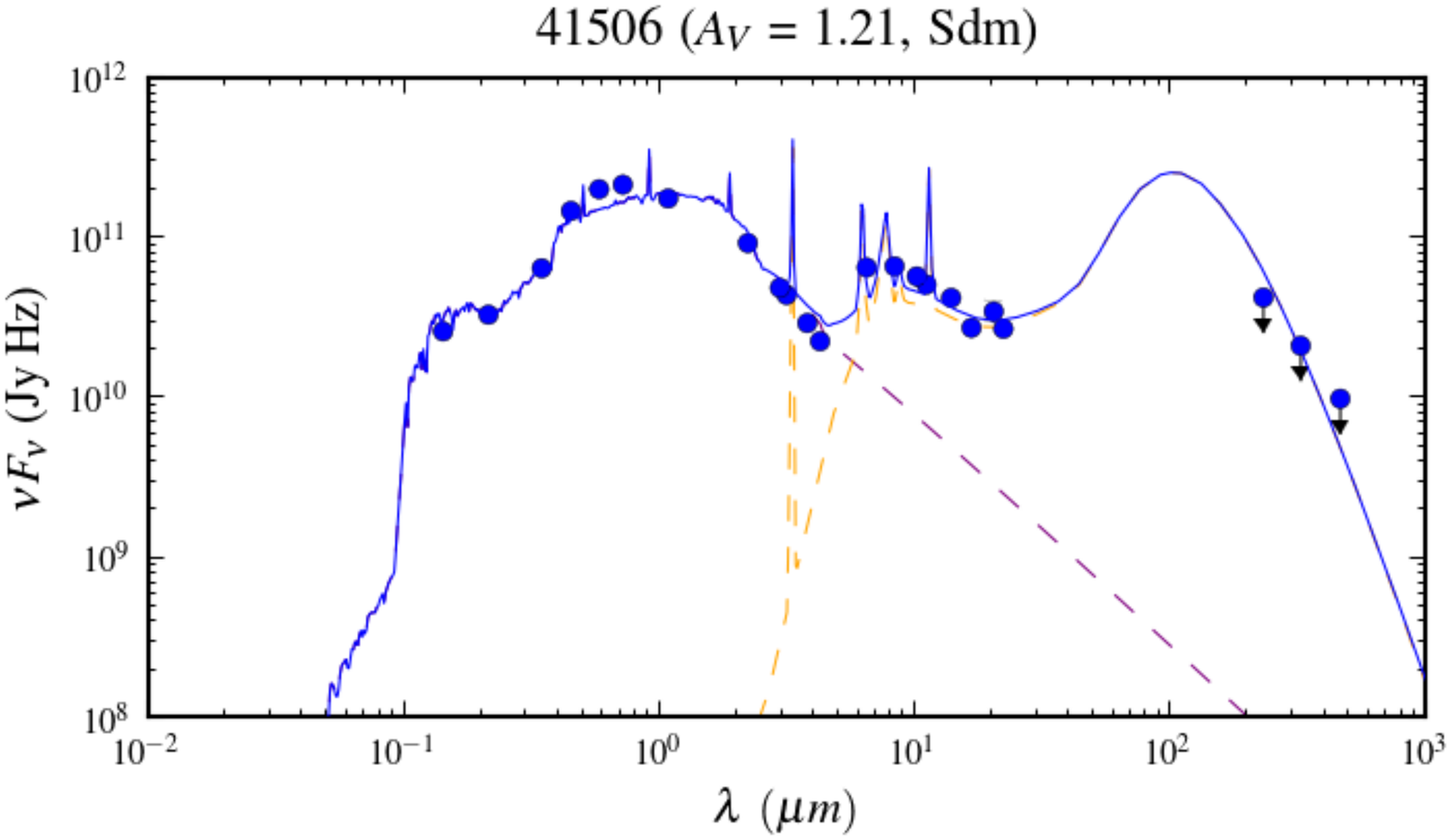}
\includegraphics[width=0.245\textwidth,angle=0]{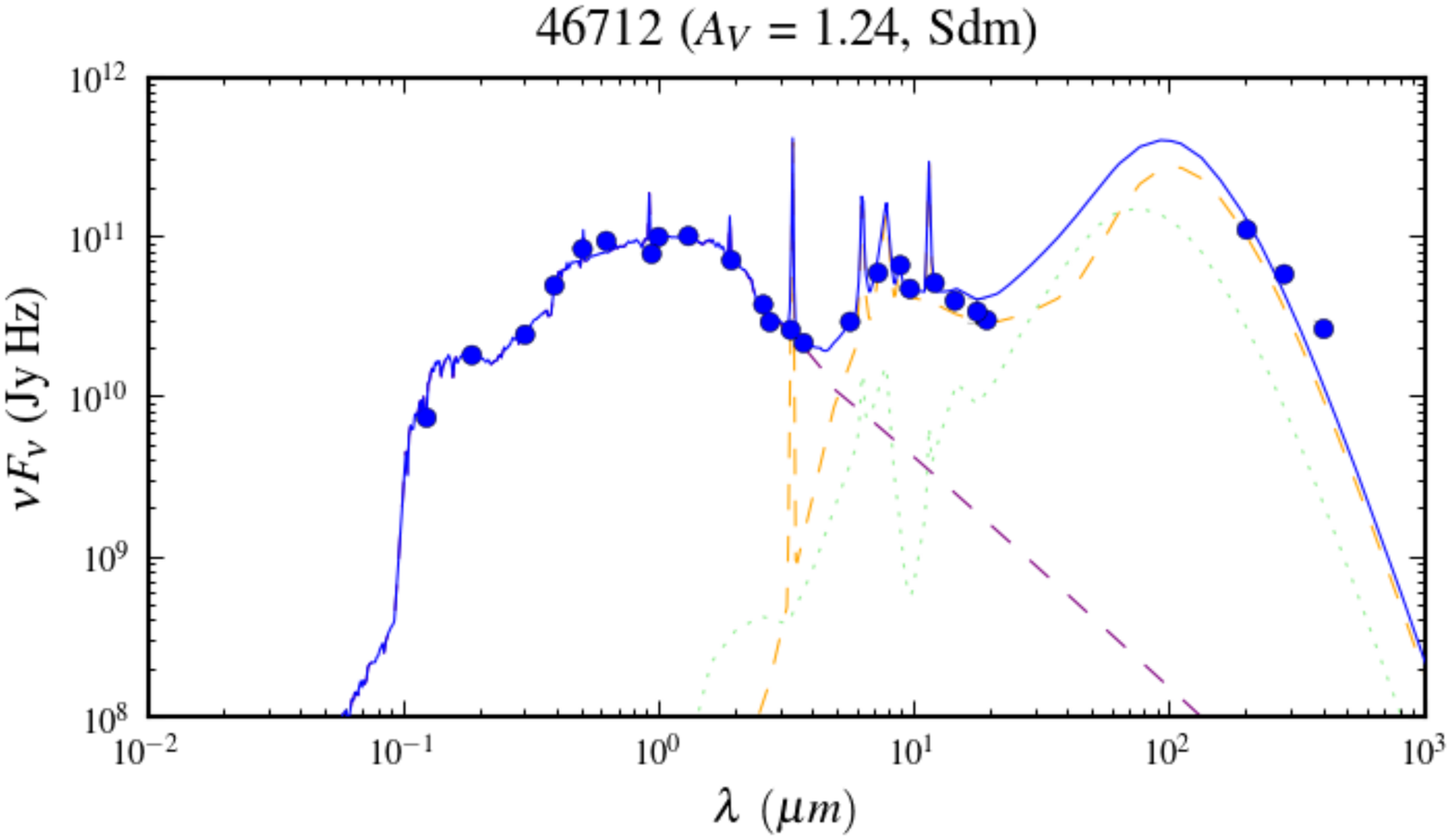}
\includegraphics[width=0.245\textwidth,angle=0]{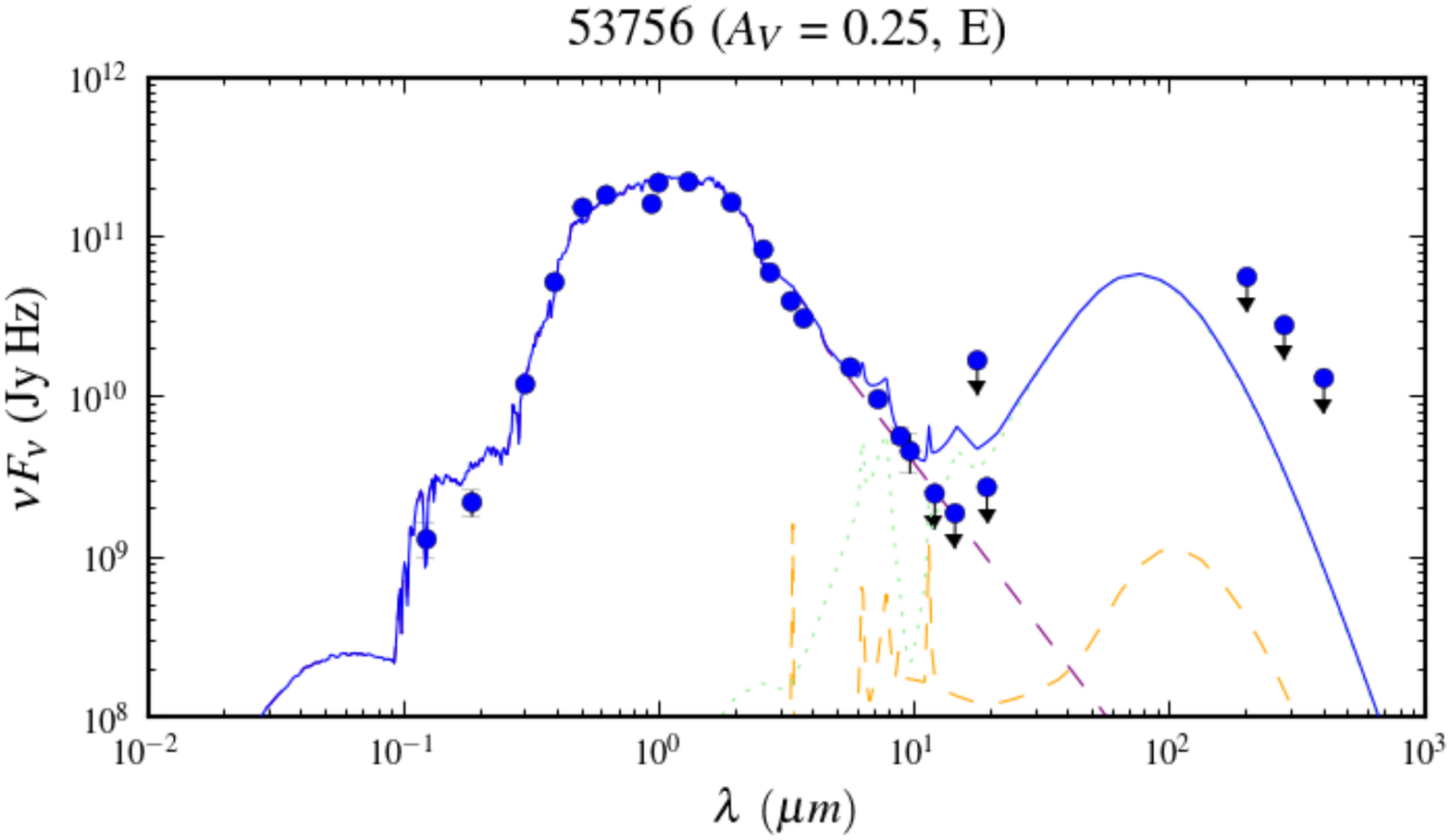}
\includegraphics[width=0.245\textwidth,angle=0]{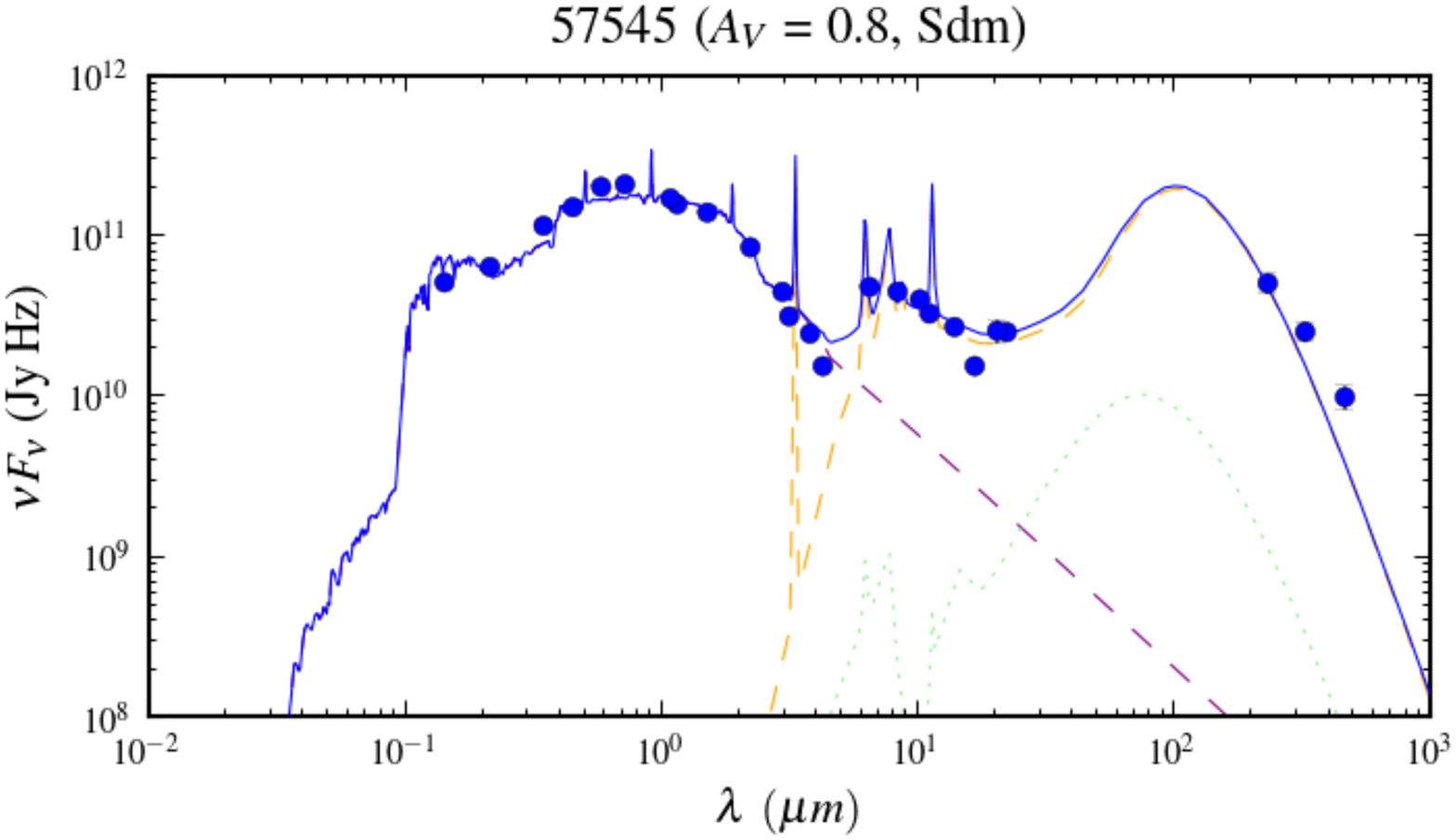}\\
\includegraphics[width=0.245\textwidth,angle=0]{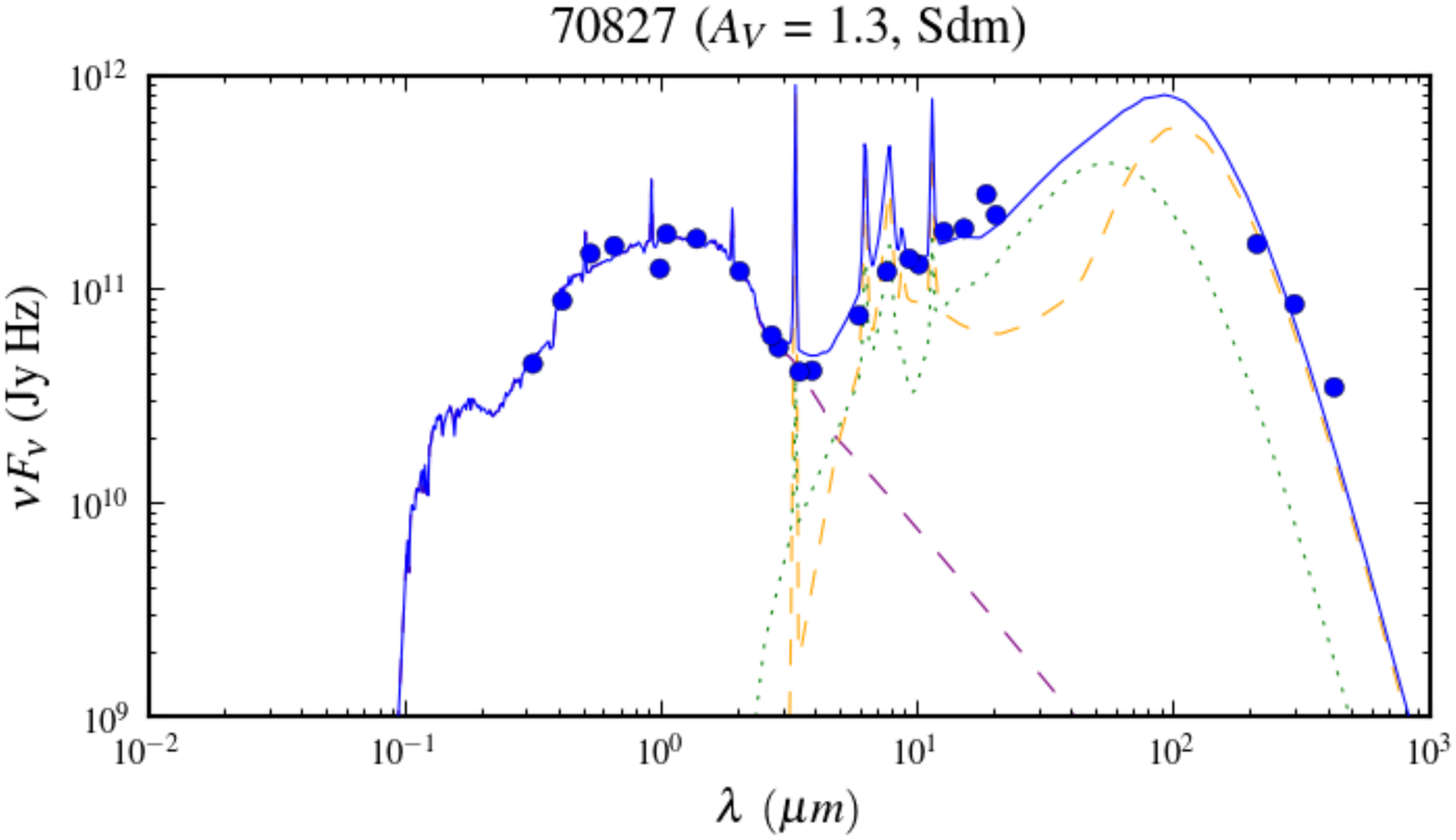}
\includegraphics[width=0.245\textwidth,angle=0]{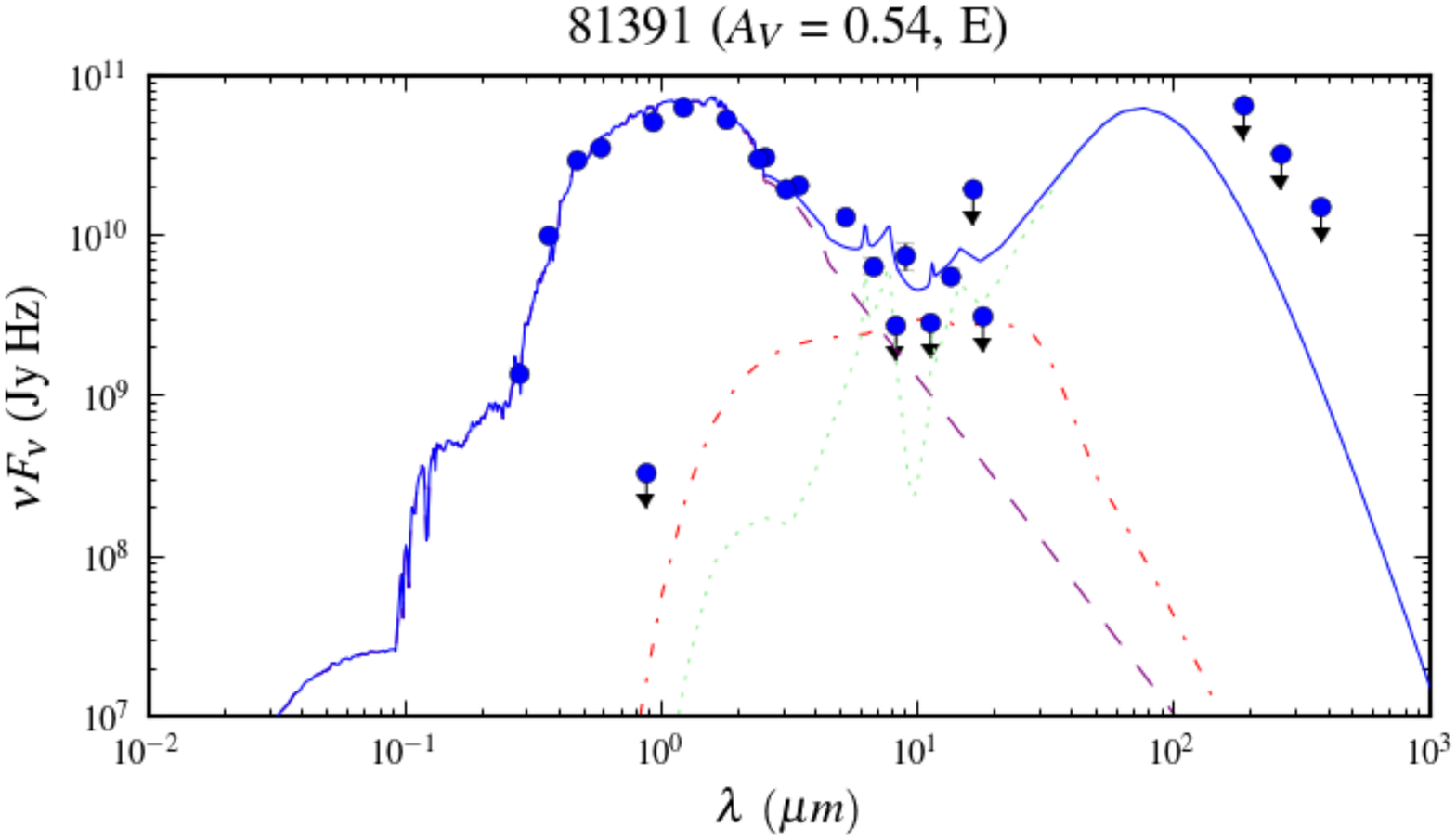}
\includegraphics[width=0.245\textwidth,angle=0]{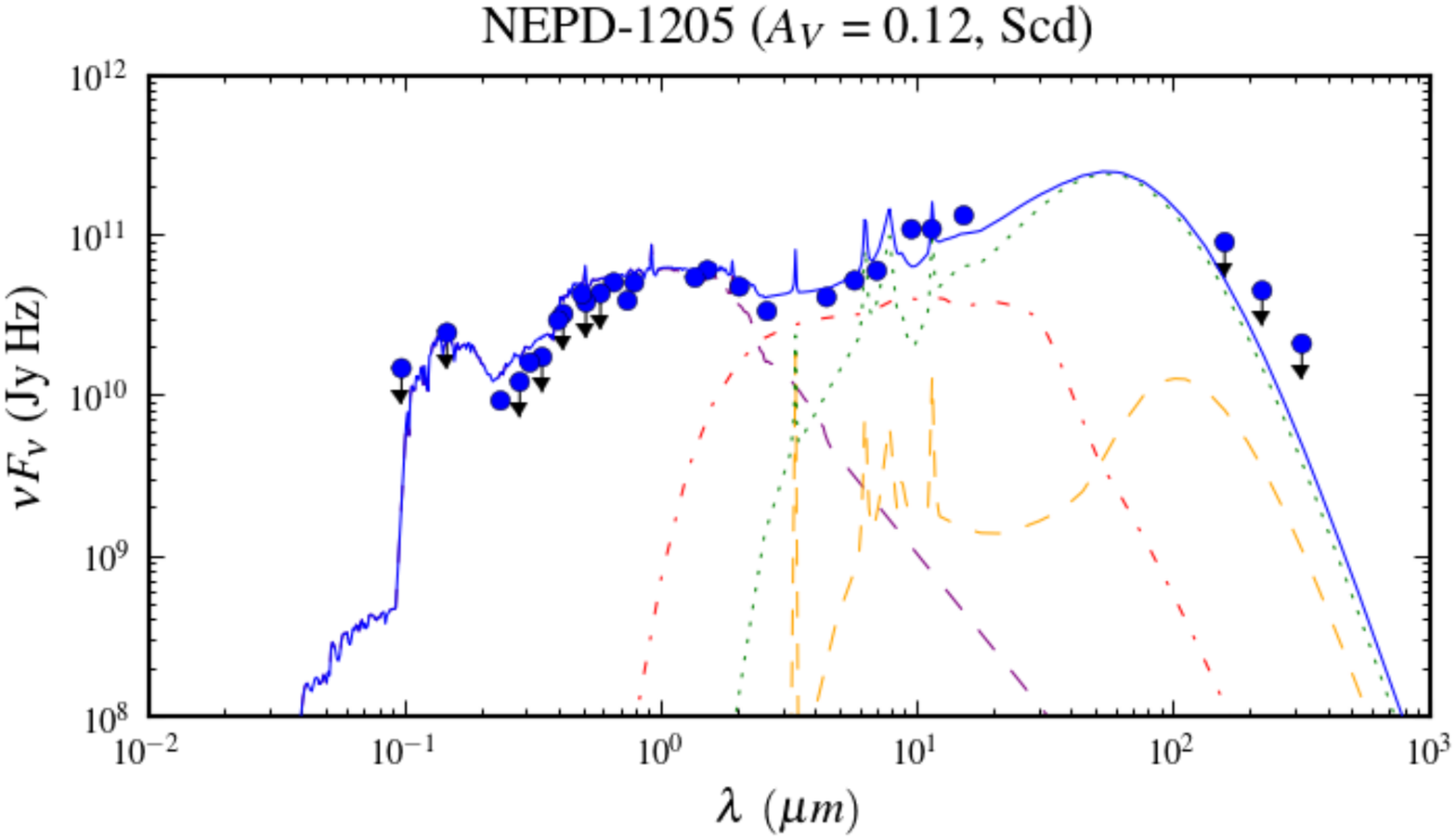}
\includegraphics[width=0.245\textwidth,angle=0]{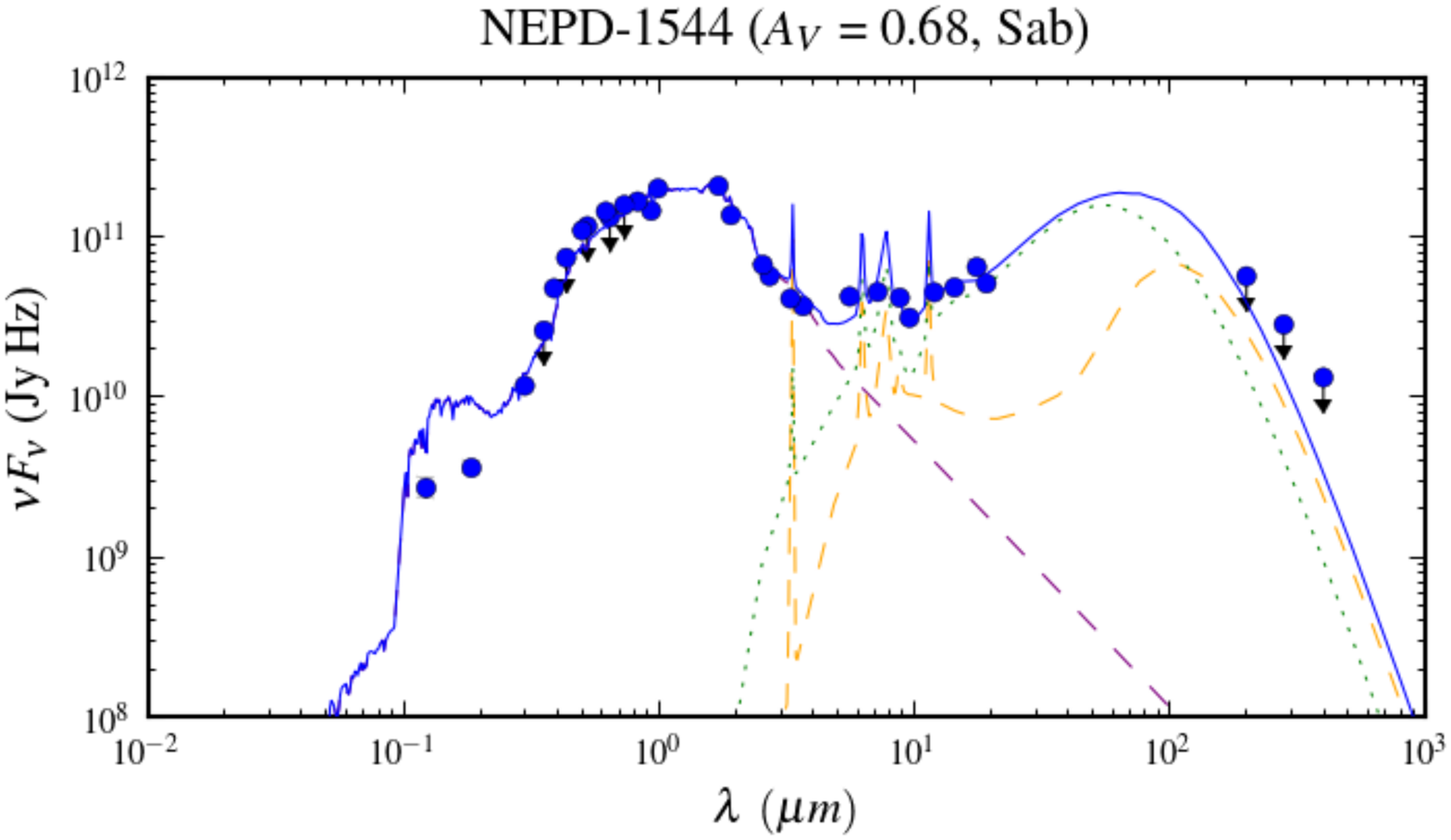}\\
\includegraphics[width=0.245\textwidth,angle=0]{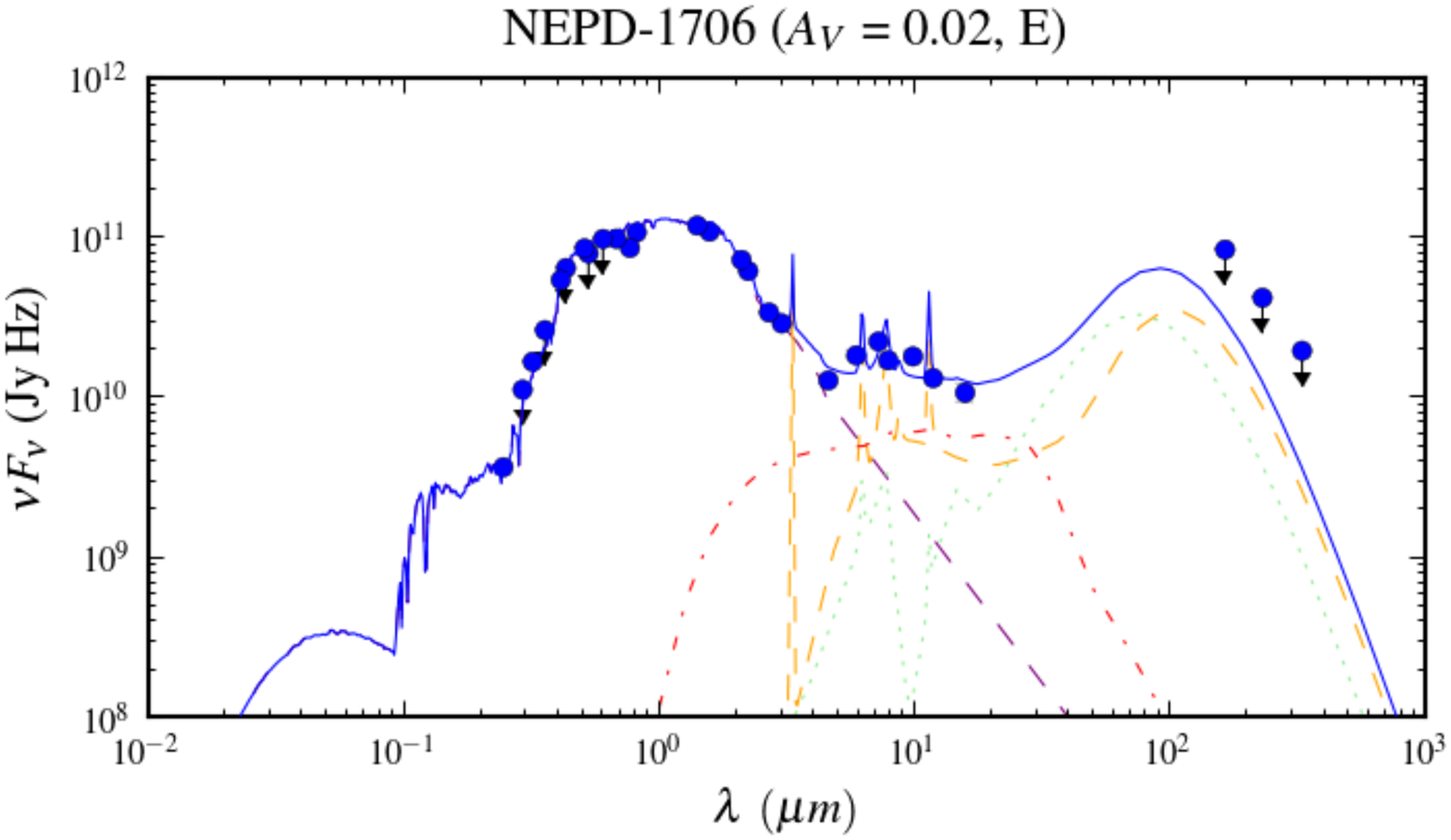}
\includegraphics[width=0.245\textwidth,angle=0]{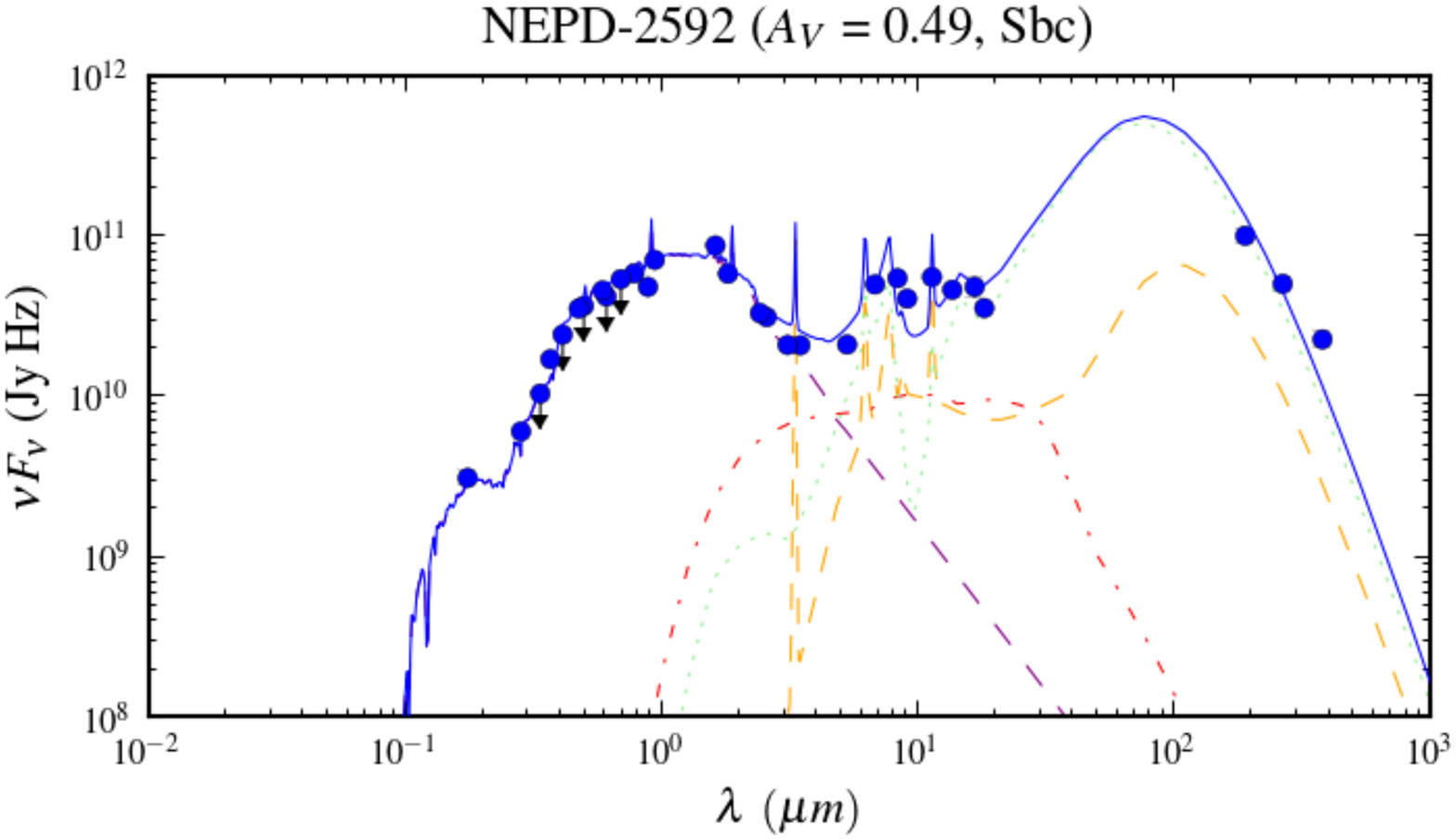}
\includegraphics[width=0.245\textwidth,angle=0]{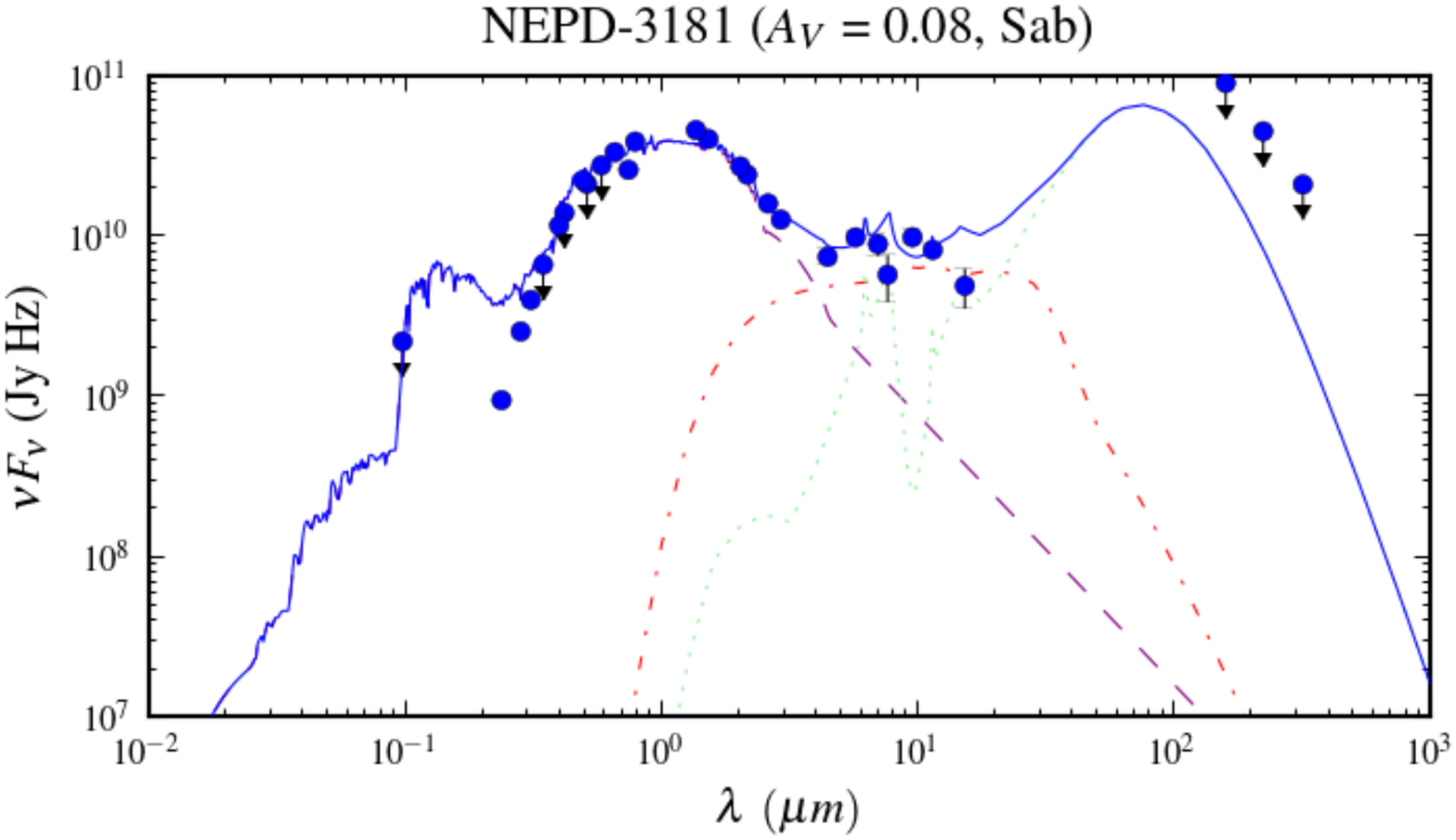}
\includegraphics[width=0.245\textwidth,angle=0]{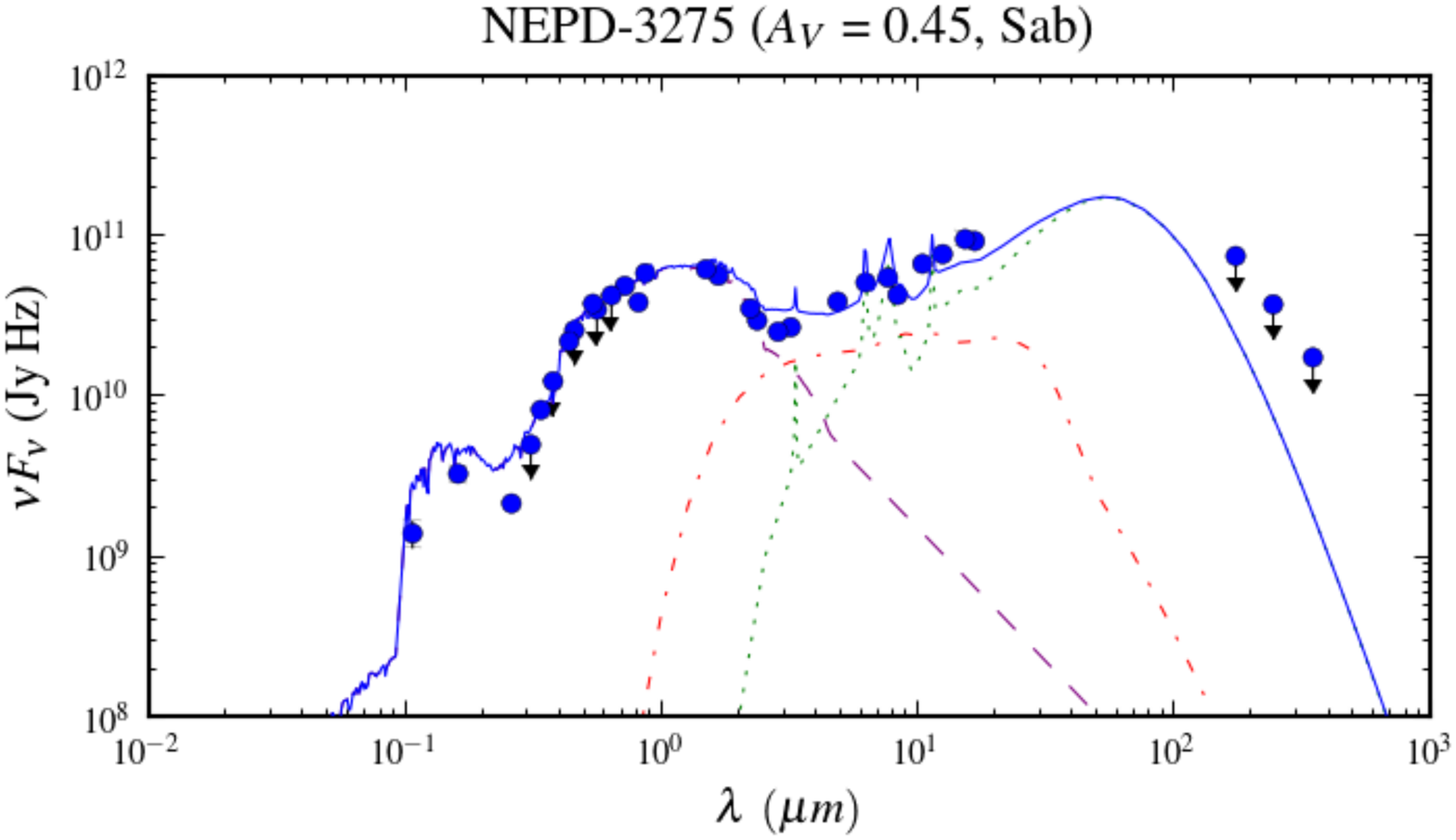}\\
\includegraphics[width=0.245\textwidth,angle=0]{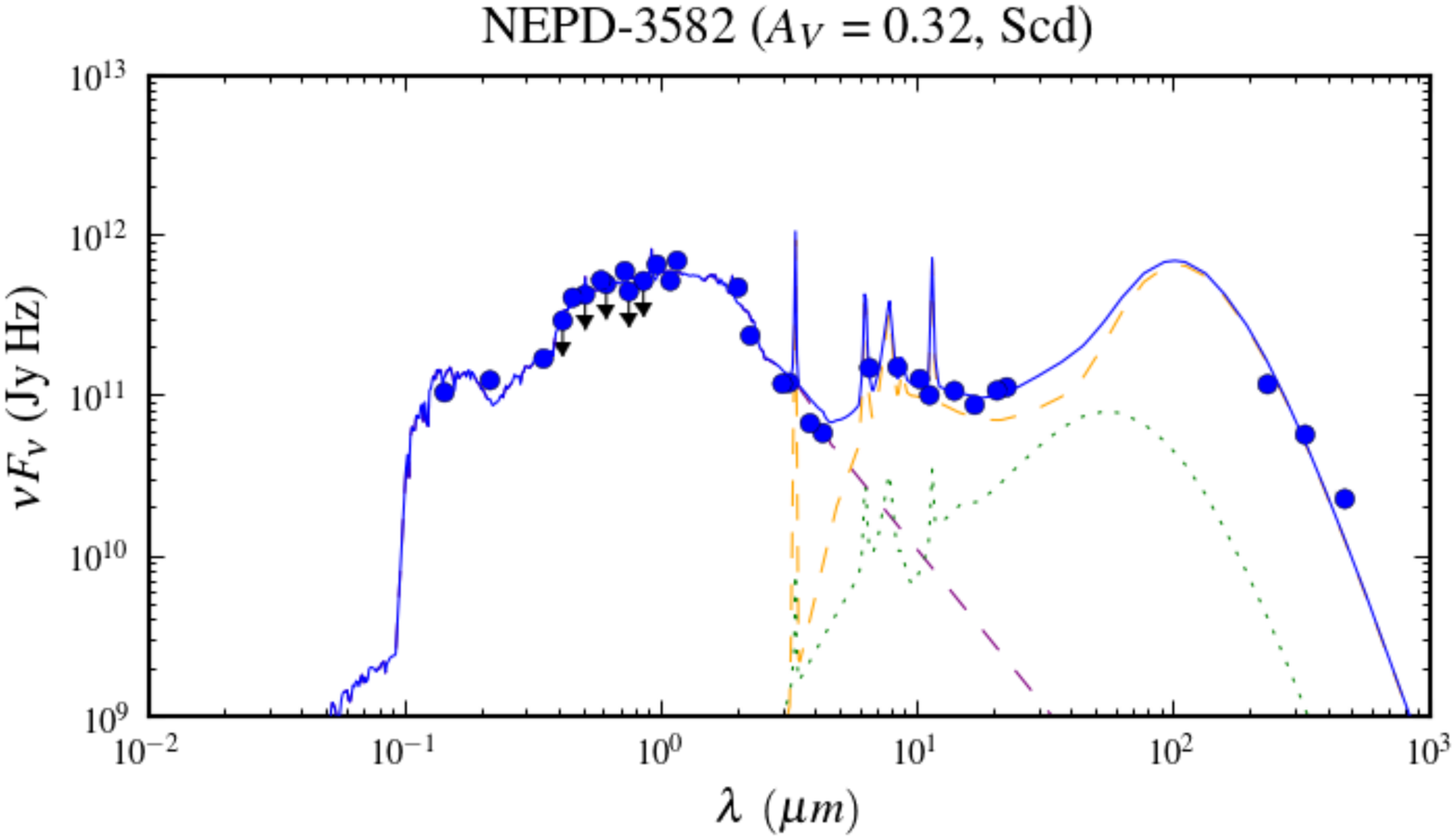}
\includegraphics[width=0.245\textwidth,angle=0]{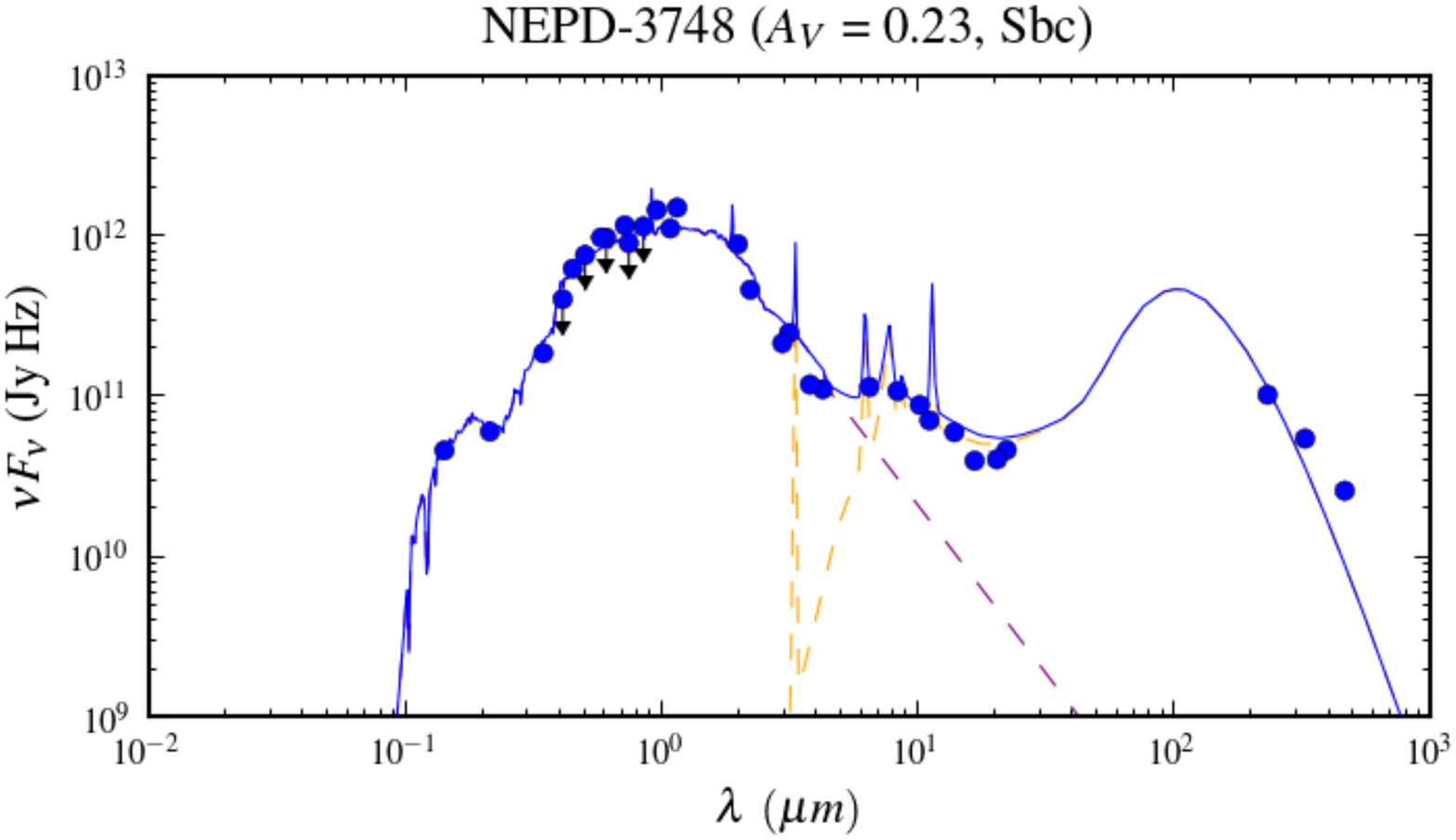}
\includegraphics[width=0.245\textwidth,angle=0]{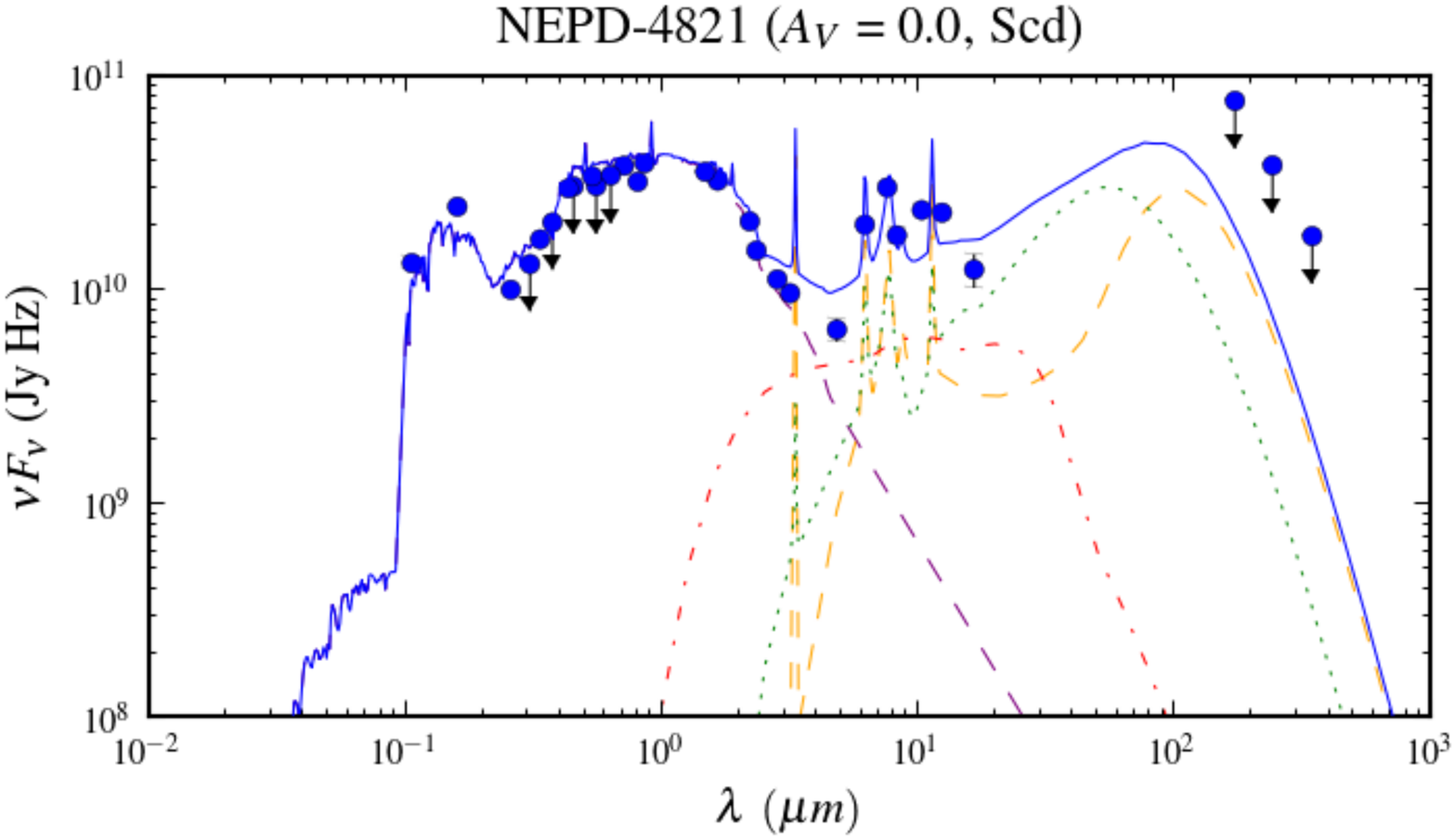}
\includegraphics[width=0.245\textwidth,angle=0]{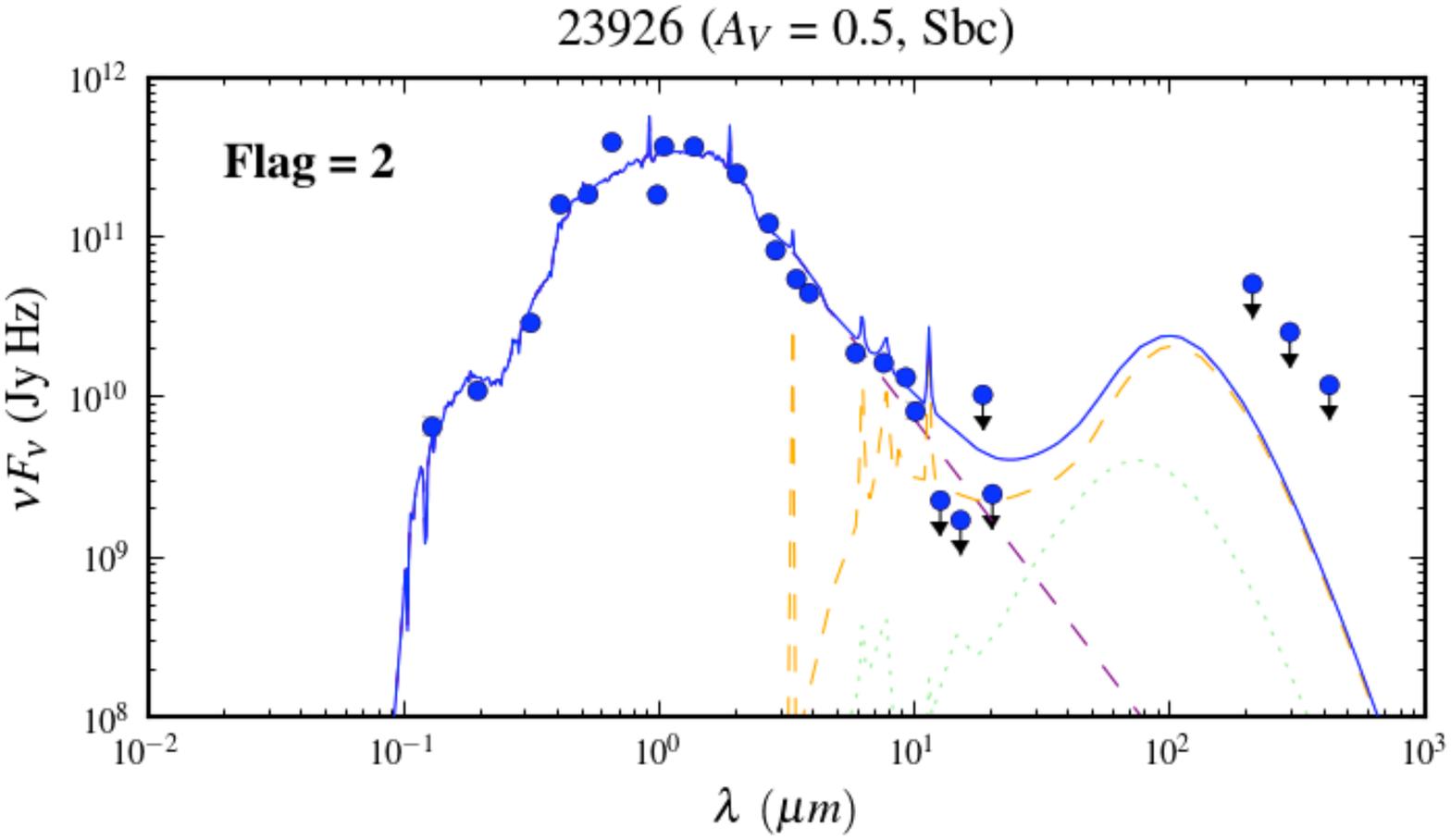}\\
\includegraphics[width=0.245\textwidth,angle=0]{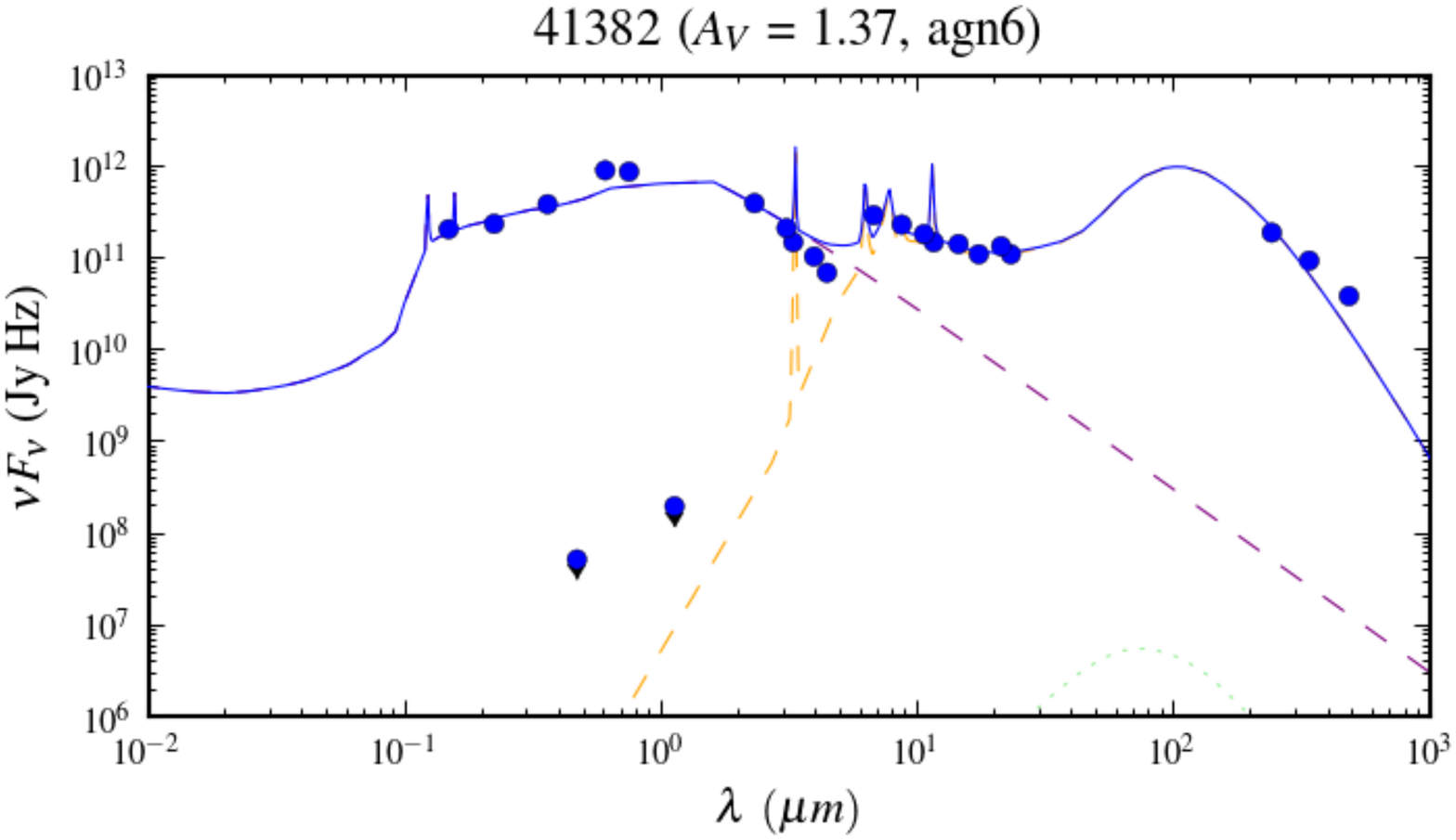}
\includegraphics[width=0.245\textwidth,angle=0]{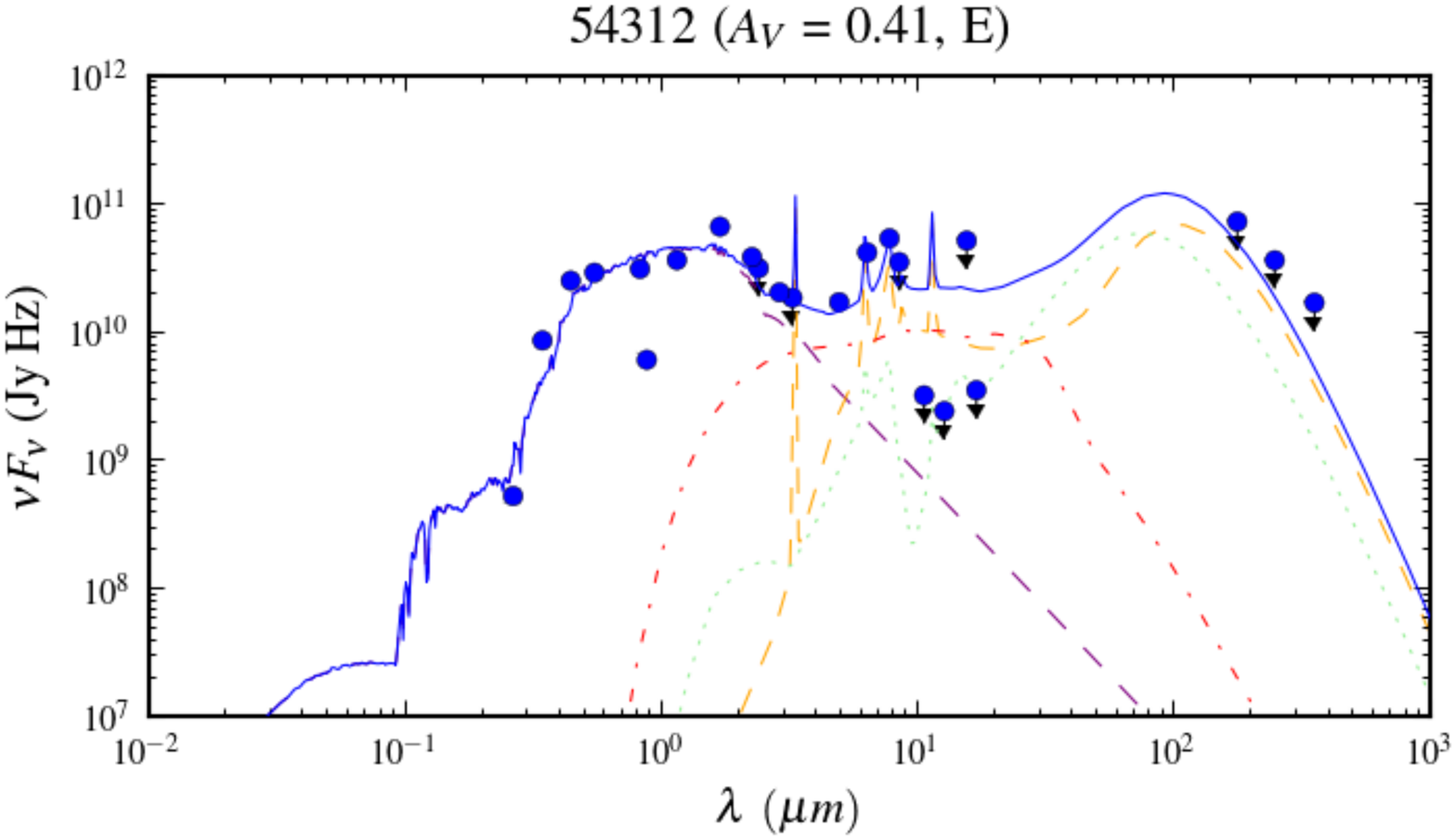}
\includegraphics[width=0.245\textwidth,angle=0]{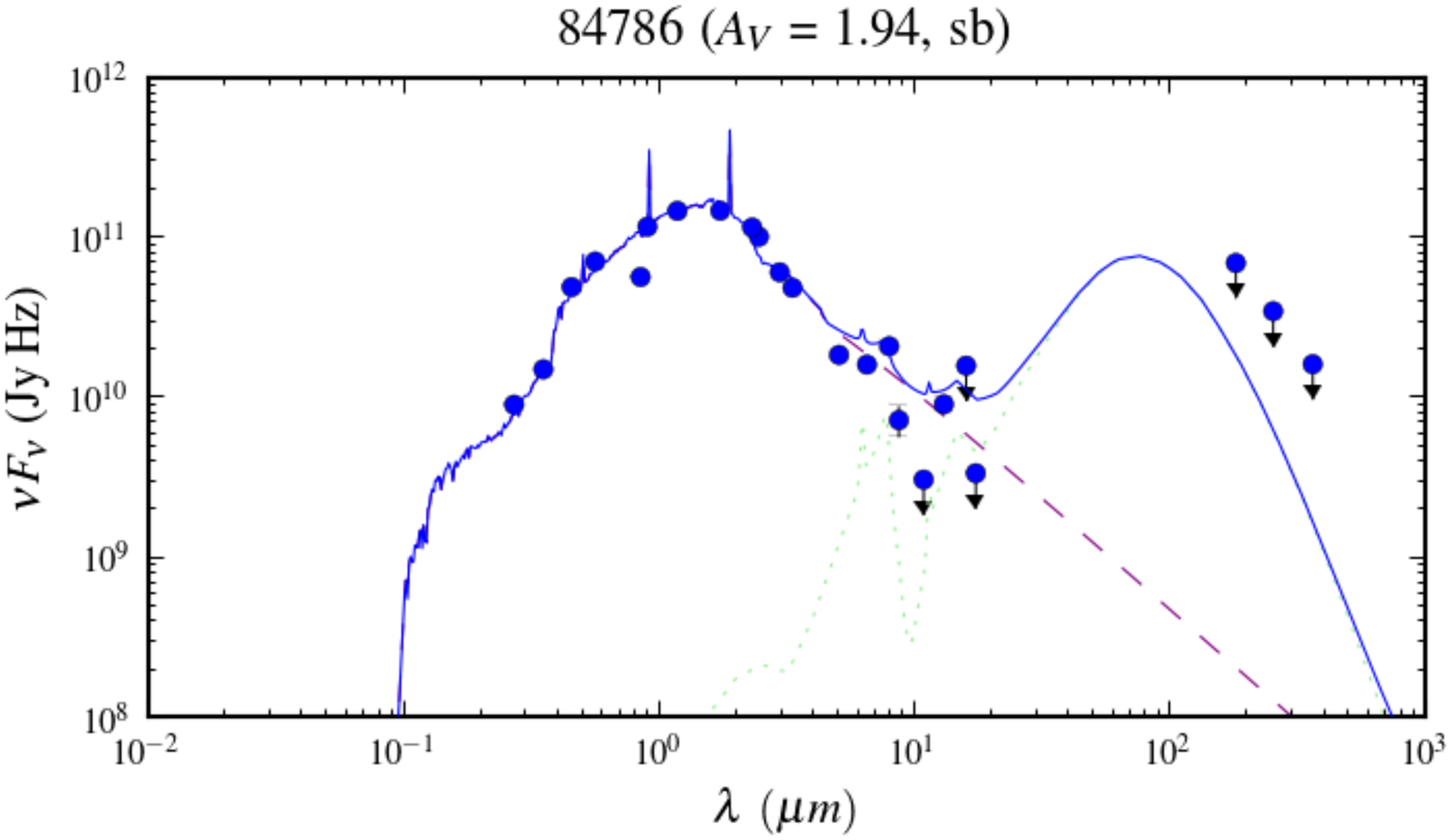}
\includegraphics[width=0.245\textwidth,angle=0]{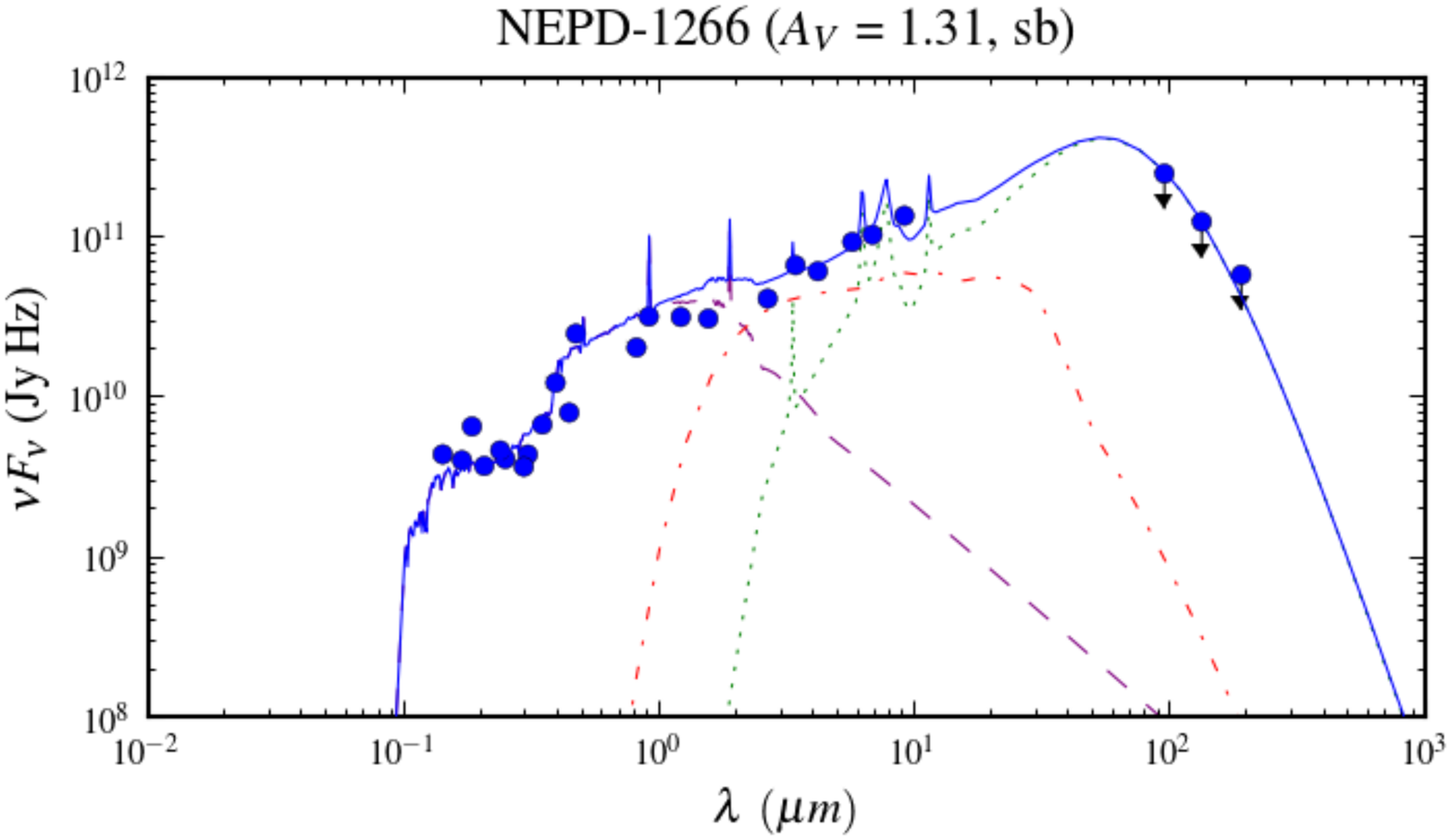}\\
\includegraphics[width=0.245\textwidth,angle=0]{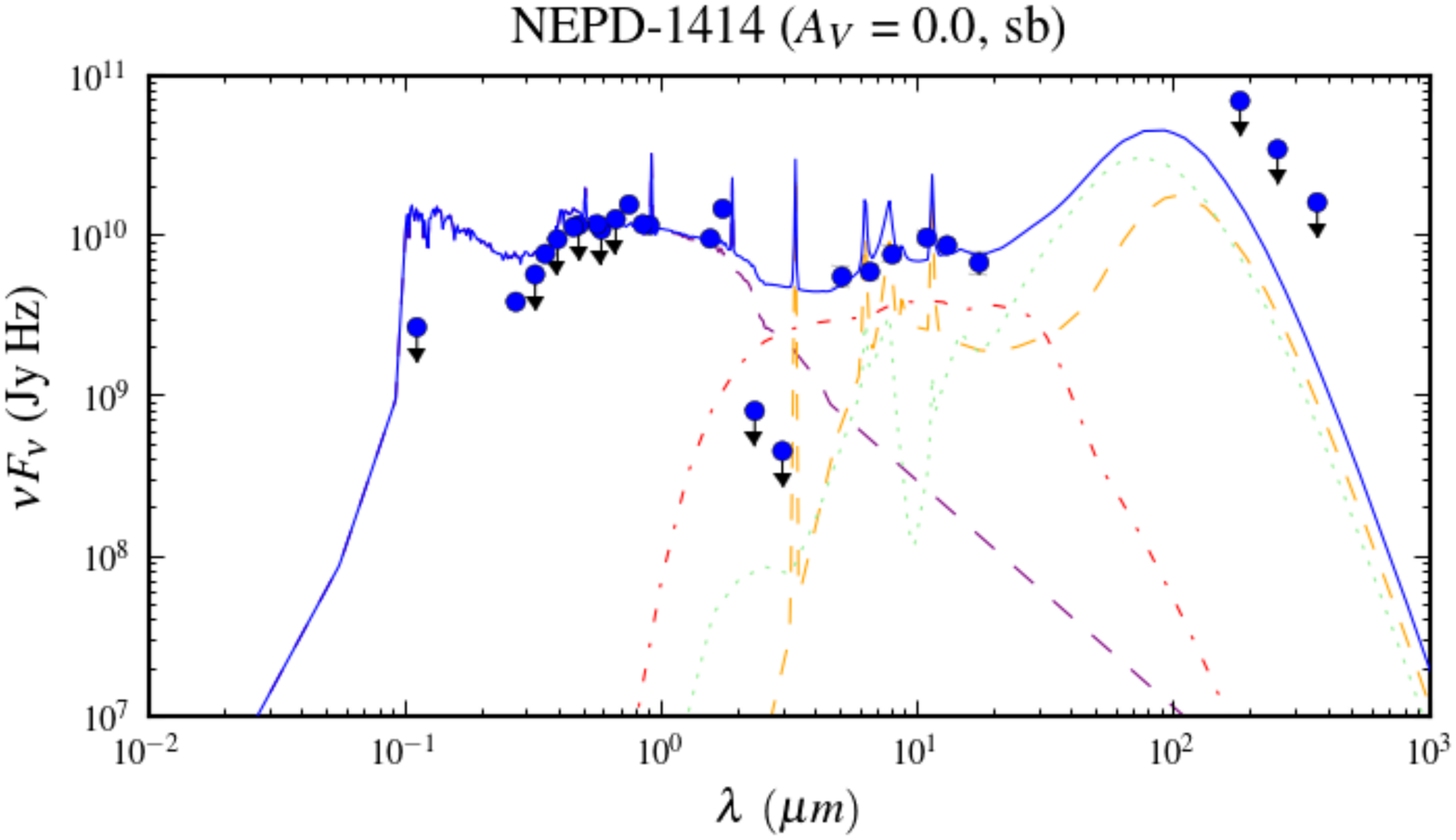}
\includegraphics[width=0.245\textwidth,angle=0]{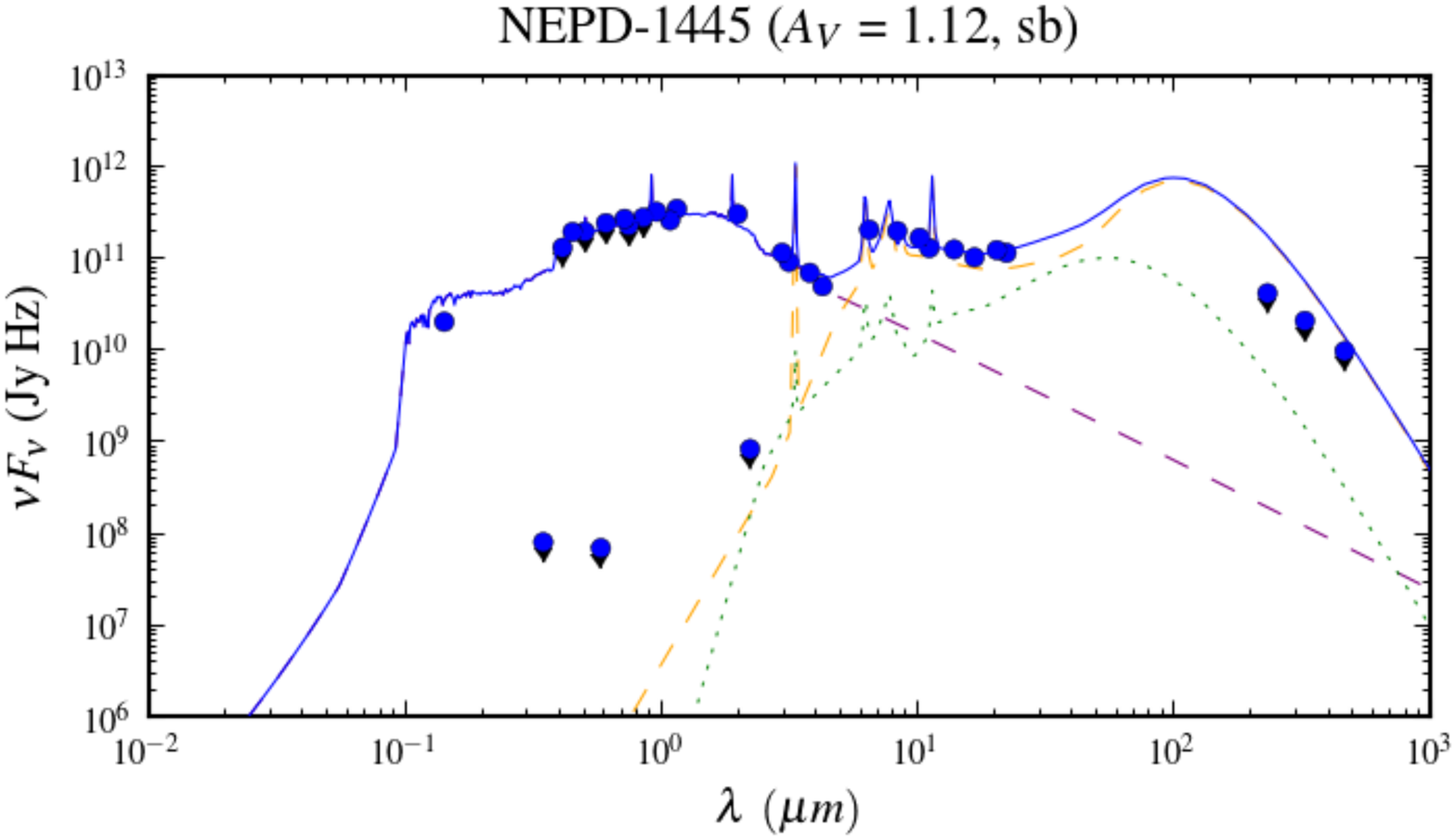}
\includegraphics[width=0.245\textwidth,angle=0]{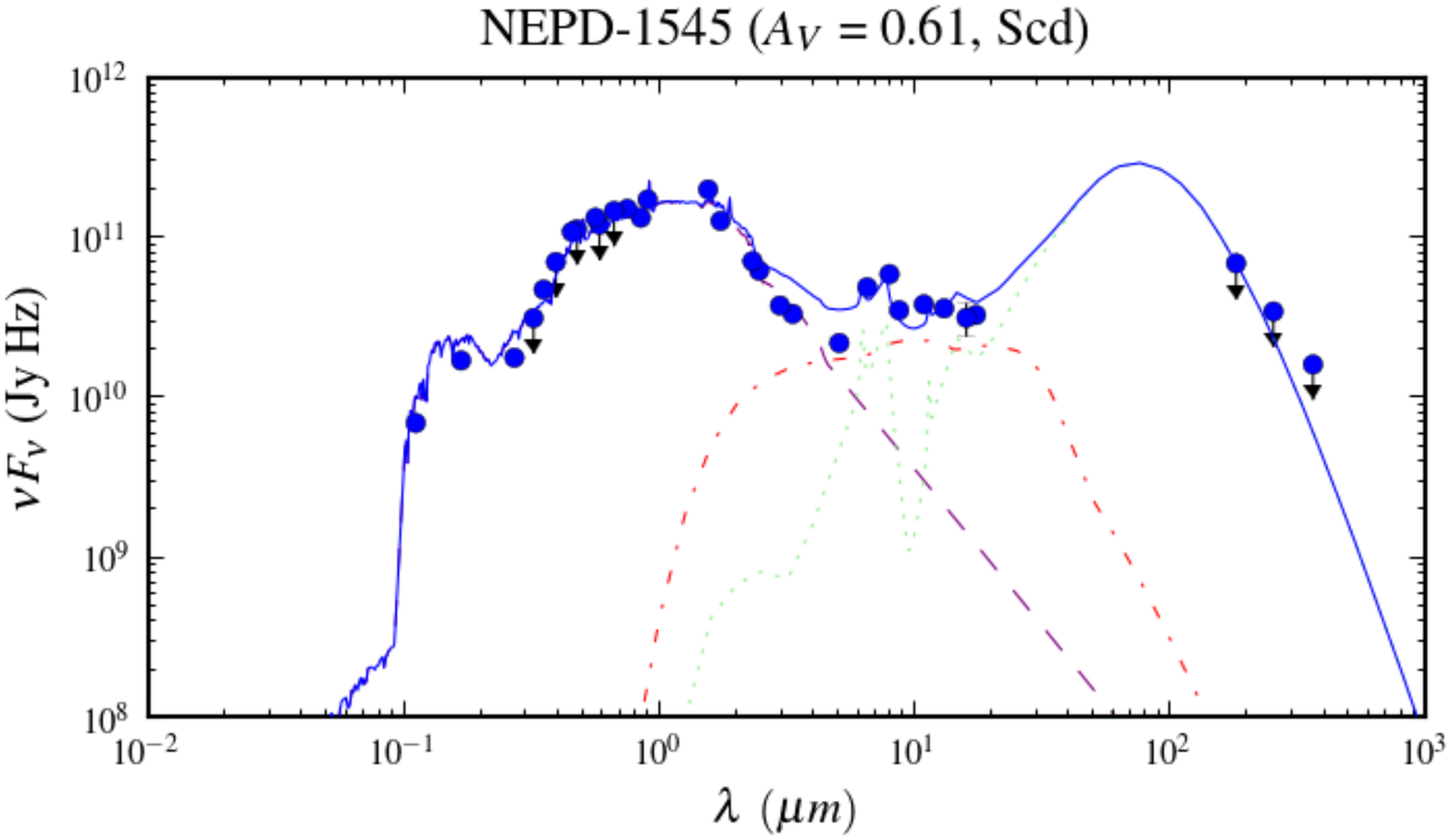}
\includegraphics[width=0.245\textwidth,angle=0]{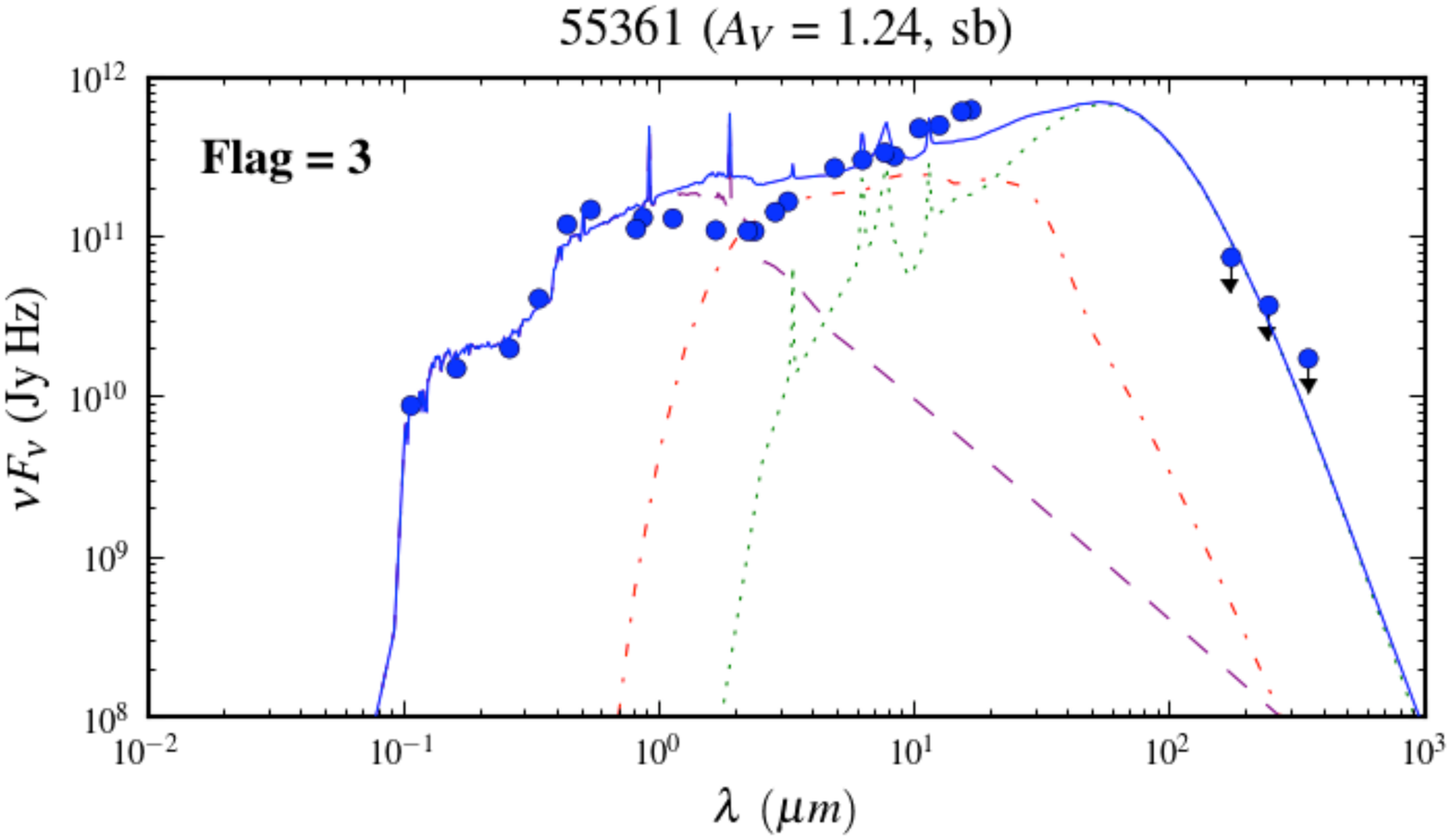}
\caption{Individual SED fits for the 28 sources with good quality spectroscopic redshifts. The different components used in the fit are shown: optical component (dashed purple), cirrus template (dashed orange), starburst component (dotted green), and AGN dusty torus component (dot-dashed red). The total SED is shown with the solid blue line. The name, optical extinction value $A_v$, and optical template are shown on the top of each plot. The SED plots are arranged in terms of their visual flag, from left to right and from top to bottom. The first plot of each quality group (1, 2 ,3) is denoted as such.}
\label{fig:SED_fit_example}
\end{center}
\end{figure*}

We employed the set of IR templates from \citet{Rowan2008}. It includes a cirrus template (IR emission from a quiescent galaxy), an AGN dust torus and two SB templates (M82 and Arp 220). Our IR model is a combination of three components: cirrus + SB + AGN. We tested this model with each SB template, and selected as best-fit the one with the lowest $\chi^2$. In Fig. \ref{fig:SED_fit_example} we show examples of template SED fittings (including the source also shown in Fig. \ref{fig:sed}). After selecting the best fit (lowest $\chi^2$ value) for each source, we also performed a visual check of each SED and assigned quality flags to each fit. Good fits were assigned a quality flag of 1 (the fitted SED follows all the data points closely), satisfactory fits were assigned a quality flag of 2 (one or two data points are not fitted well by the best-fit SED), while a flag value of 3 was given to bad fits (the fitted SED fails to fit more than two bands and/or large deviations are observed overall). For the following, only quality 1 (95 sources) and 2 (86 sources) fits will be considered. SED fits with flag 3 are most probably a result of either (a) wrong photometric redshift estimation or (b) a problem with the cross-matching between the different datasets. Despite the adequate range of templates employed for our fits, it is also plausible that for a subset of these sources no combination of our templates can explain their total emission leading to a bad quality fit. Finally, as was demonstrated through the process of the photometric redshift estimation, we expect that there are photometric offsets affecting each of our bands differently. Given that these can be constrained only in terms of a given template library and a training sample of spectroscopic sources, we can not directly apply the photometric offsets calculated from \textit{LePhare} to the IR-radio sources. As a result the quality of our SED fits can in some cases be further degraded.

The following parameters are derived through the SED fitting process:
\begin{itemize}
\item optical extinction,
\item relative contribution of each of the IR templates (cirrus, M82, Arp 220, AGN torrus) to the 60 $\mu$m rest-frame emission,
\item predicted flux at 60 $\mu$m rest-frame,
\item extinction corrected optical luminosity (0.1 - 3 $\mu$m),
\item IR luminosity of each IR component,
\item total IR luminosity (L$_{8-1000\mu m}$).
\end{itemize}

The above can be combined to derive further values such as:
\begin{itemize}
\item total AGN luminosity ($L_{AGN;tot}=L_{AGN;opt}+L_{AGN;IR}$),
\item total (bolometric) source luminosity ($L_{SED;tot}=L_{SED;opt}+L_{SED;IR}$),
\item total AGN fractional contribution ($L_{AGN;tot}/L_{SED;tot}$),
\item IR AGN fractional contribution ($L_{AGN;IR}/L_{SED;IR}$),
\item star-formation luminosity ($L_{SF}=L_{M82}+L_{Arp220}+L_{cirrus}$).
\end{itemize}

\subsection{Spectroscopic Sub-sample as a Benchmark}
There is an overlap between our IR-radio sample and the sources targeted for spectroscopic investigation in the NEP field. This sub-sample of our sources consists of 32 objects that have spectroscopic redshifts. For 28 these sources a good redshift quality flag has been given (\citealt{Shim2013} and Takagi et al., in prep.), while in addition visual inspection of the spectra and use of the BPT diagram (\citealt{Baldwin1981}) has led to optical classification for these objects into Type 1 and  Type 2 AGN and non-active galaxies. Out of these 28 objects, 1 object is assigned a quality flag of 3 for its SED fit and is thus excluded from the following comparison, 8 are assigned a quality flag of 2 and the rest 19 have a visual quality flag of 1. We use the 27 sources with quality flag 1 or 2 to get a handle on the accuracy of our SED fitting both in terms of the type of AGN templates used for the fits as well as the derived AGN bolometric fractional contribution.\\
In Table \ref{tab:comparison} we present the fraction of spectroscopically identified AGN recovered by our SED fitting method. Here we define as AGN-dominated systems those whose SED fit give an AGN component bolometric luminosity fractional contribution higher than 50$\%$. For AGN contribution between 1$\%$ and 50$\%$ a source is defined as an AGN-composite. Although the lower limit for AGN-composite systems of 1$\%$ appears low, for several very luminous objects even 1$\%$ contribution to the bolometric luminosity results to considerable AGN luminosity. All 5 broad emission-line spectrum objects (assumed to be Type 1 AGN) are fitted with AGN templates, although they are not classified as AGN-dominated according to our definition above. There are two type 2 AGN identified through the BPT diagram and of these one is fitted with an AGN template, as an AGN-composite source, while one does not require an AGN for its SED fit. Finally, out of the 19 sources identified as non-AGN from their spectra, 12 are indeed fitted with no AGN template, while 1 of them is fitted as an AGN-dominated system and 6 as AGN-composites. Of the sources fitted with an AGN component but lacking AGN signatures in their optical spectra, only two are fitted with an optical AGN component (both of which show high extinction A(V) values of 3.9 and 1.4 mag) while the remaining 5 are the AGN component is identified through their mid-IR photometry. One spectroscopic source lacks optical classification and from its SED fit, it is classified as a composite-AGN.\\
Although our SED fits and spectral information do not match completely, there is a qualitative agreement, in that most AGN are not missed, while a sizable fraction of potentially optically ``hidden'' AGN are recovered through their IR emission.

\begin{table}
\caption{Tabular comparison between results of spectroscopic classification and classification from the SED fitting presented in this work. This pertains to a sub-sample of the IR-radio sources of 29 objects observed spectroscopically and with ``Good'' or ``Moderate'' quality SED fits. Column (1) gives the spectroscopic classification either visual or through the BPT diagram. Columns (2-4) give the fraction of AGN-dominated, AGN-composite, and non-AGN classified objects from our SED fits. Column (5) gives the number IDs of the templates used for the fits (1: elliptical template, 2-5: spiral templates, 6: starburst template, 7-9: AGN templates). Spectroscopic classifications are taken from \citet{Shim2013} and Takagi et al. (in prep.).}
\begin{center}
\begin{tabular}{|c|c|c|c|c|}
\hline
Spectra	&	AGN-dom		&		AGN-comp		&		Non-AGN			&	Templates\\
\hline
T1		&		0/5 			&		5/5 				&		0/5			&	2-6		\\
T2		&		0/2 			&		1/2 				&		1/2 			&	1,5		\\
G		&		1/19			&		6/19 				&		12/19 		&	1-6,9		\\
\hline
\end{tabular}
\end{center}
\label{tab:comparison}
\end{table}

\subsection{Importance of far-IR data for Constraining AGN and SFR}
\label{sec:AGNSPIRE}

\begin{figure}[htbp]
\begin{center}
\includegraphics[width=0.49\textwidth,angle=0]{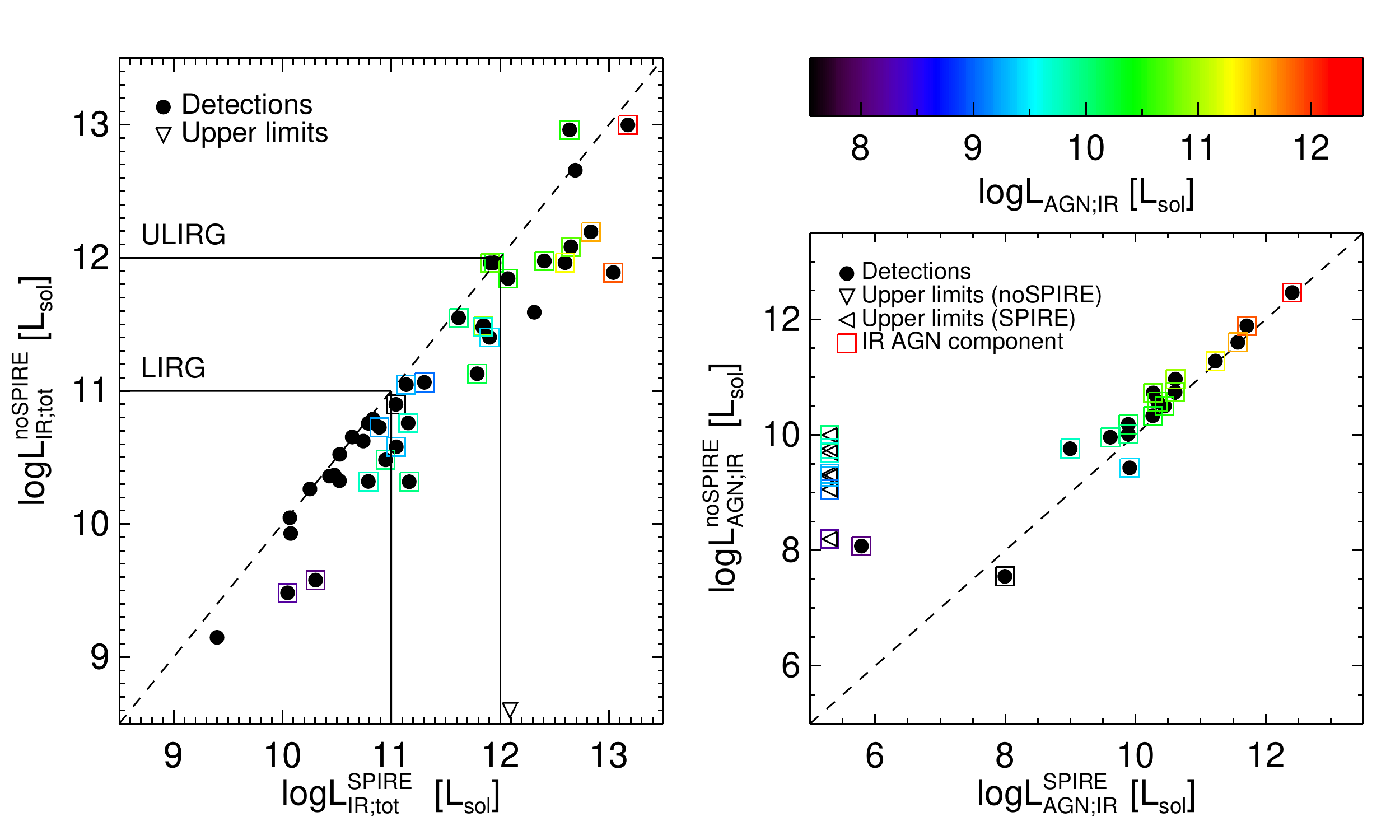}
\caption{Left: Total IR luminosity from the SED fitting without Herschel-SPIRE bands compared to total IR luminosity from the SED fitting when considering the Herschel-SPIRE photometry. We also include one source as an upper limit for which no IR luminosity could be calculated in the absence of Herschel data. Separate regions of LIRGs and ULIRGs are shown. Right: IR AGN luminosity from the SED fitting without Herschel-SPIRE bands compared to the IR AGN luminosity for the SED fitting when considering the Herschel-SPIRE photometry. Squares show sources fitted with the AGN torus templates, with the color-scale showing the IR luminosity of the AGN-component. Upper limits for quantities with and without SPIRE bands are shown too. For both panels, the diagonal dashed line shows the one-to-one relation for reference.}
\label{fig:SPIREcomp}
\end{center}
\end{figure}

In the previous section we investigated the AGN recovery success rate using the spectroscopic sub-sample as a benchmark. Given the rich multi-wavelength dataset at our disposal, we study how the availability of certain wavelengths affects the outcome of our SED fitting, especially in terms of the AGN component. In particular we are interested in the usefulness of far-IR data from Herschel. As we already mentioned, the contribution of the AGN component to the far-IR part of the SED is expected to be minimal. On the other hand, far-IR emission is crucial for constraining the instantaneous star-formation rate of a galaxy. To that end, for sources with available SPIRE data, we compare SED fits with and without the Herschel-SPIRE data points. In Fig. \ref{fig:SPIREcomp} we show the comparison between the two cases for the total IR luminosity (L$_{8-1000\mu m}$;left) and the AGN component IR luminosity. As the SED fitting process involves separate fitting of the optical/NIR and mid-IR/far-IR parts of the broadband SED, the determination of an optical AGN is excluded from this comparison. As expected, the total IR luminosity is significantly under-estimated in the absence of far-IR points, despite the presence of mid-IR measurement up to 24 $\mu$m. The difference is of the order of half a magnitude, apparently independent of total IR luminosity itself. The dusty AGN torus component is more significantly affected by the inclusion of far-IR measurements. The most obvious difference concerns a sizable fraction of sources that, while in the absence of far-IR points are fitted with an IR AGN component, in actuality appear to not be AGN at all. These sources are shown as leftward facing arrows in the right panel of Fig. \ref{fig:SPIREcomp}. These sources are exclusively found at the lower AGN luminosity range. Interestingly enough however, those sources that do appear to be genuinely hosting an AGN component, have their luminosities accurately measured even in the absence of far-IR data points. From this we infer that the far-IR emission coming from the AGN is negligible and therefore SEDs covering up to mid-IR wavelength are sufficient to reliably constrain the AGN luminosity. In particular, especially at AGN luminosities higher than $10^{10}L_{sol}$, the availability of far-IR data does not seem to affect the determination of AGN properties, neither in terms of whether one is present or not, nor about its luminosity. \\
As a final remark here, it should be noted that given that we are using empirical templates to estimate AGN luminosities and that AGN, especially at the lower end of the AGN luminosity function, are poorly studied in the far-IR, we can not draw any robust conclusion concerning the contribution of an AGN component to the far-IR. In a following paper we shall study this question in more depth and assess the importance of any AGN contribution at these very long IR wavelengths.

\subsection{Identifying AGN and Constraining SF}
We have described how we treat the broadband SEDs during the fitting process and we have also assessed both the success rate of AGN identification and how the availability of far-IR data affects our results. It is interesting to briefly note the main SED features that appear to dominate the identification of AGN and the constraining of star-formation in our sources. In terms of the optical and near-IR part of the SED, AGN are usually characterized by a power-law like continuum (e.g., source 41382 in Fig. \ref{fig:SED_fit_example}), modulated by broad emission lines for the case of Type 1 AGN. In practice optical AGN are characterized by bluer colors, due to the influence of the accretion disk that dominates the emission. Given our radio-IR selection we do not recover many of these type of AGN. On the other hand, a prominent near-IR bump (at around 1.6 $\mu m$) is a result of the emission of old stellar populations and characterizes passive (non star-forming) galaxies. For the IR part of the SED, again AGN are somewhat characterized by a power-law that extends well into the near-IR, which is complemented by the warm/hot dust emission from the obscuring torus surrounding the accretion disk (e.g., \citealt{Antonucci1993}, \citealt{Urry1995}). Assuming a homogeneous torus that emits isotropically, such emission is usually approximated by a black body that dominates the emission in the mid-IR (around $\sim 24\,\mu m$, e.g., source NEPD-3275 in Fig. \ref{fig:SED_fit_example}). Beyond the mid-IR and into the far-IR the contribution of the AGN becomes negligible, while emission from a cold dust component heated by newly formed stars dominates. This results to a prominent far-IR bump, characteristic of star-forming galaxies, peaking at wavelengths $\sim150-200$ $\mu m$ (e.g., source 70827 in Fig. \ref{fig:SED_fit_example}). The SPIRE data therefore, although missing the actual peak of the far-IR bump, can constrain the fall-off of the bump and hence the level of star-formation in a source. Given the homogeneous coverage of the whole of NEP field with the SPIRE instrument, we can place constraints on the instantaneous SFRs of non-detected sources through upper flux limits (e.g., source 53756 in Fig. \ref{fig:SED_fit_example}).

\section{AGN content}
\label{sec:agn}

We are interested in investigating the AGN content of our radio sources and in particular in identifying those systems that show a composite behavior, their SEDs revealing signatures of both nuclear and star-forming activity. This is done purely and explicitly in terms of our SED fitting results, unlike previous studies where individual colors where employed to the same end (e.g., \citealt{Takagi2007}, \citealt{Hanami2012}). Out of the 237 sources fitted, 47 are fitted with one of the optical AGN templates, 111 are fitted with the IR AGN template, while 106 sources do not require any AGN contribution for their SED fits. There are 27 sources for which both an optical and IR AGN template is used for their SED fit. The majority of the sources has its optical emission fitted by one of the spiral galaxy templates (135), while only 33 sources are fitted with an elliptical template galaxy. 

\subsection{AGN Contribution in Radio-sources}
\begin{figure*}[ht]
\begin{center}
\includegraphics[width=0.45\textwidth,angle=0]{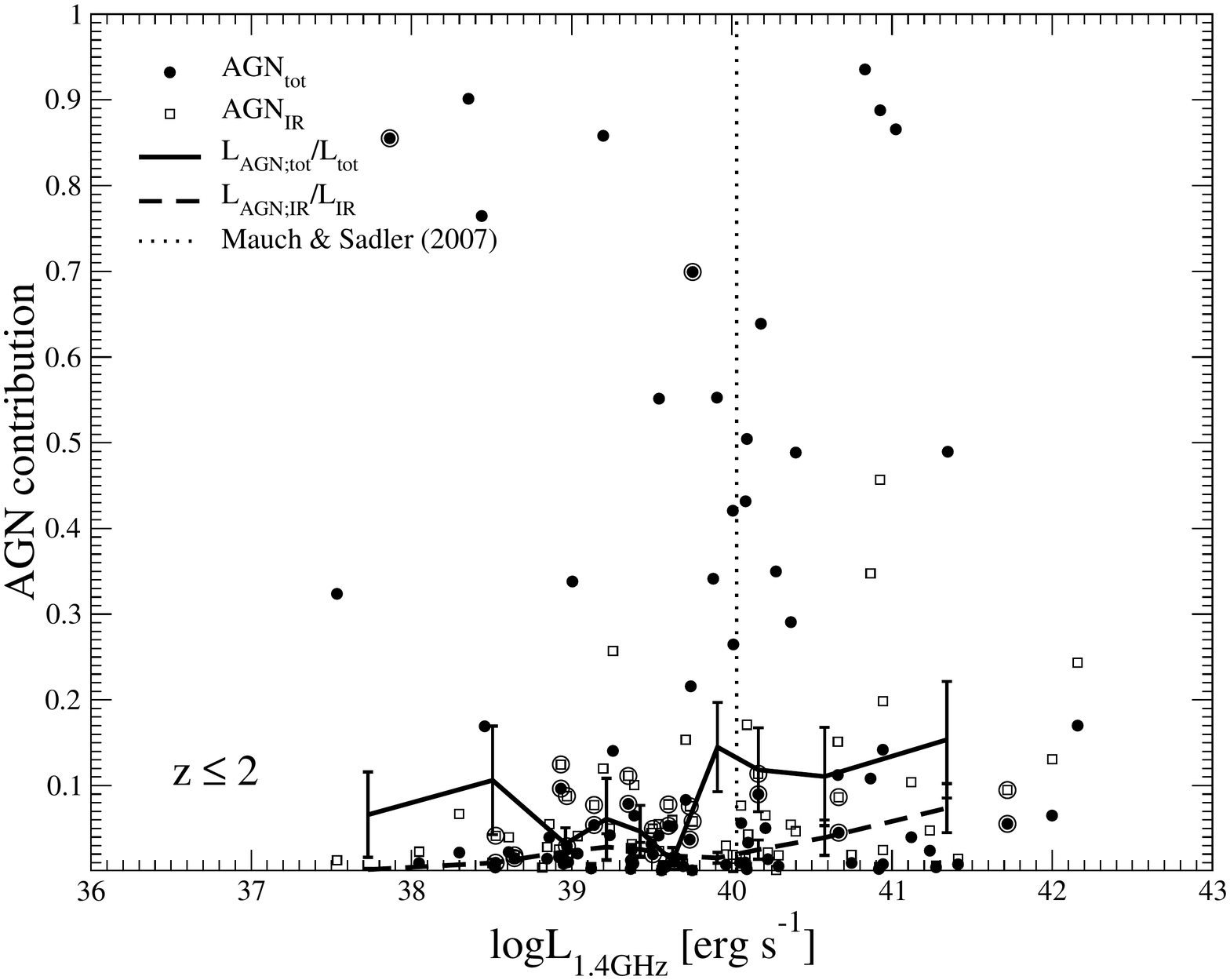}
\includegraphics[width=0.45\textwidth,angle=0]{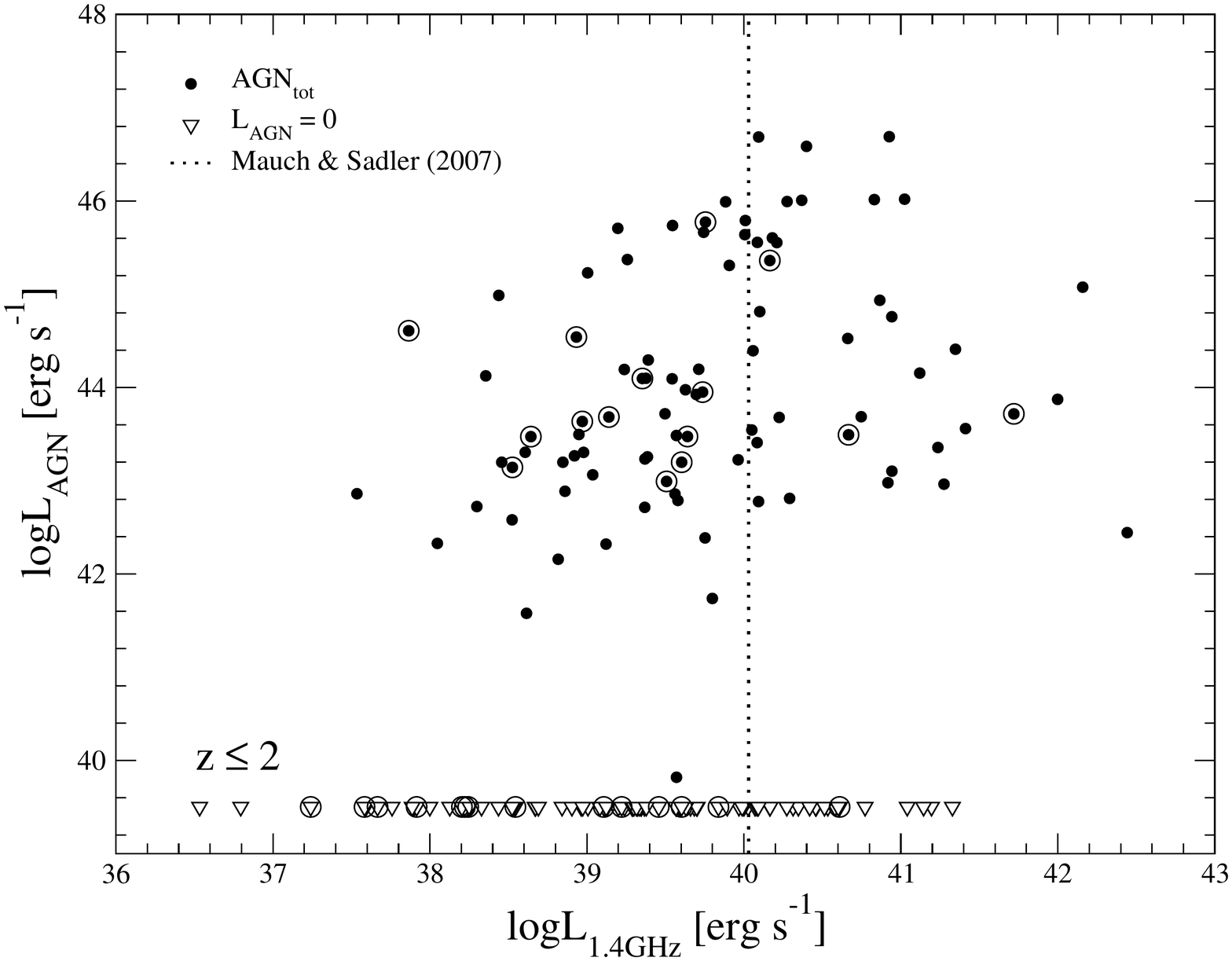}
\caption{Left: The fractional contribution of the AGN component to the bolometric luminosity as derived from the SED fits as a function of total radio luminosity at 1.4 GHz (0 denotes a non-AGN fit, while a value of 1 would denote a purely AGN-fitted SED). Points denote individual source values while the lines show averaged values over radio luminosity bins. Both the total AGN contribution (optical + IR; filled circles) and the IR AGN contribution (open squares) are plotted. Right: Absolute bolometric AGN luminosity (optical + IR) as a function of total radio luminosity at 1.4 GHz. With open downward triangles we show sources fitted without an AGN component. The vertical black dashed line shows the rough luminosity cut where the luminosity function of local radio-sources gets dominated by AGN systems (from \citealt{Mauch2007}). Sources with spectroscopic redshifts are denoted with an additional red circle.}
\label{fig:AGN_Lr}
\end{center}
\end{figure*}

\begin{figure*}[ht]
\begin{center}
\includegraphics[width=0.45\textwidth,angle=0]{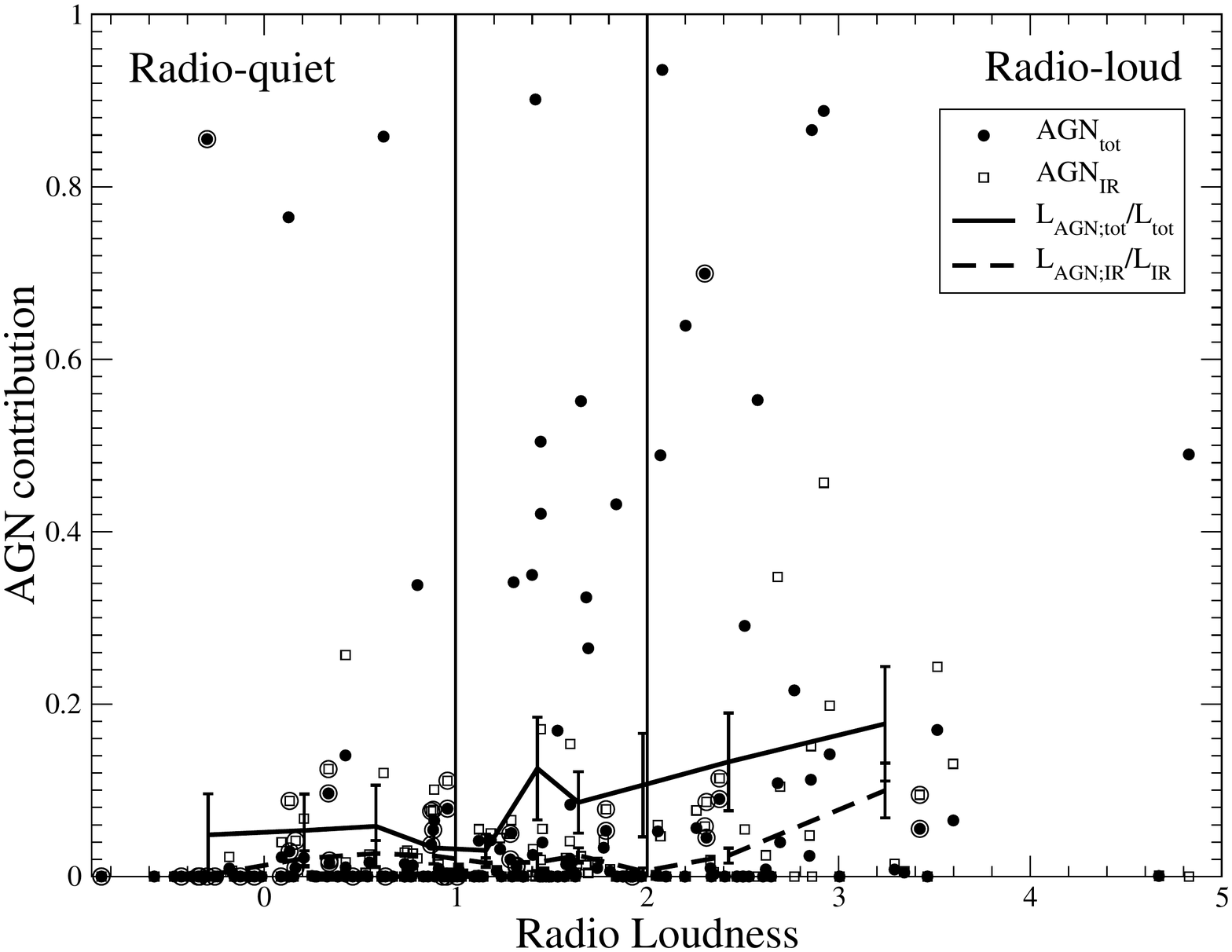}
\includegraphics[width=0.45\textwidth,angle=0]{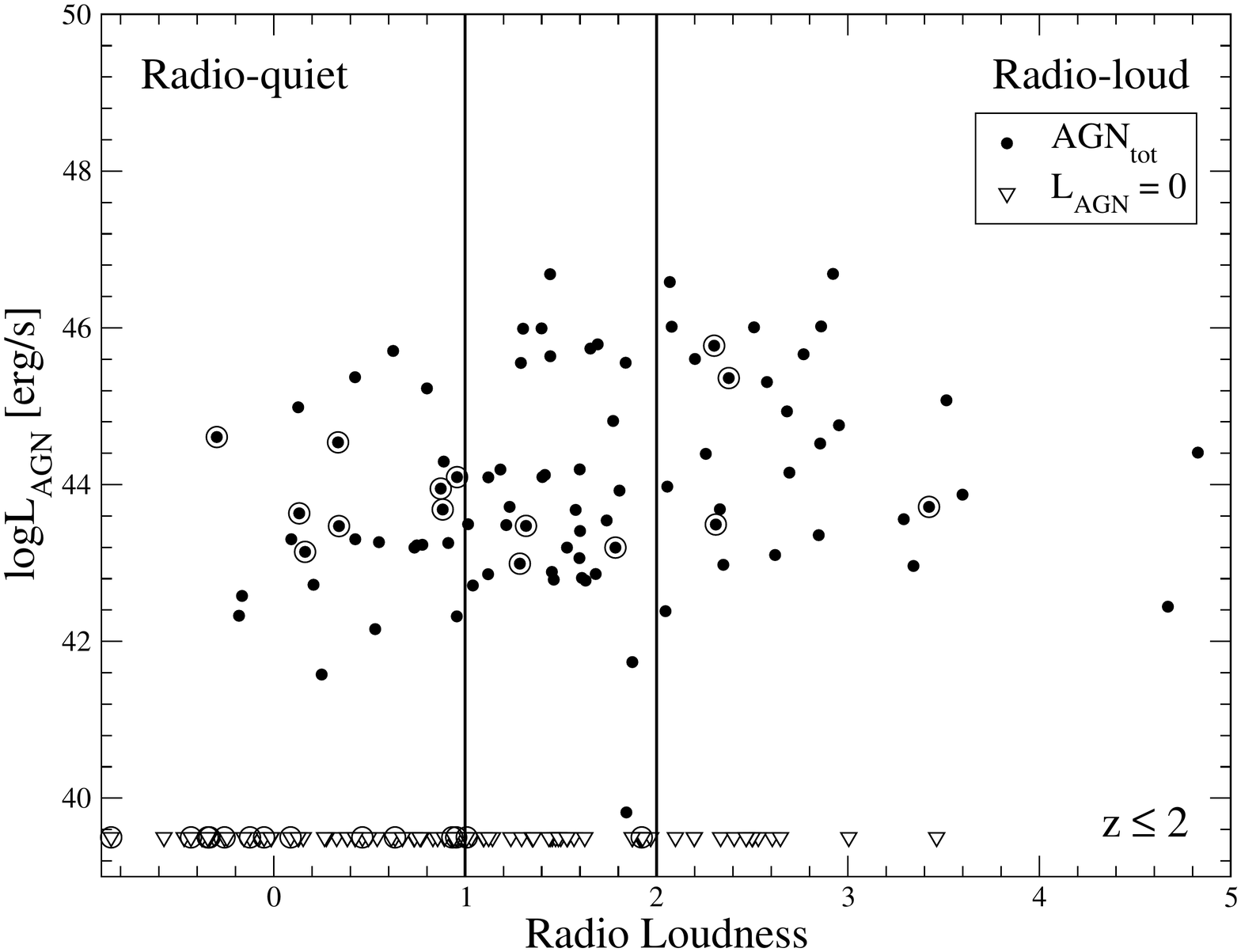}
\caption{Same as in Fig. \ref{fig:AGN_Lr} but as a function of radio loudness R$_i$. The vertical black lines denote the limits for radio-quite and radio-loud regime.}
\label{fig:AGN_Ri}
\end{center}
\end{figure*}

In using the SED fittings results we can firstly see how the AGN luminosity fractional contribution as well as absolute luminosity behaves as a function of both radio luminosity and radio-loudness. The results are shown in Figs. \ref{fig:AGN_Lr} and \ref{fig:AGN_Ri}.\\

We observe that both for the highest radio luminosities, as well as for the highest radio loudness values, there is a moderate excess of sources with a significant contribution to their bolometric luminosity by an AGN component. These are the radio-AGN systems in which we are interested in terms of their star-formation properties. We also see that the rise in the AGN contribution in terms of bolometric luminosity agrees well with previous studies and it starts roughly at around $10^{40}$ erg s$^{-1}$ (e.g., \citealt{Mauch2007}). However, we also see significant scatter, with AGN identified at the lowest radio luminosities (presumably radio-quiet AGN) as well as non-AGN found in the highest radio-luminosities (presumably vigorously star-forming galaxies). This drives the point that at the faintest radio fluxes the fraction of AGN is comparably smaller to that found in the bright radio surveys (like the FIRST and the NVSS), while the radio population is mainly dominated by star-forming (e.g., \citealt{Seymour2004}, \citealt{Mainieri2008}) or potentially composite galaxies (e.g, \citealt{Strazzullo2010}).

Several other points of interest arise from Figs. \ref{fig:AGN_Lr} and \ref{fig:AGN_Ri}. Even at the lowest radio luminosities, where star-formation should dominate, we find on average 10$\%$ contribution to the energy output from an AGN component. It should be noted however that despite the constraints coming from the Herschel-SPIRE upper flux limits, lack or scarcity of far-IR detections might be leading to an over-estimation of the AGN component for some of these sources (as was shown in Section \ref{sec:AGNSPIRE}). Similar behavior is seen for the most radio-quiet sources. Conversely, even for the highest radio-luminosities for most of these sources the bolometric luminosity is actually dominated by their stellar component. It should be noted that for both bright and faint ends of the radio luminosity and radio-loudness, we suffer from low-number statistics combined with possible photometric redshift effects. Sources misidentified as high-z galaxies, would be artificially placed at the high luminosity end of these plots, affecting the implied trends. Given the shallowness of the NEP-Wide survey, the cosmic volume probed is relatively small (compared to surveys likes FIRST and NVSS) and hence the probability of these brightest sources to be true is small (for the local Universe we have $\log{\Phi}={-7}\, mag^{-1}Mpc^{-3}$ for a radio 1.4 GHz luminosity of $\sim10^{42}$ erg s$^{-1}$).

Another way to look at the AGN content of our radio sample is by looking at the absolute AGN luminosity as a function of either $L_{1.4GHz}$ or $R_{i}$. This is shown in the right panels of Figs. \ref{fig:AGN_Lr} and \ref{fig:AGN_Ri}. We can see that for radio luminosities above $10^{40}$ erg s$^{-1}$ and radio-loud objects we have average AGN luminosities of the order of $10^{44}$ to $10^{45}$ erg s$^{-1}$, typical of luminous AGN. In addition we can see that the fraction of sources not fitted with any AGN component (shown as upper limits with downward triangles in the plots) decreases with increasing radio-luminosity and radio-loudness. At radio luminosities $L_{1.4GHz}<10^{40}$ erg s$^{-1}$ the fraction of sources with zero AGN component is 55$\%$, while at $L_{1.4GHz}>10^{40}$ erg s$^{-1}$ this falls to 40$\%$. It should be noted that radio-loudness exhibits significant scatter of $L_{AGN}$ and AGN fraction even well within the radio-loud regime. In particular it appears that there are some galaxies at the highest radio-loudness values, but with moderate to low AGN luminosity. We expect that radio-loudness, as defined here, might be contaminated by the emission coming from the SF component. This is potentially a result of (a) the low radio frequency (1.5 instead of 5 GHz, with star-formation decreasingly important at higher radio frequencies) and (b) the relatively narrow but very deep survey, which misses the most bright but also most rare radio-loud AGN that characterize the high end of the radio-luminosity function as measured, e.g., by \citet{Mauch2007}. Keeping the above in mind we will focus more on the radio-luminosity of a source as a tracer of its jet activity rather than its radio-loudness value.

\begin{figure}[ht]
\begin{center}
\includegraphics[width=0.49\textwidth,angle=0]{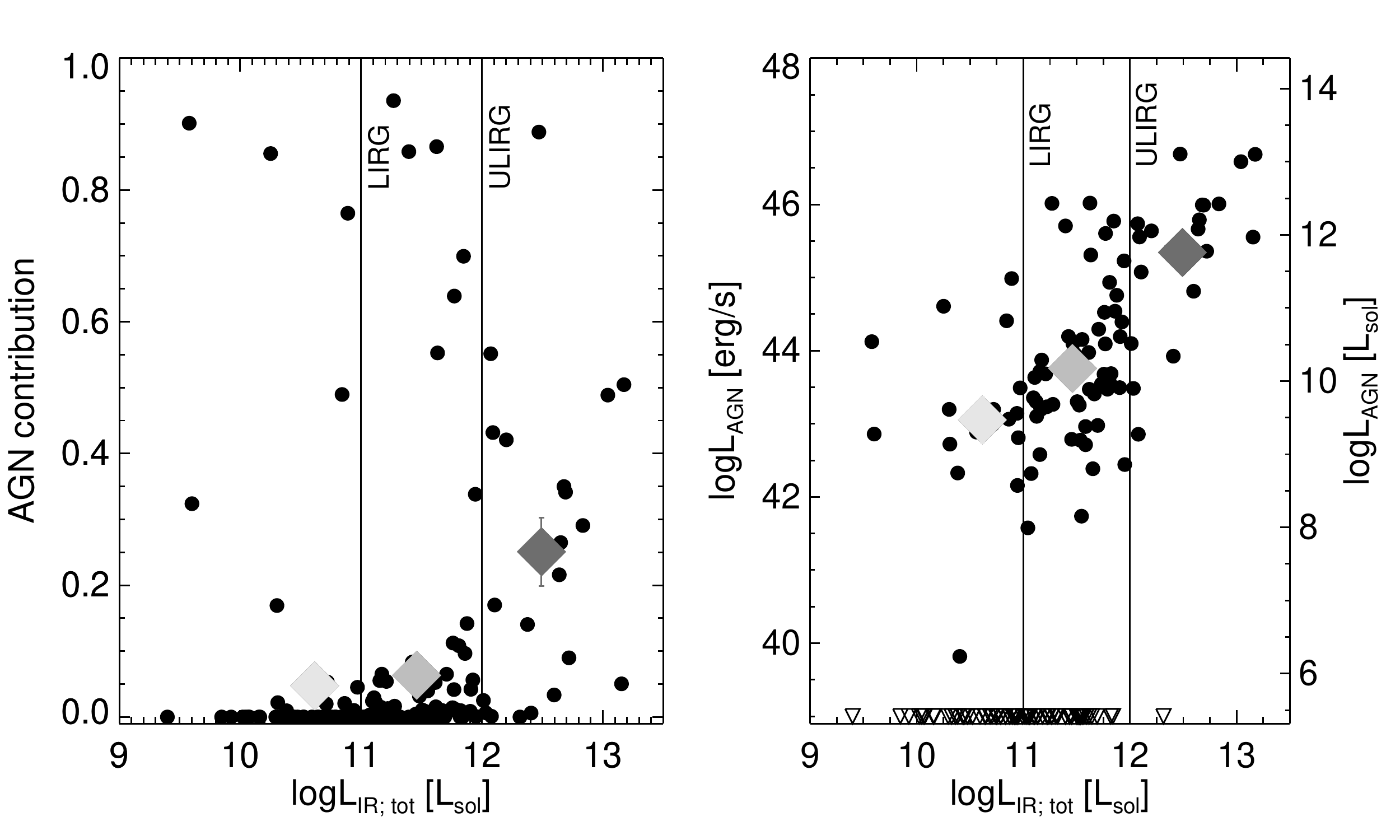}
\caption{AGN fractional contribution to a source's bolometric luminosity (left) and absolute AGN total luminosity (optical + IR; right) as a function of the total IR luminosity derived from the SED fits. Vertical lines denote the regions of luminous infrared galaxies (LIRGs; $L_{IR}>10^{11}$) and ultra-luminous infrared galaxies (ULIRGs; $L_{IR}>10^{12}$). For each IR luminosity regime we also plot average values of the AGN fraction contribution and AGN absolute luminosity. In the right panel we also note with downward triangles sources with no AGN component in their fitted SED.}
\label{fig:AGN_LIR}
\end{center}
\end{figure}

In Fig. \ref{fig:AGN_LIR} we plot the AGN fractional contribution to the bolometric luminosity (left panel) and the absolute AGN luminosity (right panel) as a function of the total IR luminosity of a source, as derived from the SED fit. We see that both these two values increase with IR luminosity, with sources classified as ULIRGs ($L_{IR}>10^{12}L_{sol}$) showing on average $\approx 30\%$ AGN contribution to their bolometric luminosities. On the other hand, as it is showcased in the right panel of the same figure, only one source classified as a ULIRG has no AGN component, with all other sources showing AGN luminosities of the order of $10^{46}$ erg s$^{-1}$. Although these are not new results (e.g., \citealt{Lutz1998}, \citealt{Veilleux1999}, \citealt{Imanishi2010}, \citealt{Pozzi2012}), they allow for a consistency check of both our IR luminosity determination and the identification of AGN features in our broadband SEDs.

\subsection{Mid-IR Colors and AGN Content}
\begin{figure}[htbp]
\begin{center}
\includegraphics[width=0.49\textwidth,angle=0]{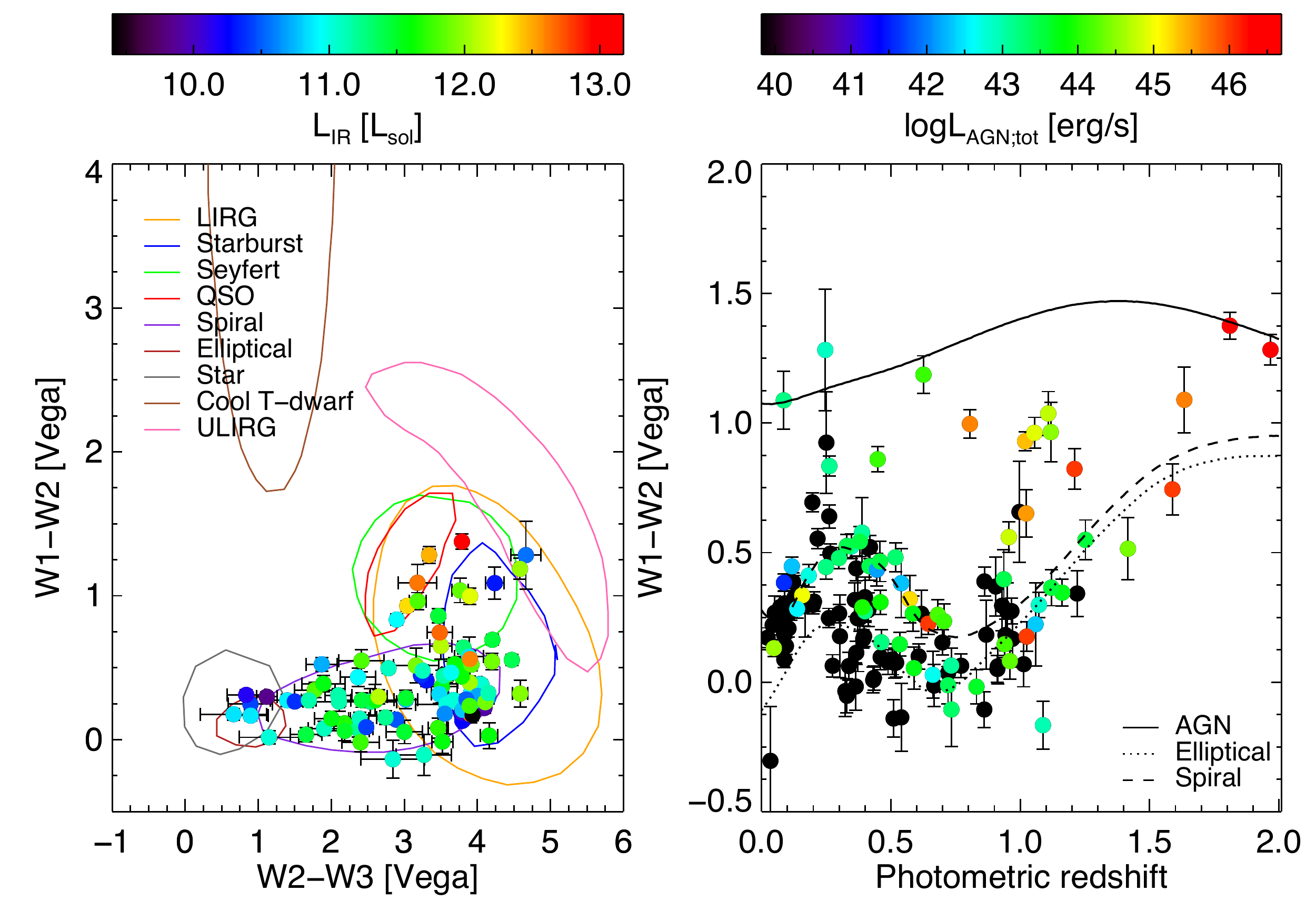}
\caption{Left: W1-W2 color as a function of W2-W3 color for sources detected in all three WISE bands. The color-coding shows the total IR luminosity from the SED fitting. We adopt the same WISE color axis ranges for direct comparison to Fig. 12 of \citet{Wright2010}. We only plot sources with $z\leqslant2$. Right: W1-W2 color as a function of redshift for sources detected in the two WISE bands. The color scale shows the total AGN luminosity in logarithmic scale. The curves show the color evolution for three different types of objects, spiral (dashed), elliptical (dotted), and AGN galaxies (solid). They are reproduced from the templates of \citet{Assef2010}.}
\label{fig:WISEcolors}
\end{center}
\end{figure}

We are interested in comparing our sample with other AGN selected in the IR. To this end, we use the near- to mid-IR colors of our sources from WISE. In Fig. \ref{fig:WISEcolors} (left panel) we plot a color-color diagram utilizing the WISE detected sources in our sample. Different regions of the (W1-W2) vs. (W2-W3) color-color plot have been found to be populated by different classes of objects (e.g., \citealt{Wright2010}; shown with differently colored contours in Fig. \ref{fig:WISEcolors}). We see that our most IR luminous sources are found within the QSO and ULIRG regions, however avoiding the ULIRG colors. This implies a significant contribution from an AGN component for these sources, resulting in a relatively flatter near-IR spectrum compared to star-formation dominated systems with strong PAH features in the W1 band. Apart from these brightest sources, the rest of the sources cover the whole range of WISE colors for spiral galaxies, with a significant fraction falling within the starburst region. Our sources do not exhibit strong red W1-W2 colors, indicating the absence of heavily extinct objects. It should be noted here that the position of a source on the color-color space is not only related to its SED shape but also to its redshift. This is exhibited in Fig. \ref{fig:WISEcolors} (right).

In the right panel of Fig. \ref{fig:WISEcolors}, we plot the (W1-W2) colors as a function of redshift. We compare these with the expected colors from empirical templates for different classes of objects from \citet{Assef2010}. We find that very few of our objects fall in the color region expected by the AGN template. Most of our sources seem to follow the spiral template at redshifts $\lesssim0.3$, while for higher redshifts points cluster closer to the elliptical template prediction. The AGN bolometric luminosity contribution, as calculated from our SED fits, does show some correlation with the expected magnitudes from the empirical templates. Sources with the most luminous AGN components (shown in yellow to red colors in this figure) appear to deviate on average the most from the spiral/elliptical templates towards the AGN template. However, given that the AGN template plotted in Fig. \ref{fig:WISEcolors} is based on mainly optically selected AGN, with an original assumption of a Type 1 AGN, it is no surprise that our originally infrared/radio-selected sample, mostly consisting of systems that are not AGN-dominated, does not follow the AGN template. This is also in agreement with Fig. \ref{fig:WISEcolors} (left), where only a few sources appear to fall in the QSO region of the diagram. Returning to the color-color plot in the left panel of Fig. \ref{fig:WISEcolors}, we can see that, although both spiral and elliptical galaxies can potentially reach W1-W2 colors of $\sim1$ and a redshift of z=2, most of our sources lie at $z<1.4$, where the distinction between the AGN and non-active galaxies is relatively clear. Therefore we conclude that the effect of the redshift range of the sample should not affect the distribution of sources on the WISE color-color plot significantly.

\section {Star-formation in the host galaxies of radio-AGN}
\label{sec:SFR}

We now turn to the star-formation in the host galaxies of our IR-radio sources. To calculate the star-formation rate we employ the infrared luminosity-to-SFR relation, as described in \citet{Kennicutt1998}:
$$SFR_{IR}[M_{sol}\,yr^{-1}]\,=\,4.5\,\times\,10^{-44}L_{IR}\,[erg\,s^{-1}],$$
where L$_{IR}$ is usually defined as the integrated luminosity in the wavelength range 8-1000 $\mu$m. It is known that this relation refers to starburst galaxies, for which relative young stellar populations created continuously within the last 10-100 Myr dominate the emission between 10 and 120 $\mu$m. Although this is a theoretically derived relation, assuming a Salpeter initial mass function (IMF), empirical calibrations of the relation between SFRs and L$_{8-1000\mu m}$ exist in the literature. These are usually within $30\%$ of the relation from \citet{Kennicutt1998} and therefore broadly agree with each other. This will be taken into consideration for the following. 

\begin{figure}[htbp]
\begin{center}
\includegraphics[width=0.49\textwidth,angle=0]{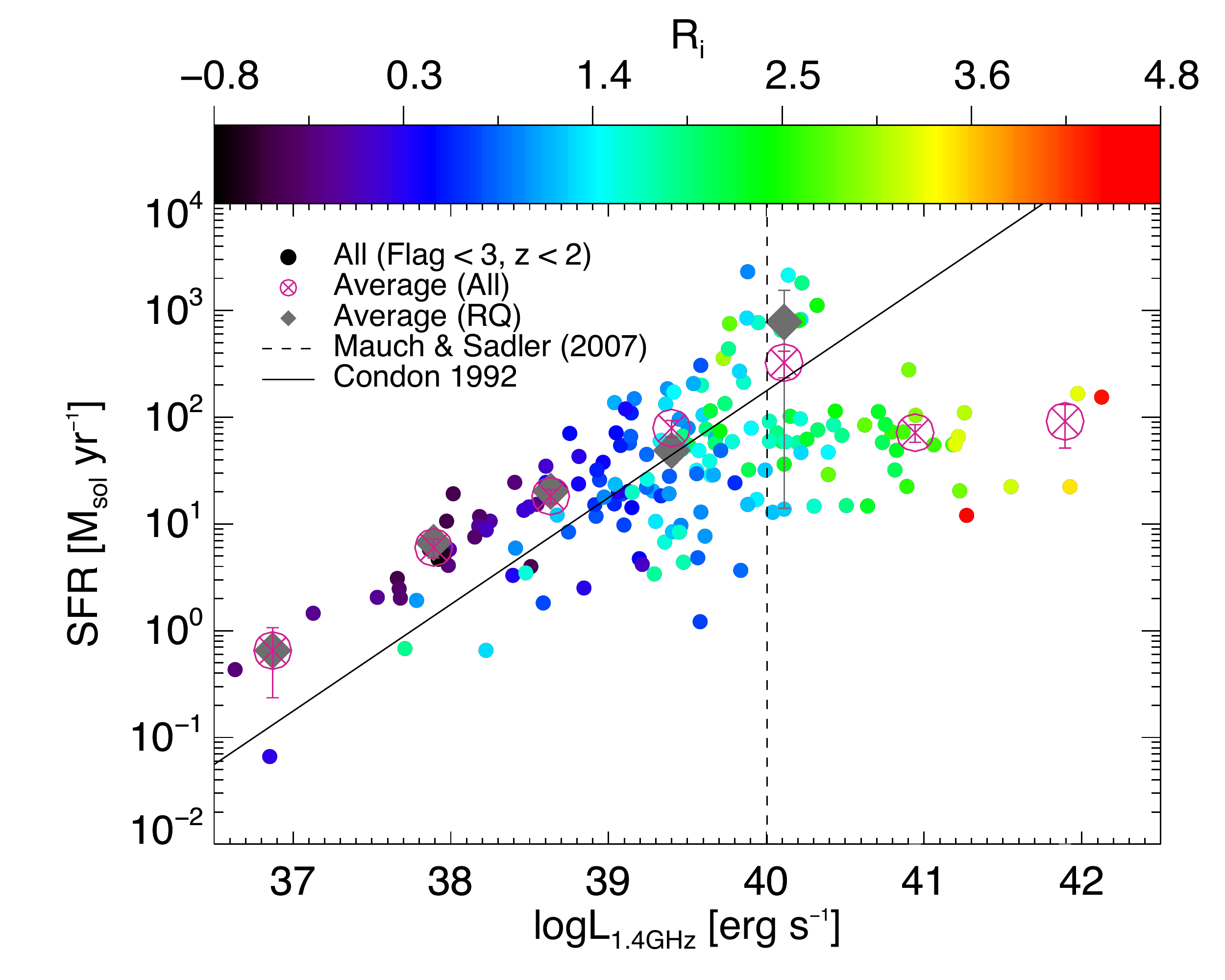}
\caption{Star-formation rate as a function of radio-luminosity at 1.4 GHz (L$_{1.4GHz}$). We use color scale for the radio-loudness of each source, with blue being the most radio-quiet objects and red the most radio-loud. Average values of all sources (crossed-circles) and RQ sources (diamonds) over radio-luminosity bins are also plotted. The solid line denotes the relation between SFR and radio emission from supernovae taken from \citet{Condon1992}. As in previous plots, the vertical dashed lines marks the radio-luminosity limit above which sources are usually classified as radio-AGN.}
\label{fig:SFR_Lr}
\end{center}
\end{figure}

Although the actual fraction is still under debate, it is known that the AGN can significantly contribute to both the mid-IR and, to lesser degree, to the far-IR luminosity of a source. It is therefore crucial to distinguish between the star-formation and the AGN components in our SEDs. Therefore, in order to calculate the SFR for these sources, we first calculate the IR luminosity due to star-formation. This is defined as the sum of the infrared luminosities of the M82, Arp220 and cirrus templates used for the SED fit of each source. Most sources are fitted with either one or the other, with a few using neither one. As was described in Sect. \ref{sec:sed}, in addition to these starburst templates, an AGN torus template and one of a quiescent galaxy are used. The AGN torus luminosity is not considered in the calculation of the SFR. Having calculated SFR for each object we plot these as a function of radio-luminosity in Fig. \ref{fig:SFR_Lr}. In the same figure we also use a color scale to show the radio-loudness of each source. We can observe that radio-faint and radio-quiet sources follow the theoretically expected relation between radio-emission and SFR closely. This relation arises due to the dominating non-thermal synchrotron emission of supernovae within a star-forming galaxy, that peaks at radio wavelengths (e.g.,\citealt{Condon1992}). 

However, as we cross the luminosity boundary above which radio-AGN should dominate, a drop in the SFR is seen with respect to the expected relation. This is due to the increasingly important contribution of AGN radio-jets to the radio luminosity at 1.4 GHz. Sources that fall significantly off the expected relation should be dominated by their radio jets. It is interesting to note that although the SFR appears to taper off at the highest radio-luminosities (with a decreasing tendency), it is still in absolute terms higher than the SFR of pure star-forming galaxies at lower radio-luminosities (and implied lower redshifts). The differentiation between RQ and RL radio-sources is clearly seen through average values of the RQ sources and those of the entire sample.

\begin{figure}[htbp]
\begin{center}
\includegraphics[width=0.49\textwidth,angle=0]{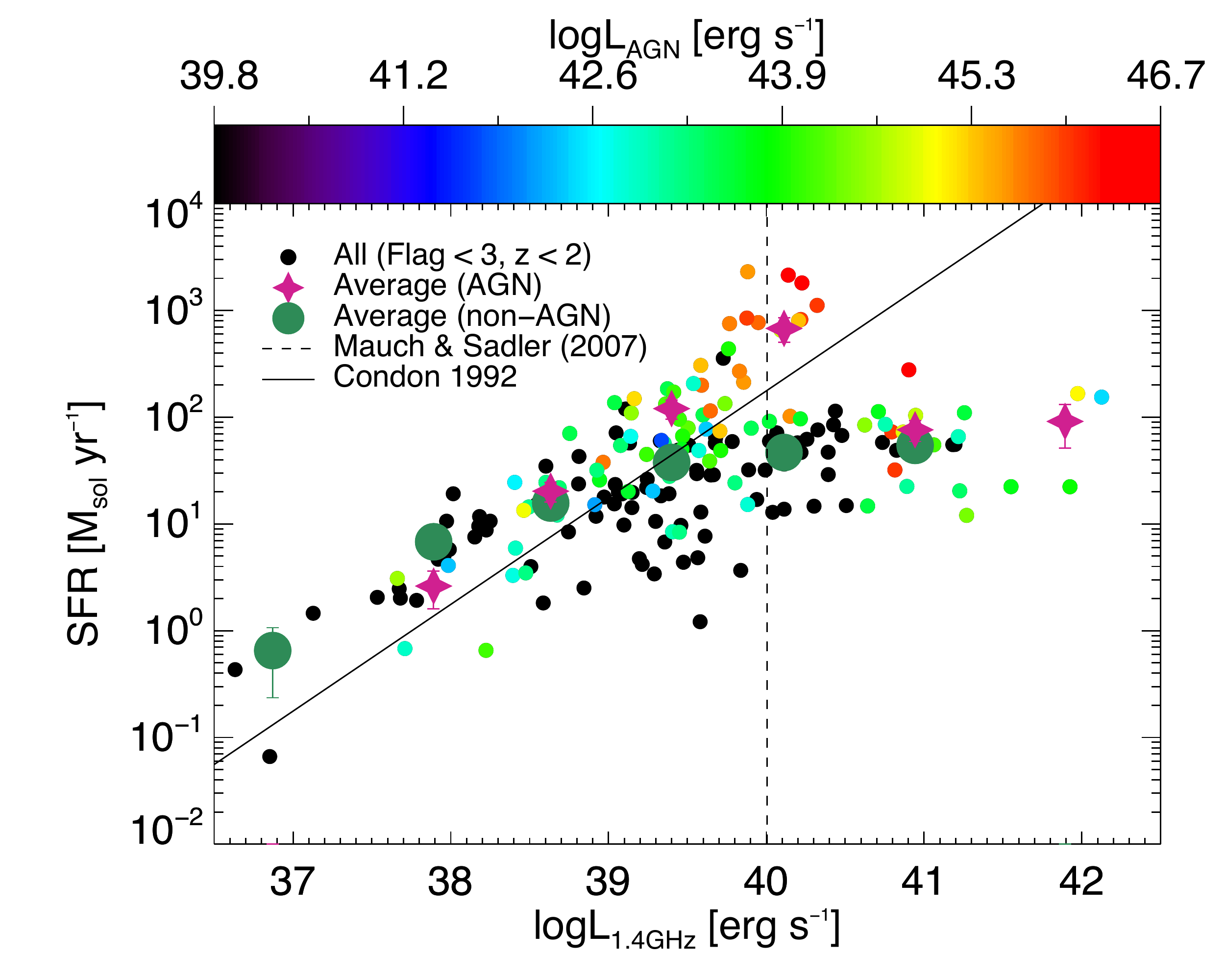}
\caption{As in Fig. \ref{fig:SFR_Lr} but the color scale corresponds to total AGN luminosity from the SED fitting. Dark blue denotes AGN luminosities $\leqslant 39.8$ erg s$^{-1}$ (therefore including sources with zero AGN component). The averages over radio-luminosity bins of sources fitted with an AGN component (magenta stars) and those with no-AGN component (dark green circles) in their best SED fit are also plotted.}
\label{fig:SFR_Lr_LAGN}
\end{center}
\end{figure}

We are now interested in looking at the same relation but now with respect to the actual AGN luminosity, as it is derived from the SED fitting. This is plotted in Fig. \ref{fig:SFR_Lr_LAGN}, where the x- and y-axis are the same as in Fig. \ref{fig:SFR_Lr} but the color scaling now reflects the bolometric luminosity of the AGN (here defined as L$_{AGN;opt}$+L$_{AGN;IR}$). We observe a disconnect between the most radio-loud/radio-luminous sources and those with the most luminous AGN component in their SEDs. Interestingly, sources with the most luminous AGN components also show the highest SFR compared to all other sources. In addition it also appears that at any given radio-luminosity bin AGN-fitted sources show higher SFR than non-AGN ones. Focusing on the higher radio-luminosity end, it becomes apparent that there are sources with considerable radio excess (i.e., their emission is dominated by a radio-jet) that are not picked up as AGN from the SED fitting (these are the sources included in the dark green circle point at L$_{1.4GHz}=10^{41}$ erg s$^{-1}$). These sources therefore exhibit optical and IR emission dominated by their stellar components rather than from their active nucleus. We will return to this point later on.

\begin{figure*}[htbp]
\begin{center}
\includegraphics[width=0.79\textwidth,angle=0]{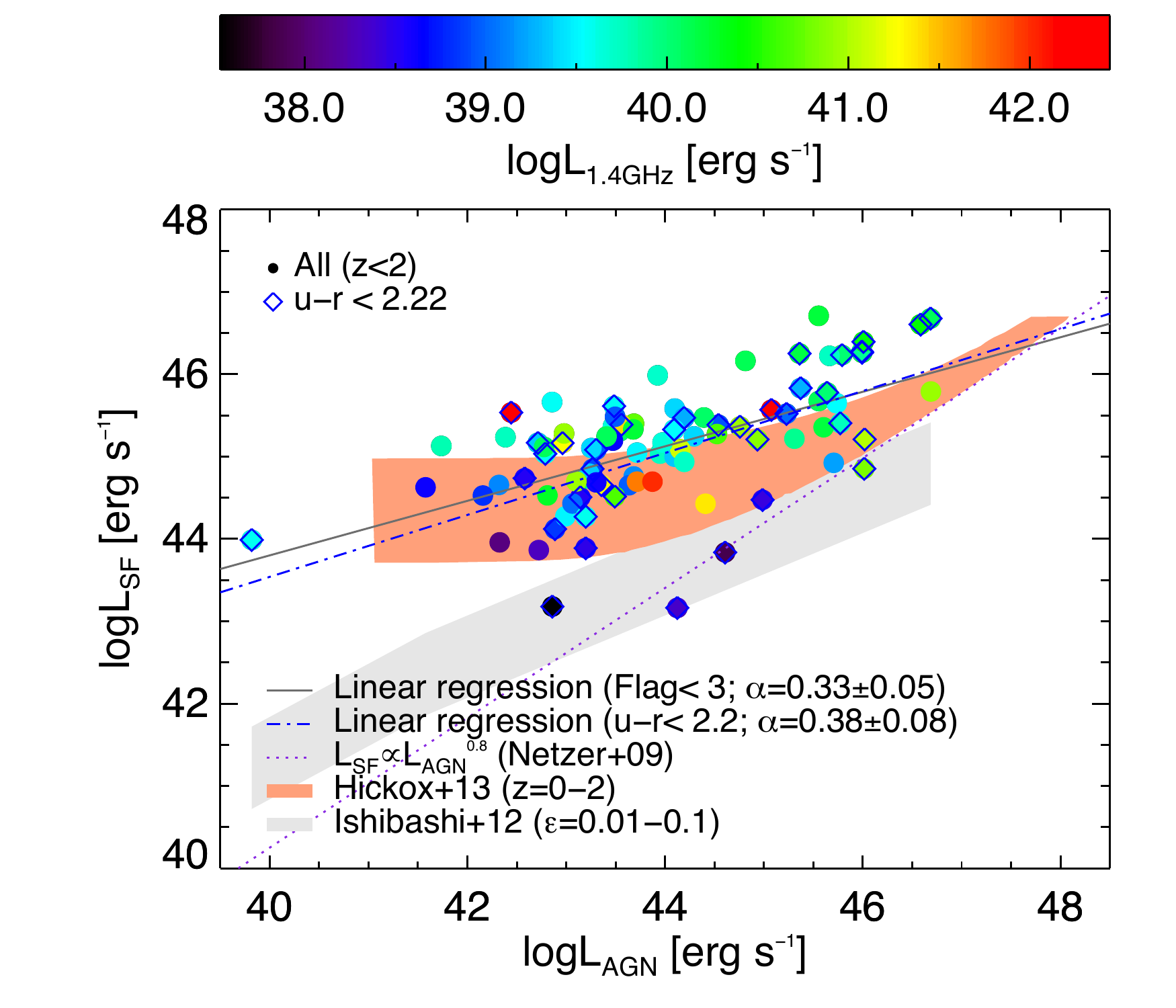}
\caption{IR luminosity from star-formation as a function of the total AGN luminosity (optical + IR). The color scales represents the radio luminosity of the sources at 1.4 GHz. We differentiate between early- (u-r$>2.22$) and late-type galaxies (u-r$<2.22$; marked here with blue diamonds). The solid black line and the dot-dashed blue line are linear fits to all sources and late-type sources, respectively. The purple dotted line represents a slope of 0.8, taken from \citet{Netzer2009}. The salmon shaded locus is taken from Fig. 3 (right panel) of \citet{Hickox2013} and covers a redshift range of 0 to 2 (with higher redshift leading to higher $L_{SF}$). The gray shaded locus is derived from the model of \citet{Ishibashi2012} for a typical galaxy velocity dispersion of 200 km s$^{-1}$, a gas mass fraction $f_{g}=0.16$, and a star-formation efficiency range of 0.01 to 0.1 (with higher SF efficiency leading to higher $L_{SF}$).}
\label{fig:SFR_LAGN}
\end{center}
\end{figure*}

We elaborate further on the apparent very high SFR for the most luminous AGN in our sample in Fig. \ref{fig:SFR_LAGN}, where the IR luminosity from star-formation is plotted versus total AGN luminosity. We see a clear correlation between the two quantities. Sources with higher AGN luminosity also show stronger star-formation. Fitting a line through all the sources we get a slope of $0.33\pm0.05$ for all the sources. If we fit a line only to sources with observed u-r colors typical for late-type galaxies (u-r$<$2.2, e.g., \citealt{Strateva2001}) we get a slope of $0.38\pm0.08$. The Pearson correlation coefficient for all sources and for late-type galaxies are 0.61 and 0.63, respectively. Although not expressed in terms of positive AGN feedback, a positive relation between the two luminosities is put forward by \citet{Netzer2009} through the study of a large sample type 2 AGN. This kind of relation appears to grossly underestimate the star-formation of our sources especially at the lower end of AGN luminosities. A similar, slightly steeper, relation is predicted from \citet{Zubovas2013}, who model the positive feedback of an AGN interacting with and inducing SF in a dense molecular gas disk within which the AGN is embedded. Finally, we also compare our data points with the positive AGN feedback model of \citet{Ishibashi2012}, which again assumes the compression of the ISM from an AGN outflow. In that model the SFR depends on the central AGN luminosity (which drives the outflow), the SF efficiency, and the gas fraction of the galaxy. In Fig. \ref{fig:SFR_Lr_LAGN} we plot the solutions of this model for a range of SF efficiencies (ranging from 0.01 to 0.1, typical of normal SF galaxies and starbursts, respectively). Although at the low-AGN-luminosity end the relation appears flatter than those of both \citet{Netzer2009} and \citet{Zubovas2013}, it still falls significantly below the SFRs inferred from our data. In addition, given that this correlation is recovered for all AGN in our sample (not only the radio-loud ones), we cannot interpret Fig. \ref{fig:SFR_LAGN} as evidence for radio-jet positive feedback.

On the other hand, we see a good agreement with the simple theoretical model from \citet{Hickox2013} (shown within the gray shaded locus for a range of redshift from the local Universe to z$\sim2$) that assumes a one-to-one positive relation between SF and AGN accretion. Indeed our sources, which extend to z$\sim$2 appear to be included within the two lines, with however a significant fraction of sources (and namely those at intermediate radio luminosities) appear to lie above the z$\sim$2 line of the model. It should be noted however that the normalization of this model is done assuming a constant SFR to accretion rate ratio, derived for host galaxies hosting X-ray AGN (e.g., \citealt{Rafferty2011}, \citealt{Mullaney2012}). Whether such a normalization is valid for our sample of radio-selected sources is not obvious.

\section{Stellar masses and specific SFR}
\label{sec:sSFR}

\begin{figure}[htbp]
\begin{center}
\includegraphics[width=0.4\textwidth,angle=0]{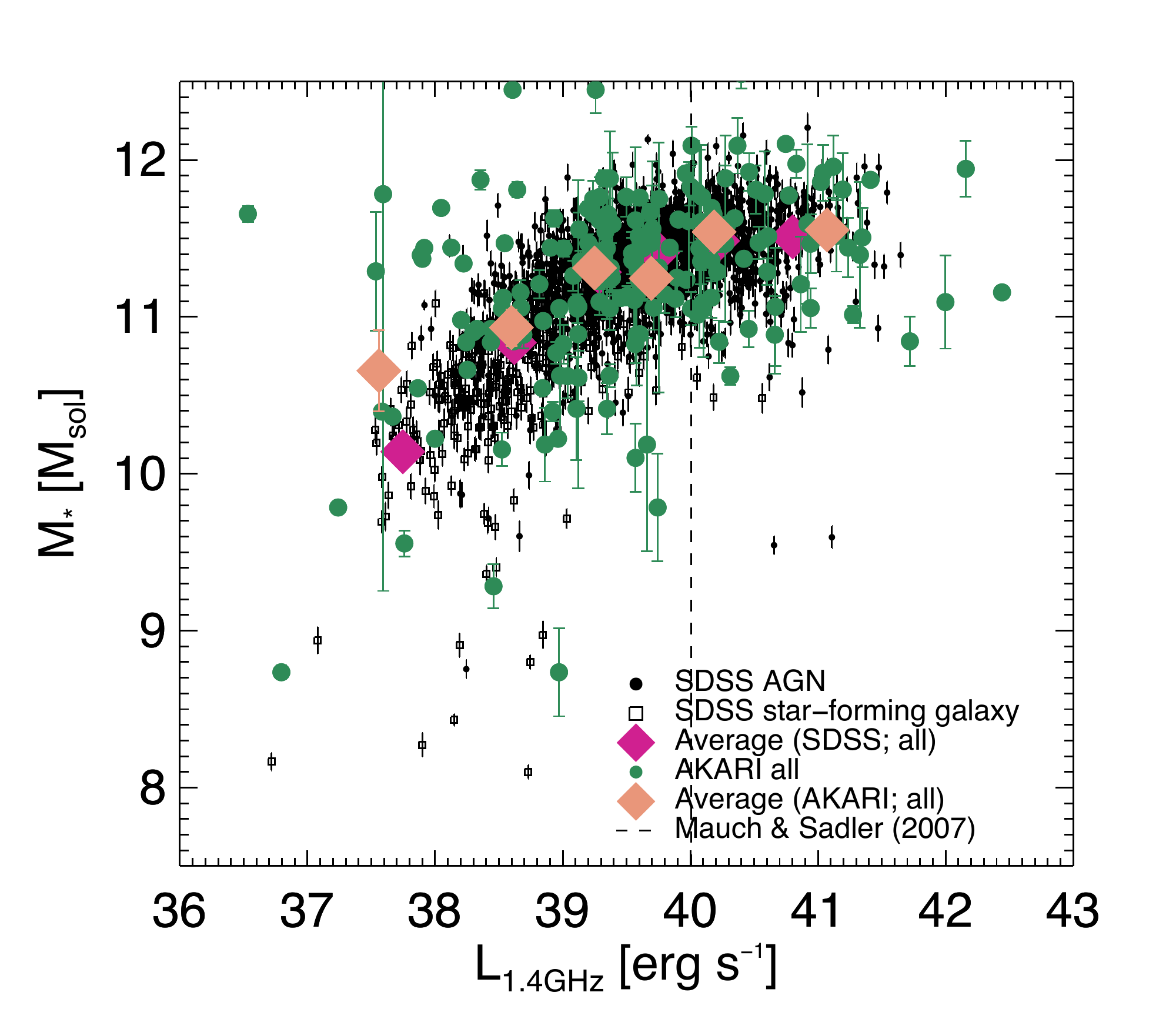}
\caption{Stellar mass as a function of $L_{1.4GHz}$ for the AKARI-WSRT sample (green circles and salmon diamonds) and the SDSS-FIRST sample (black circles and magenta diamonds). Individual values (circles) and averaged values over luminosity bins (diamonds) are plotted. For the SDSS sample star-forming galaxies (squares) and optical emission line AGN (circles) are plotted separately.}
\label{fig:mass}
\end{center}
\end{figure}

It is important to differentiate between the different host galaxies of AGN. In particular it is known that radio-galaxies tend to live in more massive galaxies (e.g., \citealt{Kauffmann2003}, \citealt{Best2005}) than other AGN. Moreover, there is a known dependence between SFR and stellar mass of a galaxy (also known as the main sequence of star-formation), as well as evolution of the SFR per unit mass with cosmic time (e.g., \citealt{Elbaz2011}). We thus need to constrain the stellar masses of our sample. To do this we use the observed SEDs to interpolate the rest-frame luminosity of each source at 2.2 $\mu$m. Old stellar populations emit the bulk of their light at the NIR wavelengths and thus 2.2 $\mu$m rest-frame luminosity should be a good proxy for stellar mass. To do the conversion between the 2.2 $\mu$m luminosity and the stellar mass we use a constant light-to-mass ratio value of 0.85 (for a compilation of references of observationally constrained mass-to-light ratios see, e.g., \citealt{Portinari2004}). In Fig. \ref{fig:mass} we plot the distribution of stellar masses in our sample as a function of L$_{1.4GHz}$. For comparison we also show the stellar masses from the SDSS-FIRST sample from \citet{Best2005} (these are derived from spectral features, e.g., D4000 break, combined with stellar population synthesis models). The stellar mass of our sample are on average similar to those of the SDSS, despite the much brighter optical selection of the SDSS galaxies compared  to the AKARI-WSRT sample. This can in part be attributed to the near-IR selection of our basic sample, that preferentially picks up older, more massive, galaxies. We do recover the positive correlation between the $L_{1.4GHz}$ and $M_{stel}$, in that more radio-luminous sources reside in heavier host galaxies. It is interesting to note that the luminosity distributions of the two samples are almost the same. The AKARI-WSRT sample extends out to high redshifts while probing a relatively faint radio population. On the other hand the SDSS-FIRST sample probes a significant fraction of the entire sky in a very shallow manner, recovering very bright radio-sources at relatively low redshifts.

\begin{figure}[htbp]
\begin{center}
\includegraphics[width=0.49\textwidth,angle=0]{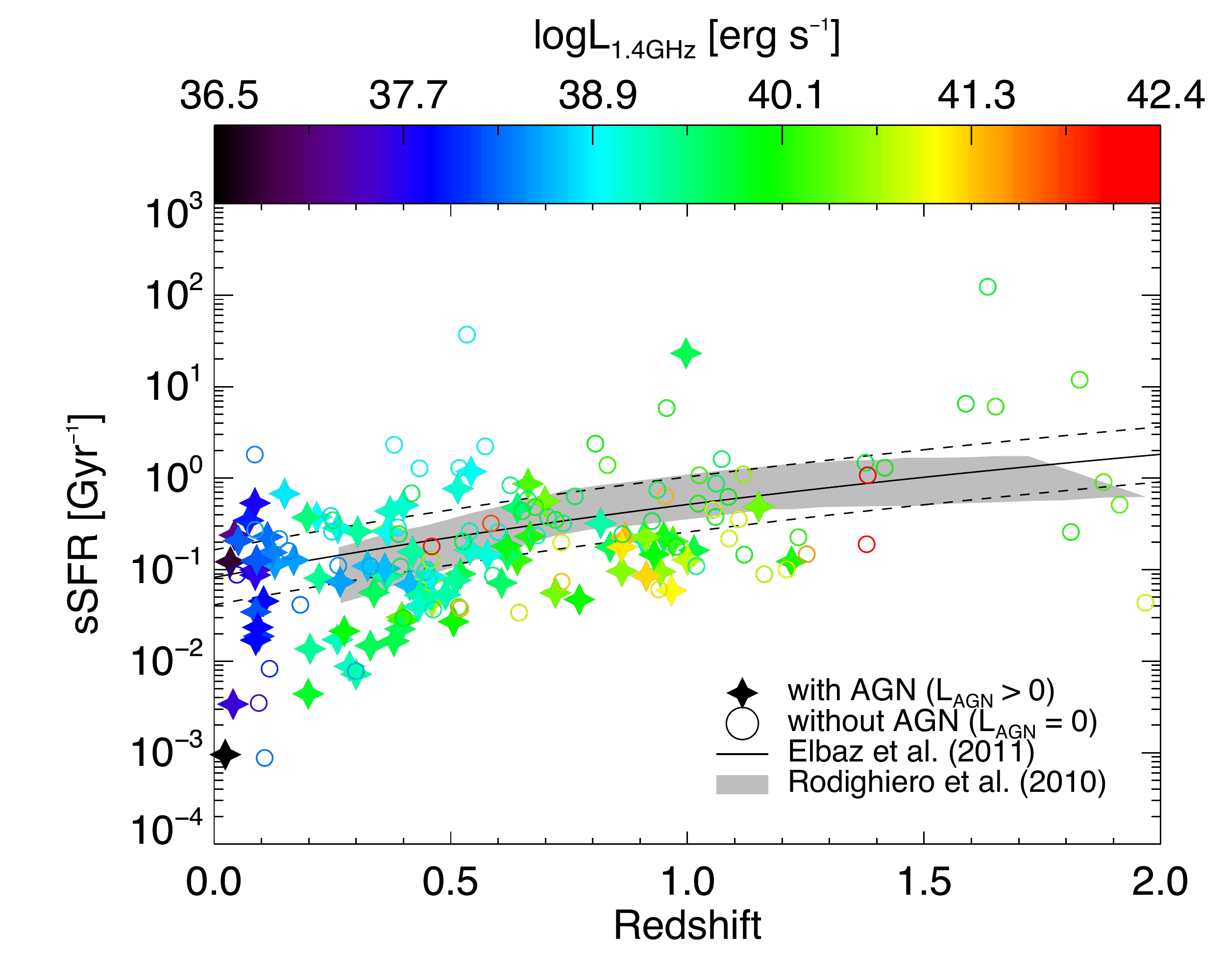}
\caption{Specific SFR as a function of redshift for the AKARI-WSRT sources. The solid line shows the main sequence of star-forming galaxies as defined by \citet{Elbaz2011}, while the dashed lines show the 3$\sigma$ offset from that relation. The shaded gray locus shows the same relation as defined by \citet{Rodighiero2010b}. The color-scale follows the radio luminosity at 1.4 GHz. AGN-fitted and non-AGN fitted sources are shown with filled stars and open circles, respectively.}
\label{fig:sSFR_Lr}
\end{center}
\end{figure}

The specific SFR (sSFR) describes the SFR per unit of stellar mass and is mass-independent measure of star-formation activity in galaxies. We can calculate this measure of the sources in our sample. As a result we plot in Fig. \ref{fig:sSFR_Lr} the sSFR of the AKARI-WSRT sources as a function of their redshift. In effect we are placing our sample on the main sequence of star-forming galaxies as defined by \citet{Elbaz2011}. We see that for the local Universe (z$<0.4$) there is a considerable scatter about the main sequence with both vigorous star-forming galaxies, as well as ``dead'' galaxies present. This trend continues at higher redshifts, with however the observational effect of our flux-limited surveys leading to sources with the lowest sSFR being missed at redshifts above 0.5. Finally, there is evidence for more radio-luminous sources lying closer to and below the passive-galaxy regime than less radio-luminous ones. 

\begin{figure*}[htbp]
\begin{center}
\includegraphics[width=0.8\textwidth,angle=0]{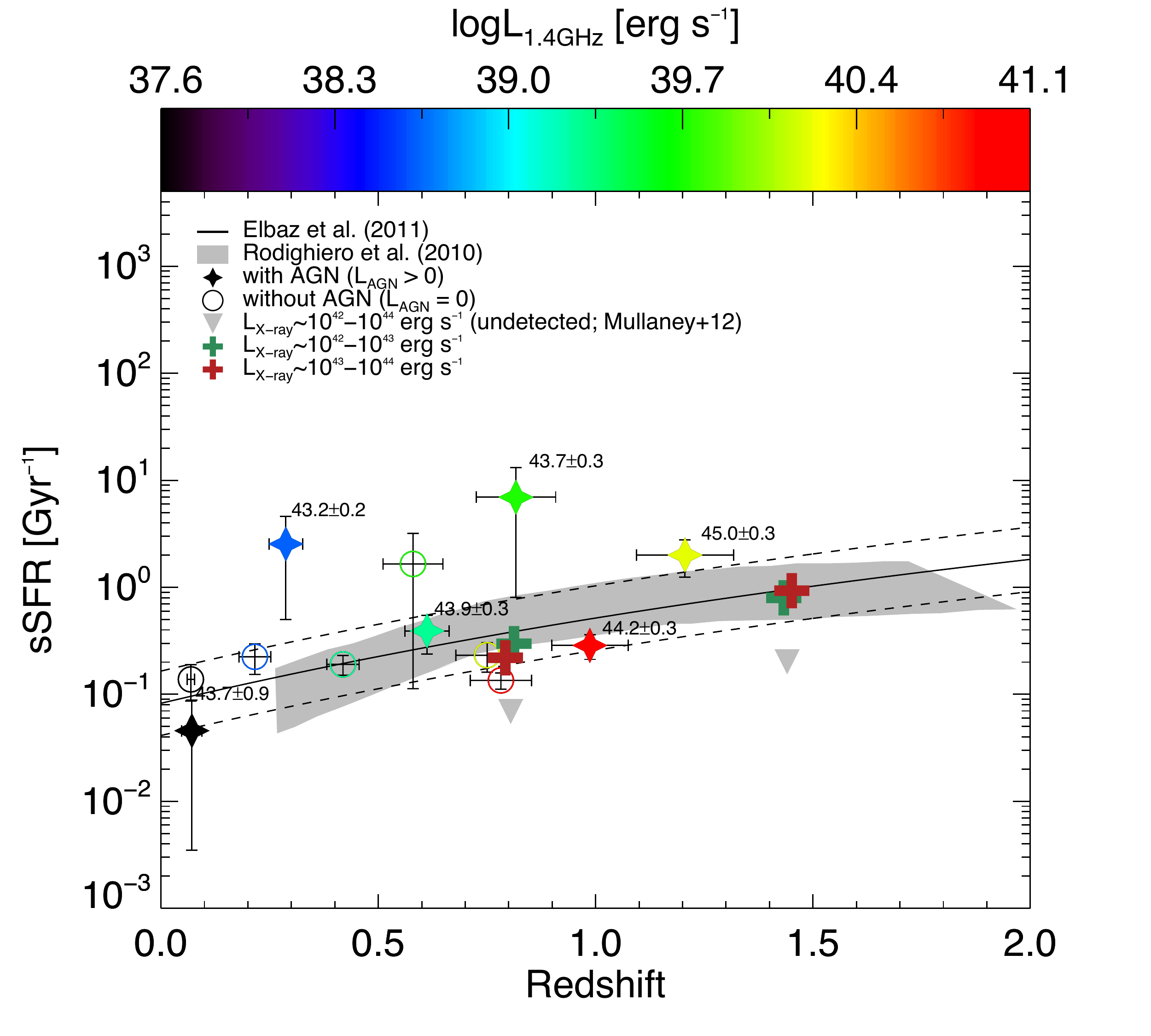}
\caption{As in Fig. \ref{fig:sSFR_Lr} but for average values over radio luminosity at 1.4 GHz instead of individual sources. We separate again sources with an AGN component in their SED fit (filled stars) from sources with no AGN component (open circles). Next to each star the average value of the AGN-component luminosity within that radio-luminosity bin is given. For comparison we also plot the average values of sSFR for the X-ray selected AGN sample of \citet{Mullaney2012} in the different X-ray luminosity bins. The star-formation ``Main Sequency'' from \citet{Rodighiero2010b} is also shown (gray shaded locus, decreasing sSFR for increasing stellar mass).}
\label{fig:sSFR_Lr_avg}
\end{center}
\end{figure*}

In Fig. \ref{fig:sSFR_Lr_avg} we again plot the sSFR as a function of redshift, this time averaging over radio-luminosity bins. In addition, for each radio-luminosity bins we separate sources with and without an AGN component in their best-fit SEDs. Thus we want to investigate both the effect on star-formation of the presence of an AGN and that of the presence of a radio-jet. Two observations can be directly made from this figure. For a given radio-luminosity bin, sources with an AGN component appear to have higher sSFR than those with no AGN component \footnote{It should be reminded here that the SFR and as a result the sSFR are calculated from the IR luminosity after subtracting the contamination from the AGN in the IR wavelengths.}. This trend appears reversed only in the first radio-luminosity bin, which also has the lower average redshift. This is probably related to the 4 points in Fig. \ref{fig:SFR_LAGN} with relatively lower SFRs than the implied trend, given their AGN luminosities. However, as the mean redshifts for AGN sources are higher than non-AGN ones, an increased sSFR due to redshift evolution is expected. Even taking into account the uncertainties in sSFR, redshift evolution can not account fully for the difference between the AGN and non-AGN samples.

The second observation is with respect to a fixed AGN luminosity. Namely, we see that for sources with AGN component in their best-fit SED, increasing radio-luminosity leads to a decrease of the sSFR at radio-luminosities higher than $10^{40}$ erg s$^{-1}$ (which is shown with light green color in Fig. \ref{fig:sSFR_Lr_avg}). As can be seen when comparing the green, yellow, and red diamonds, even though these sources have on average similar total AGN luminosities, at increasing radio-luminosities their sSFR decreases. Conversely, comparing between the black, blue, and teal squares, we observe an increase of sSFR with increasing radio-luminosity. One last remark about this plot pertains to the lowest-redshift bin,at around z=0.1. This is the only bin where sources fitted with AGN show on average lower sSFR than their non-AGN counterparts. This might indicate a change in the SFR properties in the host galaxies of AGN or AGN-composite systems in the nearby Universe.

\section{Discussion}
\label{sec:discuss}

Let us summarize briefly the main findings before discussing further.
\begin{itemize}
\item Our SED fitting can constrain the presence of nuclear activity, showing that there is an excess of AGN with increasing radio and IR luminosity, confirming the results of previous studies (Figs. \ref{fig:AGN_Lr}, \ref{fig:AGN_Ri}, and \ref{fig:AGN_LIR}). 
\item Most of the sources in our sample exhibit colors typical of late-type galaxies, with some pure AGN and early-type galaxies thrown in the mix. Given our selection criteria, we conclude that most of our sources are composite systems which are not dominated by their nuclear emission (Fig. \ref{fig:WISEcolors}).
\item We have shown that there is a strong positive correlation between star-formation IR luminosity and AGN luminosity, with the most luminous AGN in our sample exhibiting the highest SFR among the entire sample (Figs. \ref{fig:SFR_Lr_LAGN} and \ref{fig:SFR_LAGN}).
\item On average, when looking at sources of similar bolometric AGN luminosity and radio luminosities around and above $\sim10^{40}$ erg s$^{-1}$, those with stronger radio emission show significantly lower specific star-formation. In particular, radio-moderate AGN appear to inhabit starburst like galaxies at redshift of around $z\sim1$, while at the same redshift, radio-luminous AGN show specific star-formation typical of normal star-forming galaxies at that redshift (Fig. \ref{fig:sSFR_Lr_avg}).
\end{itemize}

\subsection{Comparison with Other Studies}
 Of particular interest is the study of \citet{Zinn2013}, as it can be more easily compared to our results. With a similar wavelength coverage as ours, the authors additionally use X-ray data to select their AGN sources. They find that sources with radio-jets show higher SFR compared to those without, even at fixed X-ray luminosities. In addition, they show a positive correlation between the jet power and the derived SFR. These results appear potentially at odds with our own. It should be noted however that \citet{Zinn2013} ``radio+X-ray'' sample on which the comparison relies lie at 1.4 GHz luminosities around $10^{24}-10^{26}$ W Hz$^{-1}$ (which translates to around $10^{41}-10^{43}$ erg s$^{-1}$), that is at the end of and beyond our luminosity coverage. In addition, the X-ray selection points towards higher accretion rates and effectively higher AGN bolometric luminosities. This would lead to higher SFR (as implied by our Fig. \ref{fig:SFR_LAGN}). Lacking an estimation of stellar masses combined with possible redshift effects between the different samples in their comparison (e.g., their X-ray and radio samples show markedly different redshift distributions) could lead to the differences seen in terms of SFR between X-ray AGN with and without a radio-jet. As a final side note, \citet{Zinn2013} use a single black body fit to the Herschel data to derive SFR for their sources, in comparison we employ the superior wavelength coverage of the AKARI IRC, combined with Herschel data for individual detections, to disentangle the AGN and star-formation contributions in each source.
 
On a more tangential direction, the morphologies, environments, and merger rates of radio-AGN have been studied extensively. Although not directly comparable to our work, these can give us some hints as to what processes might be relevant to radio-AGN. A high incidence of merger-induced morphological distortions have been found in the brightest radio-AGN (e.g., \citealt{Karouzos2010}, \citealt{Almeida2011}, \citealt{Almeida2012}, \citealt{Sabater2013}, although also see \citealt{Wen2012}). Moreover there are hints for radio-AGN to reside in denser environments than other ``flavors'' of AGN or non-active galaxies (e.g., \citealt{Best2004}, \citealt{Tasse2008}, \citealt{Bradshaw2011}, \citealt{Lietzen2011}, Karouzos et al. 2013 subm.). For relatively lower radio-luminosity sources, clear-cut results are not readily available. Combined with the star-formation studies mentioned previously, there is an emerging picture of radio-AGN being triggered in part by mergers, while secular processes should also play an important role (e.g., \citealt{Tadhunter2011}). Telling are also the timescales involved with these processes. Radio-jet activity is characterized by relatively short timescales ($\sim10^{6-7}$ yrs, e.g., \citealt{Tadhunter2012}) compared to the longer star-formation and dynamical relaxation timescales connected to merger events ($\sim10^{9}$ yrs, e.g., \citealt{Lotz2008}).

Shifting to a different wavelength regime, X-ray selected AGN have also been extensively studied in terms of their star-formation properties (e.g.,  \citealt{Silverman2009},  \citealt{Trichas2009}, \citealt{Mullaney2012}, \citealt{Rovilos2012}, \citealt{Harrison2012}, \citealt{Page2012}, \citealt{Trichas2012}, \citealt{Rosario2012}, \citealt{Zinn2013}) with a wide array of outcomes. Several studies claim that higher X-ray luminosity AGN appear to suppress star-formation in their host galaxies. In particular the results of \citet{Mullaney2012} are over-plotted in our Fig. \ref{fig:sSFR_Lr_avg}. It is clear that X-ray selected AGN show significantly lower sSFR than our intermediate radio-luminosity sources. However, when compared with our most radio-luminous sources, both samples appear to be fairly consistent with normally star-forming galaxies at their respective redshifts. Taking however into account the different stellar mass distribution for X-ray and radio-selected AGN, radio-AGN would have higher absolute SFR compared to X-ray selected ones. This sheds more light on the differences with \citet{Zinn2013} which we discussed above. At higher redshifts the situation is somewhat reversed, with several studies claiming systems with strong starburst activity (usually nuclear) that is coeval with a heavily obscured, potentially Compton-thick, active nucleus buried within (e.g., \citealt{Treister2009}, \citealt{Treister2011}). Comparing our results with those from \citet{Rosario2012} (e.g., Figs. 4 and 5 in their paper) we see a qualitative agreement, in that both samples appear to over-shoot the relation between $L_{SF}$ and $L_{AGN}$ found by, e.g., \citet{Netzer2009}. Higher luminosity AGN show a tighter correlation with $L_{SF}$ than less luminous ones. Different from these authors results, we find that our intermediate radio-luminosity AGN at \textit{z}$\sim$1 show sSFR significantly in excess of the expected sSFRs of inactive galaxies at these redshifts. Given the very different selection of AGN in our study, this is probably not surprising. 

In the above context, our results agree with the general picture presented in the literature in that within our radio sample, sources with high-luminosity AGN components exhibit higher absolute star-formation rates. At the same time however, we find that our most radio-dominated AGN show lower sSFRs, although still in excess of X-ray selected AGN in the same redshift range. In absence of deep X-ray observations over the entire NEP field, we can not say more. Given the rich dataset available for the NEP, a future deep X-ray survey of the NEP field would provide significant insight to this problem.

\subsection{Causality vs. Coincidence} 
 What is perhaps unique in our study is that we are apparently tracing two different and potentially competing mechanisms, as seen in contrast in Figs. \ref{fig:SFR_LAGN} and \ref{fig:sSFR_Lr_avg}. We will address these two separately first.\\
We have to stress that the positive correlation seen in Fig. \ref{fig:SFR_LAGN} although indicative of a relation between the two components, it does not tell us whether this relation is causal, and if so in which direction. A coincident relation would simply mean that there is a third mechanism that is not seen here that induces both star-formation and activity in the nucleus of a galaxy (e.g., a major merger, e.g., \citealt{Hopkins2006}). In such a case although the two activities have matching or most plausibly overlapping timescales, they do not affect each other, at least to a zeroth approximation. Even more worryingly, there is an ongoing debate as to whether selection effects could actually lead to this kind of correlation. Given the current ``shallowness'' of FIR surveys (compared to optical/NIR surveys), it is plausible that there exist, yet undetected, AGN with very low SFR that would occupy the lower right corner of Fig. \ref{fig:SFR_LAGN} and would effectively destroy the implied correlation (e.g., Matsuoka et al., in prep.). The alternative is a causal connection. In that case, one possibility is that the star-formation (which itself has been triggered by other circumstances) effectively drives matter towards the center of the galaxy creating the right conditions for accretion to be triggered and an active galaxy to emerge. Feeding of a black hole through nuclear star-formation (usually in the form of supernovae explosions and/or winds at the later stages of stellar evolution) has been studied and there is a line of evidence that supports it (e.g., \citealt{Kawakatu2008}, \citealt{Chen2009}, \citealt{Hopkins2010}), including correlations as the one shown in Fig. \ref{fig:SFR_LAGN}. Alternatively, a positive correlation between AGN and star-formation luminosity implies that outflows driven by the central AGN create the right conditions for new stars to be formed. These scenarios are usually associated with collimated outflows, of the form of radio jets, rather than broad, uncollimated outflows, also named as AGN winds (e.g, \citealt{Gaibler2012}, \citealt{Zubovas2013}). In addition, most of the proposed models require some form of a gas-rich disky structure with which the jet is expected to interact, shock, and compress, in order for star-formation to be induced. The correlation seen for our sources does not appear to agree with such a model, especially since no particular trend with radio-luminosity is seen in Fig. \ref{fig:SFR_LAGN}.

Let us now turn to the anti-correlation showcased in Fig. \ref{fig:sSFR_Lr_avg}. Feedback from AGN has been predominantly discussed in terms of a negative rather than a positive feedback. From a theoretical aspect negative AGN feedback has been discussed in the context of the accretion rate of the AGN. As such, negative feedback can be split into two main ``regimes'', ``QSO-mode'' (e.g.,\citealt{Hopkins2010}) and ``radio-mode'' (e.g., \citealt{Best2006}) . As the names imply, the former is relevant for high accretion-rate objects, namely quasars, while the latter becomes relevant for inefficient accretors, much like radio-AGN. QSO-mode feedback employs both radiation pressure from a SMBH accreting close to its Eddington limit and QSO winds, which are launched from the surface or the corona of the accretion disk and can reach deep into the host galaxy (given sufficient column density). The radio-mode feedback is driven by radio jets and relies on their kinetic energy which is deposited into the ISM and can potentially launch molecular outflows that might remove part of the available gas from a galaxy (an extensive review of these processes is by \citealt{Fabian2012}). These modes are in turn connected to the two modes of AGN accretion, as put forward by, e.g., \citet{Hardcastle2007}.

The effect seen in Fig. \ref{fig:sSFR_Lr_avg} lends support to the negative, ``radio-mode'' or ``maintenance-mode'' feedback (e.g., \citealt{Kormendy2013}, Section 8 and 8.4 in particular). We can see that for a given AGN luminosity and within a relatively narrow redshift range ($0.9<z<1.4$), increasing radio luminosity leads to progressively lower sSFR. Given the data available to us for these sources, we are not in a position to calculate accretion rates and therefore classify them as high- or low-excitation radio-AGN. However, Figs. \ref{fig:WISEcolors} and \ref{fig:AGN_Lr} imply that we are dealing with composite AGN systems with moderate AGN luminosities rather than purely AGN-dominated galaxies. This is corroborated by the SED shapes of our sources, which in the optical/near-IR appear predominantly dominated by the stellar component, rather than exhibiting a pure power-law SED shape.

\subsection{Drawbacks and Caveats}
In this section we want to briefly outline potential issues that might deduct from the robustness of our results and discuss the degree they may affect our work. These can be split into three wide categories:
\subsubsection{Selection Effects}
Our basic selection is done in the near-IR and the radio. We start from a complete radio sample (down to a flux of $\sim100\mu Jy$), doing an additional near-IR selection, by cross-matching with the AKARI catalogue. By definition radio selection picks up a combination of star-forming galaxies, starbursts, and radio AGN. Given the high sensitivity and relatively small area coverage, we do not expect a large number of very radio-luminous AGN-dominated systems to be included in our sample. At 1.4 GHz radio emission is equally coming from star-forming galaxies and AGN radio-jets. Therefore one of the main issues that we are facing is the accurate identification and separation of star-formation and nuclear activity. This is exhibited in Fig. \ref{fig:SFR_Lr} where there appear to exist ``non-AGN'' sources, according to the SED fitting, with radio luminosities characteristic of radio-AGN. We therefore expect a degree of contamination in the sense of missed AGN. As a result, e.g., in Fig. \ref{fig:sSFR_Lr_avg} the ``non-AGN'' points (shown with squares), especially at the two highest radio luminosity bins would be somewhat contaminated by AGN sources which are missed from our SED fitting. Given the trend observed between sSFR and radio-luminosity, such a contamination would work to decrease the apparent negative-feedback effect. As such we expect that the difference between the green and red diamonds in Fig. \ref{fig:sSFR_Lr_avg} might actually be even larger, with the red diamond being shifted well within the ``Main sequence'' after taking the missed AGN into account.

A second selection bias which needs to be considered is that induced by our cross-matching with the AKARI near-IR catalogues. By definition, sources below the survey limit of the AKARI-NEPW and -NEPD in the N2 and N3 bands will not be included in these catalogues and their radio counterparts will not be cross-matched and therefore excluded from the analysis. This population of very faint near-IR sources likely represent less massive galaxies and/or high redshift sources. The former are likely to be star-forming galaxies (since local Universe radio-AGN reside in the massive end of the galaxy mass function) and their exclusion therefore should not impact our results. The latter would be a combination of starburst galaxies and intermediate power radio-AGN. At lower stellar masses, these radio-AGN would exhibit on average lower levels of star-formation (given the known star-formation
stellar mass relation) and therefore their inclusion would likely push our higher radio-luminosity bins towards lower average star-formation and specific star-formation rates.

A third selection bias concerns the exclusion of radio-loud AGN of type Fanaroff-Riley II (e.g.,\citealt{Urry1995}). These radio-loud AGN are famous for their extended radio-lobes, mainly seen at lower frequencies and at scales reaching out to Mpc. The radio power of these AGN is mainly emitted by the radio-lobes, with the central source, core, being relatively faint or undetected. In such a case our cross-matching with the AKARI catalogue would lead to a non-match. Alternatively, the radio-lobes might themselves be falsely cross-matched with an AKARI source. In the first scenario, the source is completely missed by our study. Given however the trend that we see for strong radio-emitters to quench their SF, the inclusion of these additional sources would make the observed effect stronger. In the latter case, this would introduce false radio sources in our sample, which however, given the radio spectrum of radio-lobes, would not show high radio luminosity at 1.4 GHz. As a final consideration, these sources are prevalent at fluxes above 1 mJy (e.g., \citealt{Windhorst1993}), while the bulk of our sample is found at sub-mJy levels.

\subsubsection{Observational Effects}
Although the NEP field has a wealth of multi-wavelength data, the edges of the NEPW suffer from incomplete and/or shallow optical observations. As a result we were unable to calculate reliable photometric redshifts for a significant fraction of the originally cross-matched radio sources. Although this definitely impacts the completeness and statistical robustness of this study, it does not induce any directional bias to our results. The sources missed through this observational caveat cover a wide range of radio fluxes and optical magnitudes.\\
A second effect pertains to the availability of far-IR data. The entire NEP field was surveyed by the Herschel-SPIRE instrument, however in a shallow manner. As a result, the bulk of the AKARI-WSRT sources remained undetected. As we showed in Fig. \ref{fig:SPIREcomp} the inclusion of far-IR points is crucial both for the correct identification of AGN components and for the accurate determination of the total IR luminosity of a source. Both these quantities are crucial to this analysis. For the SED fitting of sources undetected by SPIRE we used upper limits for each SPIRE band derived according to the average sensitivity of the survey. In this way we can constrain the possible far-IR emission of all sources. Given however the statistical nature of such an upper limit, we expect a certain level of contamination from areas of the sky where data might have been problematic or where confusion did not allows proper extraction of sources. For such cases we expect that we are under-estimating the upper-limits in the far-IR and as a result we are under-estimating the SFR for the sources affected by this. The fraction of non-detection in the far-IR bands increases with increasing radio luminosity. Therefore we expect that sSFR for the two higher radio-luminosity bins might be somewhat under-estimated.

\subsubsection{Analysis Effects}
By definition, SED fitting requires a number of assumptions which by and in themselves affect our results. An often debated aspect of template SED fitting is the choice of templates. It has been shown that the use of excessively many templates leads to degeneracies and affects the results negatively. On the other hand, the exclusion of a certain ``type'' of template could lead to strong directional biases in the analysis. The templates used here are drawn from \citet{Rowan2008} and have been sequentially developed and improved over several iterations (\citealt{Rowan2003a}, \citealt{Rowan2004}, \citealt{Babbedge2004}, \citealt{Rowan2005}, \citealt{Babbedge2006}). Although based on empirical templates, they have been reproduced by means of stellar population synthesis modeling and as such have both a physical meaning as well as match observations well. These templates have been used to derive photometric redshifts for the SWIRE fields resulting in a redshift accuracy better than 3.5\% and a very low outlier fraction ($\sim 1\%$), exhibiting their high quality. Given our sample selection in the radio, the inclusion of a single elliptical, ``dead'', galaxy might be decreasing our ability to effectively model the SEDs of radio-AGN which predominantly should live in early-type galaxies. However, the WISE colors of our sample, combined with the relatively low to intermediate radio-luminosities, imply that the majority of our radio sources do not actually reside in early-type galaxies.\\
A second point of interest concerns the way we derive stellar masses for our sources. As was already described, we interpolated the observed SEDs to calculate luminosities at 2.2 $\mu$m and then used a constant mass-to-light ratio to transform that luminosity into stellar mass. Several points of concern might arise:
\begin{itemize}
\item at near-IR wavelengths AGN contamination is important for sources with a strong AGN component
\item the mass-to-light ratio has been shown to not be constant and depending on several factors, including star-formation histories and the initial mass function assumed
\item SED interpolation is subject to observational uncertainties and provides a rough estimate of the K-correction term needed for accurate rest-frame emission determination
\end{itemize}
As was mentioned previously, most of our sources do not show a dominant power-law in their optical/near-IR SEDs. Instead for most cases the 1.6 $\mu$m bump can be identified leading us to conclude that for most sources old stars dominate the near-IR part of the SED. On the other hand, the light-to-mass ratio conversion is indeed highly dependent on the assumptions made and has been shown to vary among morphologically different galaxies. However, given the scope of this study, unless there is systematic difference between AGN and non-AGN and radio-faint and radio-bright sources, this would not lead to a directional bias. Although it is perceivable that there is an early-type late-type difference between bright and faint radio-galaxies, this is not supported by our data and therefore we do not expect this effect to be important. Finally, given our error budget in estimating stellar masses (light-to-mass ratio variation, AGN contamination, photometric uncertainties) we believe uncertainties due to the interpolation to not be dominant. A more physical modeling of the SEDs (using for example stellar synthesis modeling) would alleviate these issues. However currently there is no satisfactory physical SED modeling tool to account for both star-formation and AGN components in a consistent and accurate way. The fact that we recover the relation between stellar masses and radio luminosity gives us confidence about the accuracy of our method.

\section{Conclusions}
\label{sec:conc}
We have used a rich multi-wavelength dataset to construct and model the broadband SEDs of a sample of radio-sources in the NEP field. With the advantage of the excellent wavelength coverage of the AKARI IRC instrument and by fitting each SED with a combination of templates, both of star-forming and nuclear active galaxies, we constrained these two components in each of the radio sources. We studied the relation between star-formation and nuclear activity and in particular looked for evidence supporting either positive or negative feedback in radio-AGN. On face value we have found both. On closer inspection we concluded that the positive correlation between the AGN and star-formation luminosity does not appear to satisfy neither the assumptions nor the expected relation of models of radio-jet induced star-formation. This leads us to believe that such a correlation is either due to the coincidence of these two activities due to a third parameter which is not investigated here, or due to fueling of AGN through stellar feedback. Given the selection, luminosity ranges, and colors of our sources, we speculate that the latter is more probable. Visual inspection of all the optical images of our sample did not reveal, to first order, an excess of disturbed or interacting systems.

On the other hand, the signature of negative feedback appears to be rather clear, and it can be attributed to the presence of a radio jet. We can therefore conclude that we find clear evidence for suppressed star-formation in the host galaxies of radio-luminous AGN. It should be noted however that star-formation is not quenched in these systems. On the contrary, in absolute terms, the SFR exhibited by these radio-loud AGN is comparable if not in excess of lower radio-luminosity (lower redshift) non-AGN sources. As such the host galaxies of these radio-AGN are consistent with normally star-forming galaxies in their respective cosmic epochs. This is in perfect agreement with theoretical expectation of the ``maintenance'' nature of ``radio-mode'' feedback as opposed to the more invasive and radical ``QSO-mode'' responsible for the putative transformation of sources into the red sequence.

Taking this a step further and combining our results with the literature we can see an emerging picture of galaxy evolution where nuclear activity and star-formation, at least in the circum-nuclear region, is intimately connected. Although the triggering process of AGN is under debate, it is accepted that some gas reservoir is needed for both stars to be formed and SMBH to be fed. As a result of this fundamental underlying connection, we see correlations between AGN and star-formation luminosities. It appears that although for the brightest AGN sources mergers are predominantly important, lower power radio-sources and composite systems like the ones we're studying here are potentially rather associated with secular processes. We have shown that the radio-jet can suppress star-formation in these systems, bringing their host galaxies back on the ``Main Sequence'' of star-formation. At higher bolometric and/or AGN luminosities (perhaps at later evolutionary stages of the sources we study here) efficient and energetic AGN winds and/or radiation pressure from an Eddington-limited accreting SMBH maybe totally quench star-formation, transforming these galaxies into ``red and dead'' quiescent galaxies in the local Universe.

Future instruments like SPICA and the James Webb Space Telescope will be able to cover a much wider parameter space both in terms of sensitivity and wavelength range and will therefore allow the expansion of this or similar studies to both the fainter end of the optical luminosity function as well as the less massive end of the mass function of galaxies. In the more near-term, the acquisition of spectroscopic redshifts for this sample of sources and a full optical coverage of the NEPW area would lead to a clear improvement of the results. In addition to new means of AGN classification, optical spectra will allow the estimation of the accretion rates of these radio-AGN and therefore place constraints on the characteristic properties of the ``radio-mode'' feedback. In a following paper we will present the stacking analysis of the Herschel-SPIRE and PACS data of this radio sample, placing further stringent constraints on the SFR and specific SFR of our sample.

\acknowledgments{This work was supported by the National Research Foundation of Korea (NRF) grant, No. 2008-0060544, funded by the Korea government (MSIP). This research is based on observations with AKARI, a JAXA project with the participation of ESA. This publication makes use of data products from the Wide-field Infrared Survey Explorer, which is a joint project of the University of California, Los Angeles, and the Jet Propulsion Laboratory/California Institute of Technology, funded by the National Aeronautics and Space Administration. This research has made use of NASA's Astrophysics Data System Bibliographic Services. This research has made use of the NASA/ IPAC Infrared Science Archive, which is operated by the Jet Propulsion Laboratory, California Institute of Technology, under contract with the National Aeronautics and Space Administration.}

\clearpage

%% the file, uncomment the \plotone, \plottwo, and
%% If you need a layout that cannot be achieved with \plotone or
%% \plottwo, you can invoke the graphicx package directly with the
%% \includegraphics command or use \plotfiddle. For more information,
%% please see the tutorial on "Using Electronic Art with AASTeX" in the
%% documentation section at the AASTeX Web site,
%% http://www.journals.uchicago.edu/AAS/AASTeX.
%%
%% The examples below also include sample markup for submission of
%% supplemental electronic materials. As always, be sure to check
%% the instructions to authors for the journal you are submitting to
%% for specific submissions guidelines as they vary from

%% This example uses \plotone to include an EPS file scaled to
%% 80% of its natural size with \epsscale. Its caption
%% has been written to indicate that additional figure parts will be
%% available in the electronic journal.

%% Tables may also be prepared as separate files. See the accompanying
%% sample file table.tex for an example of an external table file.
%% To include an external file in your main document, use the \input
%% command. Uncomment the line below to include table.tex in this
%% sample file. (Note that you will need to comment out the \documentclass,
%% \begin{document}, and \end{document} commands from table.tex if you want
%% to include it in this document.)


\bibliographystyle{aa}
\bibliography{bibtex}

\end{document}